\documentclass[preprint,authoryear,12pt]{elsarticle}

\usepackage{amssymb}
\usepackage{graphics}

\begin{document}

\begin{frontmatter}

\title{On a Possible Giant Impact Origin for the
\newline Colorado Plateau}

\author{Xiaolei Zhang}

\address{Department of Physics and Astronomy, George Mason University,
\newline 4400 University Drive, Fairfax, VA 22030, USA
\newline E-mail: xzhang5@gmu.edu}

\begin{abstract}
It is proposed and substantiated that an extraterrestrial object of the
approximate size and mass of Planet Mars, impacting the Earth in grazing
incidence along an approximately N-NE to S-SW route with respect to the 
current orientation of the North America continent, at about 750 million 
years ago (750 Ma), is likely to be the direct cause of a chain of events 
which led to the rifting of the Rodinia supercontinent and the severing
of the foundation of the Colorado Plateau from its surrounding craton. 

It is further argued that the impactor was most likely a rogue exoplanet, 
which originated from one of the past crossings of our Solar System through 
the Galactic spiral arms, during the Sun's orbital motion around the center 
of the Milky Way Galaxy.  New advances in galactic dynamics have 
shown that the sites of galactic spiral arms are locations 
of density-wave collisionless shocks. The perturbations from such
shocks are known to lead to the formation of massive stars, which evolve
quickly and die as supernovae.  The blastwaves from supernova explosions,
in addition to the spiral-arm collisionless shocks themselves,
could perturb the orbits of the streaming disk matter, occasionally
producing rogue exoplanets that can reach the inner confines of our 
Solar System.  The similarity of the period of spiral-arm crossings of 
our Solar System, with the approximate period of major extinction events 
in the Phanerozoic Eon of the Earth's history, as well as with the (half)
period of the supercontinent cycle, indicates that the global environment 
of the Milky Way Galaxy may have played a major role in initiating Earth's 
tectonic activities.
\end{abstract}


\end{frontmatter}

\section{Introduction}

Evolving from the continental drift hypothesis of Wegener (1929) et al.,
plate tectonics is the present working paradigm to explain the 
geomorphology of our Planet Earth (for historical accounts, see, e.g., 
Sullivan 1991; as well as the many contributions in Oreskes 2003).  
The basic scenarios offered by plate tectonics had successfully 
accounted for the formation of mountain ranges at the converging and 
diverging boundaries of continental and oceanic plates. 

Within the plate tectonic paradigm, mantle convection and mantle plume 
scenarios were proposed as the main driving mechanisms for the Earth's 
tectonic movement.  This proposal, however, has trouble to account for 
the sharp change in alignment of the Emperor-Hawaii island chain (one of 
several similarly sharply-bent island chains in the Pacific), which was 
hypothesized to be produced by a stationary hotspot anchored in the mantle, 
coupled with the steady movement of the overlying Pacific Plate (Wilson 1963).  
Mantle convection would have difficulty to explain the stationary 
nature of the hotspot's anchor on the mantle, a required feature if the 
Emperor-Hawaii island chain is produced through a fixed hot spot and 
slowly moving Pacific plate above it.  The convection picture furthermore
cannot explain the sudden change in direction of the emerging
island chain half-way through the Pacific Plate's motion. This sharp-bending 
feature was sometimes proposed to be produced by the collision
of India with the Himalayan region of Asia in the middle of the Cenozoic era.  
If this is true, on the other hand, it in fact highlights the possibility 
that major changes to plate motion can be initiated by kinematic and dynamic 
events at the crustal surface level, rather than at the underlying
mantle level.  For mantle convection cell to be as large as the whole
Pacific Plate is also difficult to conceive, and one also needs a coupling
machanism between the asthenosphere and the lithospherical plates across
the Low-Viscosity-Zone to be able to drive the plate motion even if organized
large-scale mantle convection does exist (Price 2001).

Furthermore, the striking geomorphology of southern Rocky Mountains in
western United States, which is the subject of our current paper, is formed
in an intra-cratonic setting (i.e., the Colorado Plateau
is surrounded on all sides by mountains within a single ancient craton), 
and it cannot be naturally explained through the subductions of oceanic plates,
which are usually shallow-angled and do not lead to the prevailing
high-angle strike-slip faults at the Plateau's boundary (Twento 1980) 
which had incised the Colorado Plateau 
from the surrounding craton.  Our sentiments on this issue echoed
that of many past workers of the Rocky Mountains area:
``Plate-tectonic theorists pondering the Rockies have been more than
a little inconvenienced by the great distances that separate the
mountains from the nearest plate boundaries, where mountains theoretically
are built.  The question to which all other questions lead is, What
could have hit the continent with force enough to drive the
overthrust and cause the foreland mountains to rise?'' 
(McPhee 1998, p. 384); ``The Owl Creek Mountains and the Uinta
Mountains trend east-west.  Why? Why are their axes ninety degrees
from what you would expect if the tectonic force came from the west?''
(David Love, as quoted in McPhee 1998, pp. 385-386).

An alternative driver for plate motion, that of the giant impacts from Solar-System's 
asteroids and comets, had also been explored in the past few decades 
(Price 2001 and the references therein), which has the potential to
explain the sudden change of plate-motion directions, as well as
the intra-cratonic orogenic events.  
One severe limitation of the past impact proposals
is the ultimately achievable magnitude of the impacts, 
if these were produced solely by the objects of the mature Solar System.  
The giant planets of our Solar System had achieved their stable 
orbits when the Earth was still in its youth, and the asteroids
and comets of the Solar System, while possessing potentially Earth-crossing
orbits, are generally small in size and mass, which make it problematic
if we were to hypothesize that both the initial partial loss of the
crust of the Earth (assuming the terrestrial crust was formed full initially,
covering the whole Planet like that on Venus), as well as the subsequent 
cycles of supercontinent formation and dispersal, were caused mainly by 
giant impacts.

An examination into the frequency distribution of major extinction 
events within the Phanerozoic Eon (\S\ref{section4})
reveals furthermore an episodic trend of the most important of 
these extinction events, with the quasi-period of these events (especially
the clustering of major events) similar to that of the Galactic Year of 
250 million years duration.  In section \ref{section4} we will also demonstrate that the
Galactic Year is similar to the period of crossing of the Solar System across
one of the two major spiral arms at the Solar circle, and discuss in 
more detail how this periodic spiral-arm-crossing may facilitate the periodic 
production and invasion of rogue exoplanets into the inner confines of 
our Solar System, contributing both to the supercontinent cycles and to the 
periodic occurrence of major extinction events in the Phanerozoic Eon.
In the current paper we present a first example of the important role
giant impacts may have played in shaping the Earth's tectonic history.

\section{Geological Environment of the Colorado Plateau}

The Colorado Plateau is centered around the Four Corners region (i.e., the 
intersection of the states of Colorado, Utah, Arizona, and New Mexico) of 
the southwestern United States.  It encompasses an area of roughly 
337,000 km$^2$.  It is an uplifted high-desert plateau of a shallow 
bowl shape (i.e., with its rims generally of higher elevation than 
its interior).  The elevation of the Plateau ranges from 1 - 4.7 km, 
with an average elevation about 1.7 km.  

\begin{figure}
\vspace{14cm}
\includegraphics{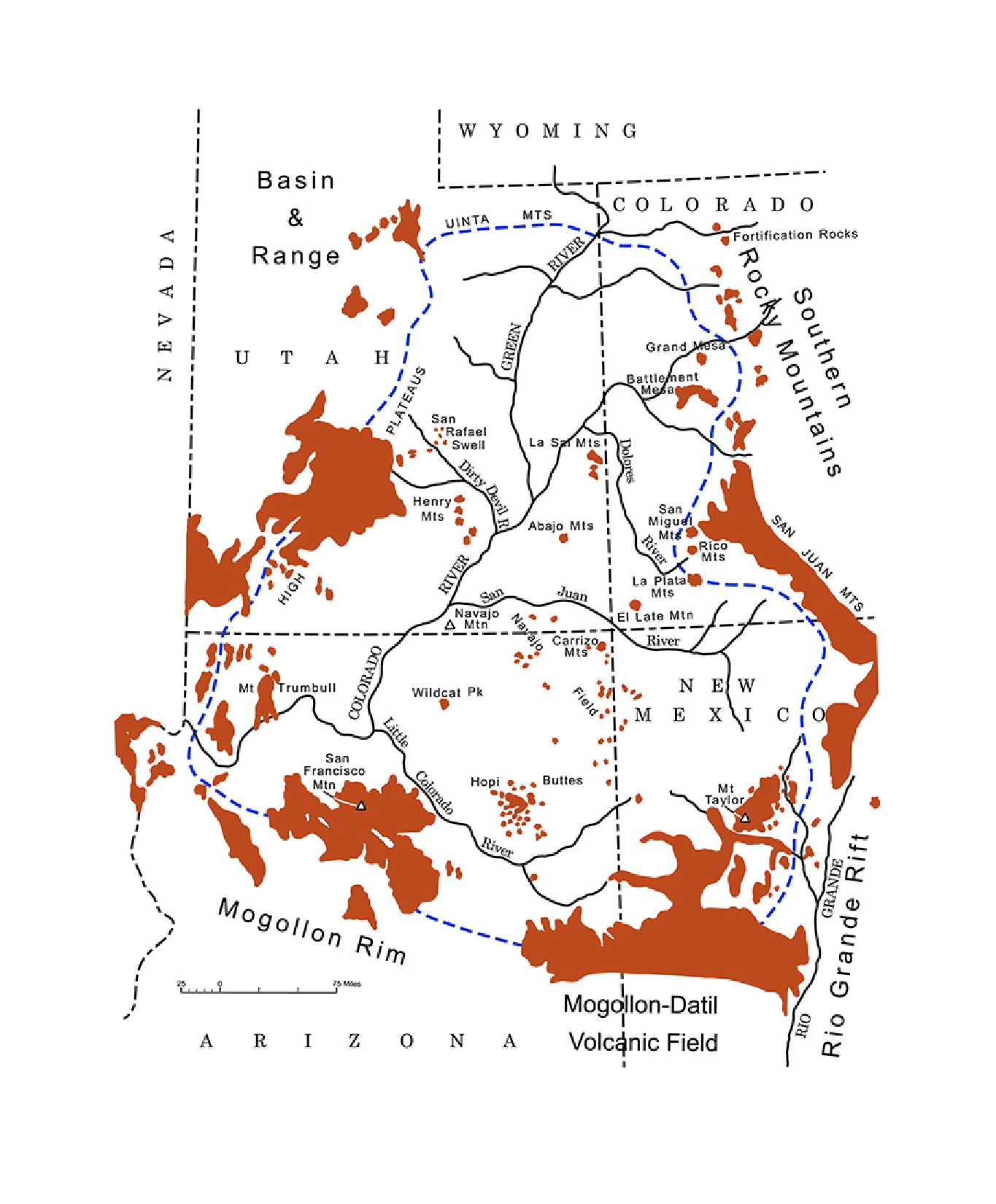}
\caption{The Colorado Plateau, enclosed within the blue dashed line,
and its Cenozoic igneous rocks, in reddish shade.  After Hunt (1956).}
\label{Fig1}
\end{figure}

One distinguishing characteristic of the Colorado Plateau is its structural 
stability (Kelley 1979 and the references therein).  It was shown to have 
been little faulted or folded during the past 600 million years, whereas 
in its surrounding area there were repeated orogenic and igneous activities 
during the Phanerozoic Eon (Figure \ref{Fig1}).  In the west, the Plateau is 
delineated by the Wasatch Line, a fault line which marks one of the innermost 
boundaries along which the Rodinia supercontinent, most recently assembled between 
1.1 - 1 Ga, split circa 750 Ma (this line coincides roughly with the Interstate-15 
freeway from Salt Lake City, Utah, to Las Vegas, Nevada). In one of the 
proposed Rodinia rifting configurations (SWEAT-Configuration), Australia and 
East Antarctica broke off from Rodinia's western side around 750 Ma
(Moores 1991), and this episode is followed by sustained periods of 
miogeocline buildup (Dickinson 1989).  Other proposed rifting configurations
and hinge lines have varying degrees of differences from SWEAT.  The fuzziness 
of the rifting hingeline in some of the proposed rifting configurations is due 
in part to the possible subsequent re-assembly of minor broken pieces along the
resulting supercontinent Laurentia (ancient North America)'s western edge.
Beyond the Wasatch Line to the west is the younger Basin and Range province,
which continues slightly beyond the rest of the three sides of the Plateau 
as well.  In the east side, the Plateau boundary is marked by the Southern 
Rocky Mountains in Colorado, and by the onset of the Rio Grande Rift Valley in 
New Mexico.  Its southern boundary is delineated by the Mogollon Rim in Arizona 
and the Mogollon-Datil Volcanic Field in New Mexico.  
Its northern boundary is lined by the Uinta Mountains in Utah, which merges
into the northern portion of the Southern Rocky Mountains.  Part of the
orogenic features of the Plateau's northern boundary in fact continues
further north to the Central Rocky Mountains in Wyoming, Idaho, and Montana.
The Grand Canyon of the Colorado River is located on the Plateau's
western side within Arizona.  The Black Canyon of the Gunnison River lies in
the eastern side of the Plateau within Colorado.

\begin{figure}
\vspace{17cm}
\includegraphics{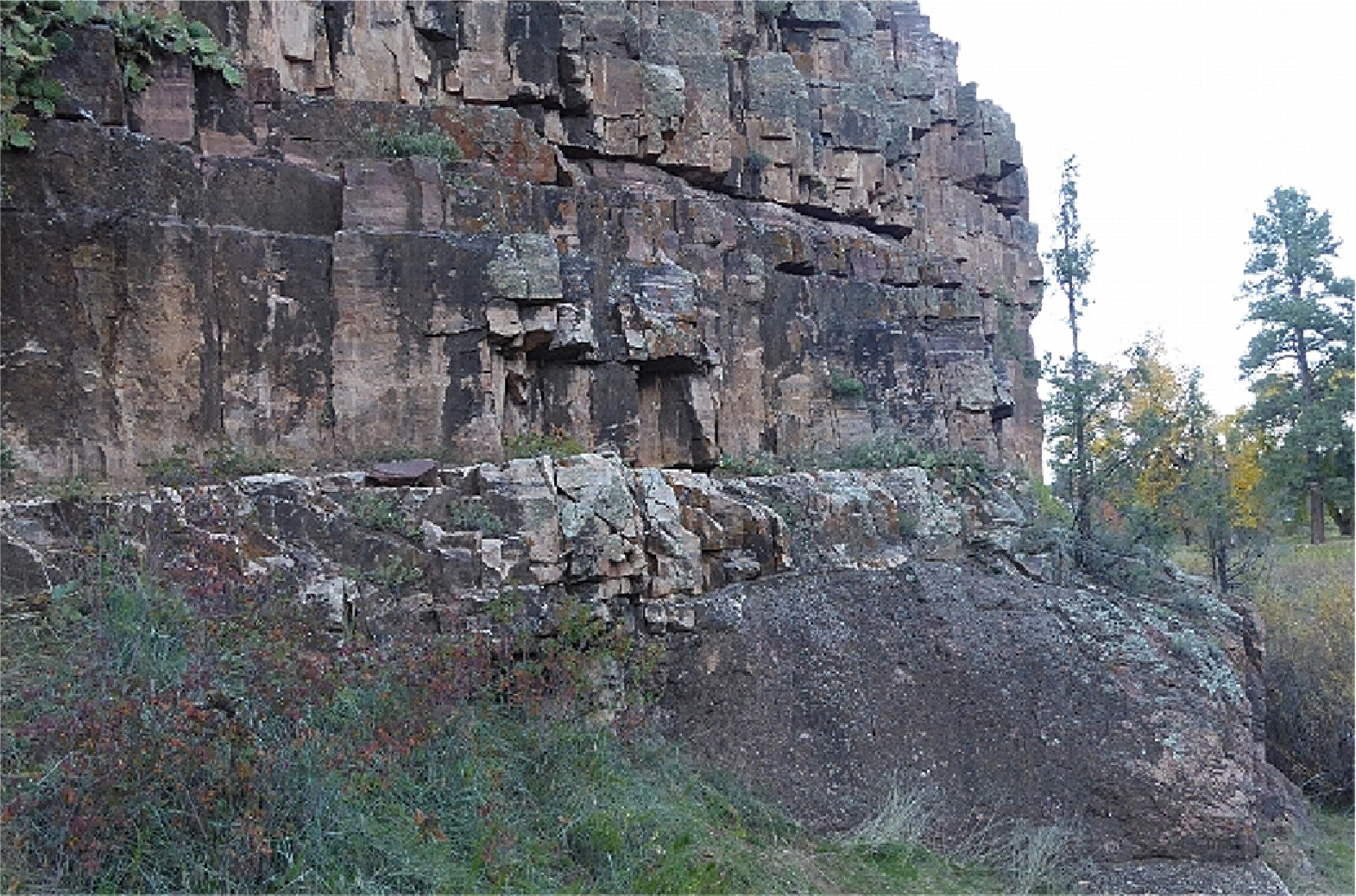}
\includegraphics{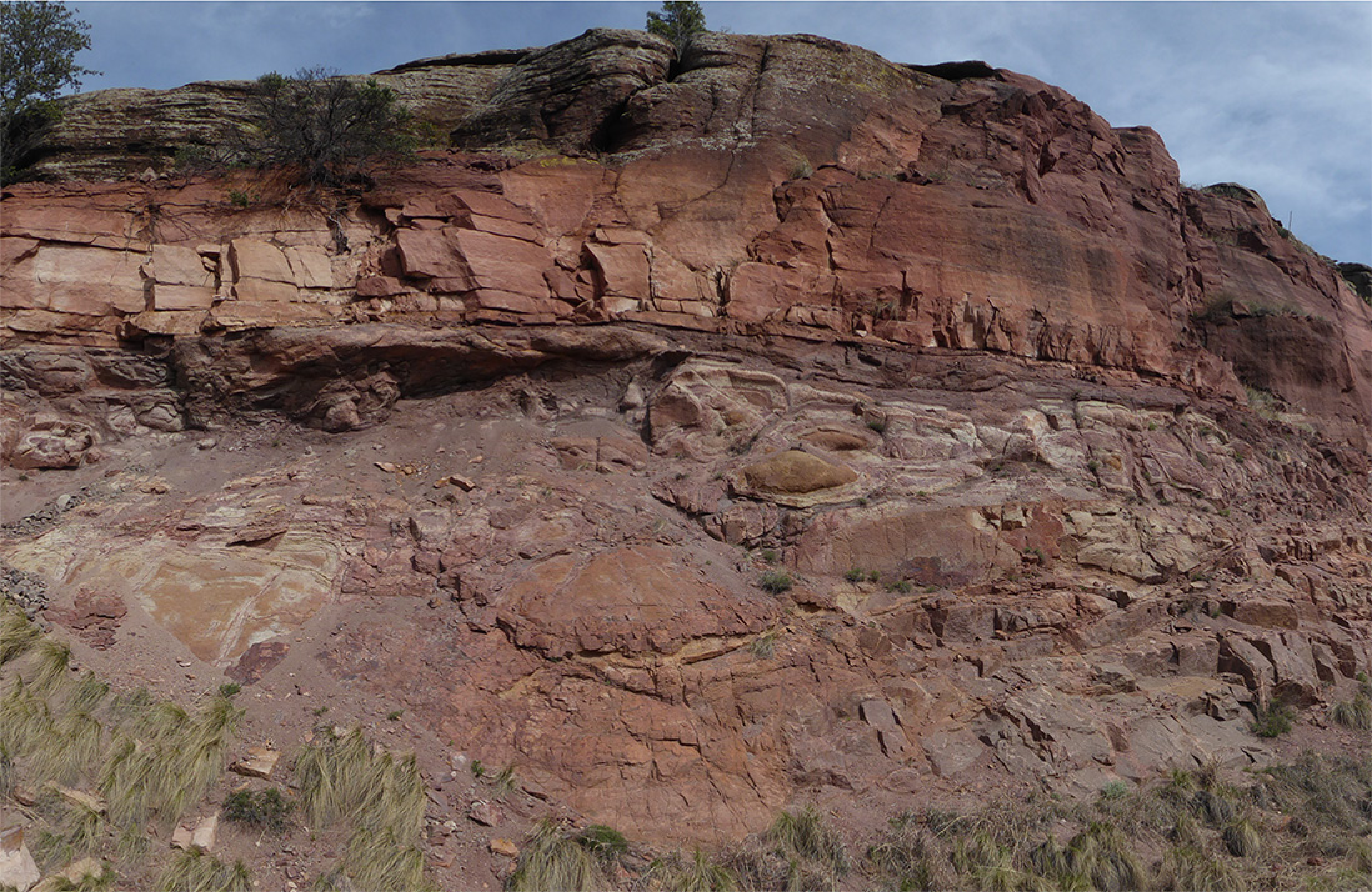}
\caption{Great Unconformity between Proterozoic and Paleozoic formations.
Top: Baker's Bridge, Colorado.  Bottom: Along Arizona state road AZ-87.}
\label{Fig2}
\end{figure}

\begin{figure}
\vspace{14cm}
\includegraphics{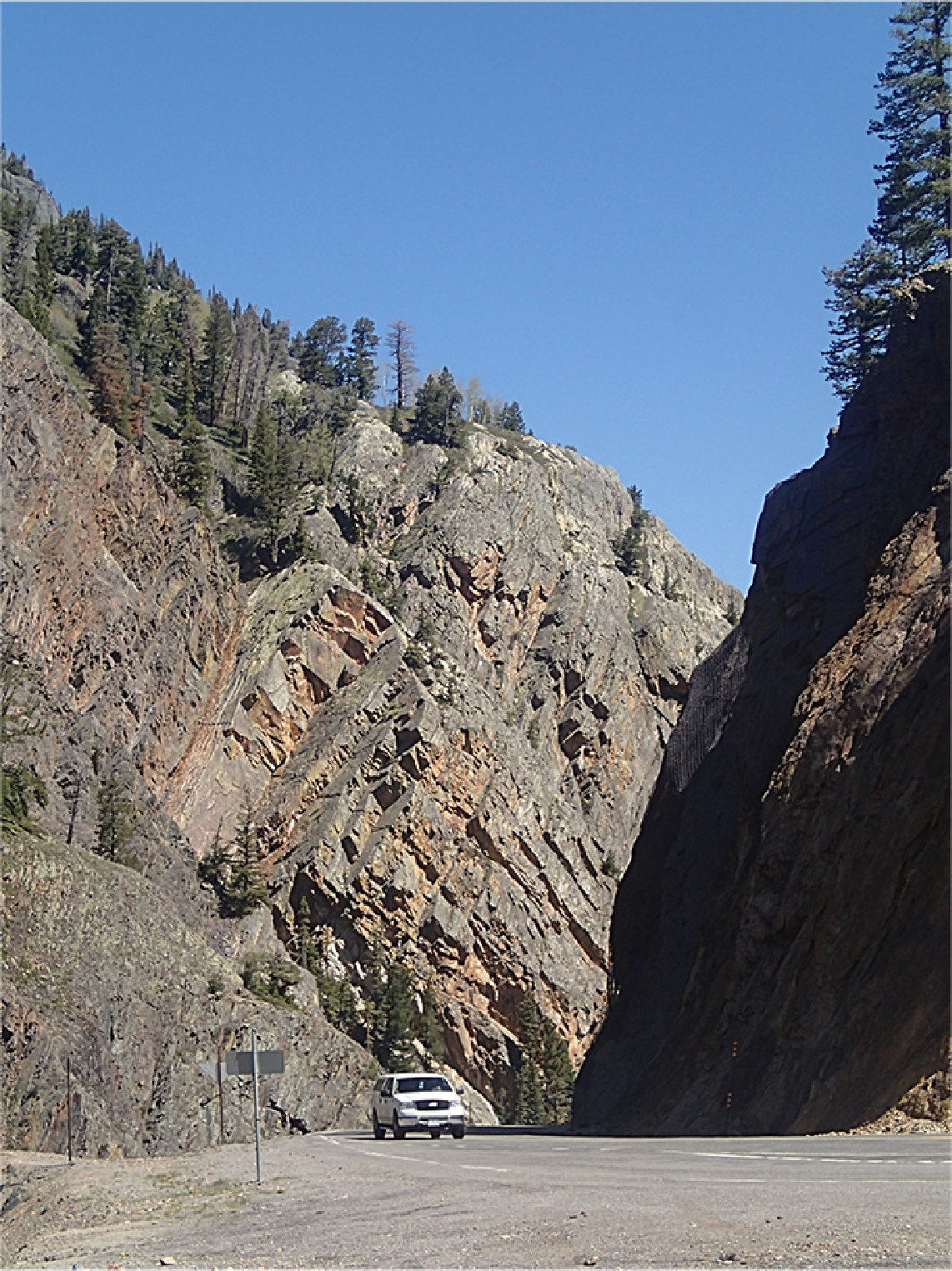}
\caption{Tilted rocks of Uncompahgre Formation along US-550
in the western San Juan Mountains (part of the Southern Rocky Mountains).}
\label{Fig3}
\end{figure}

\begin{figure}
\vspace{18cm}
\includegraphics{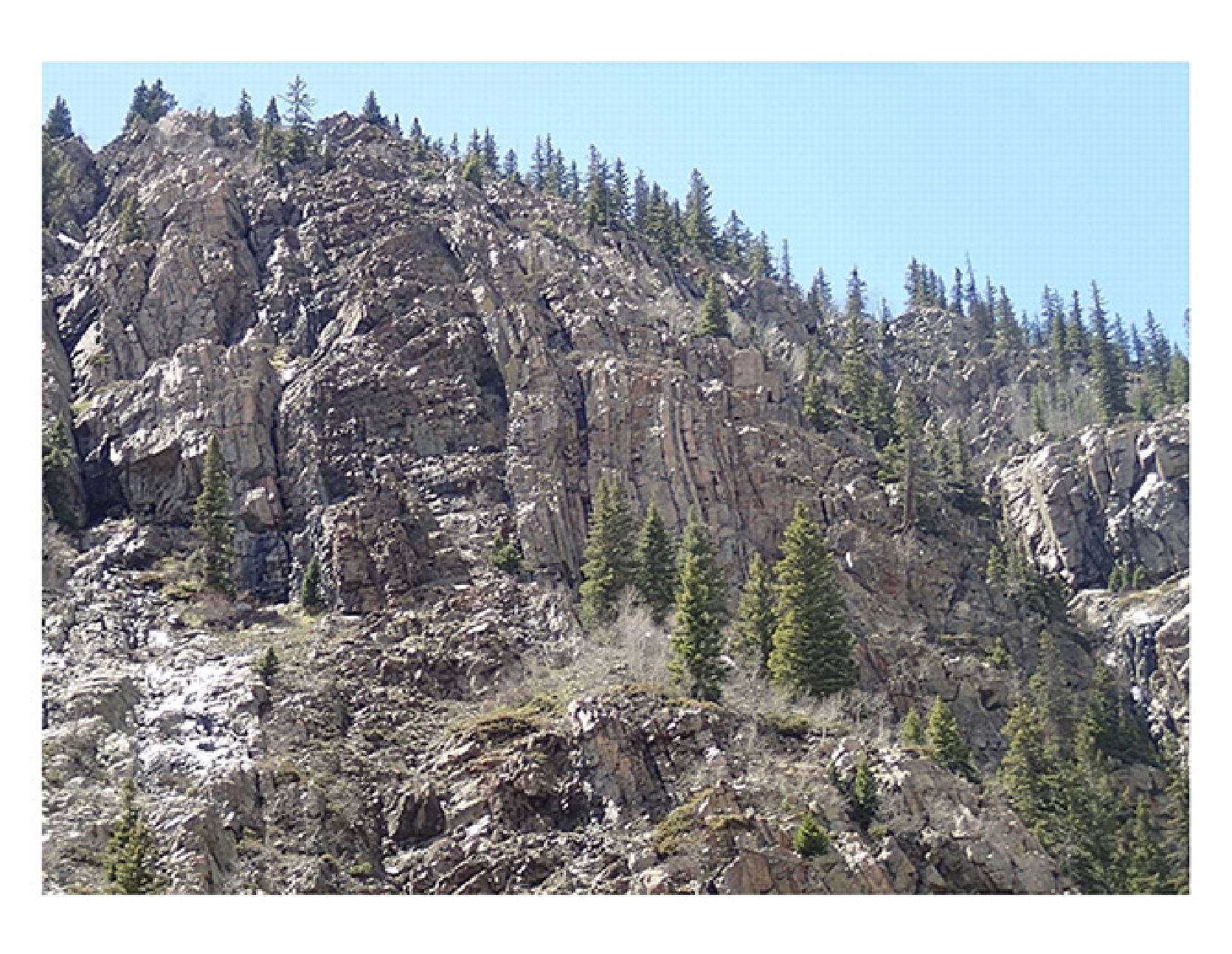}
\includegraphics{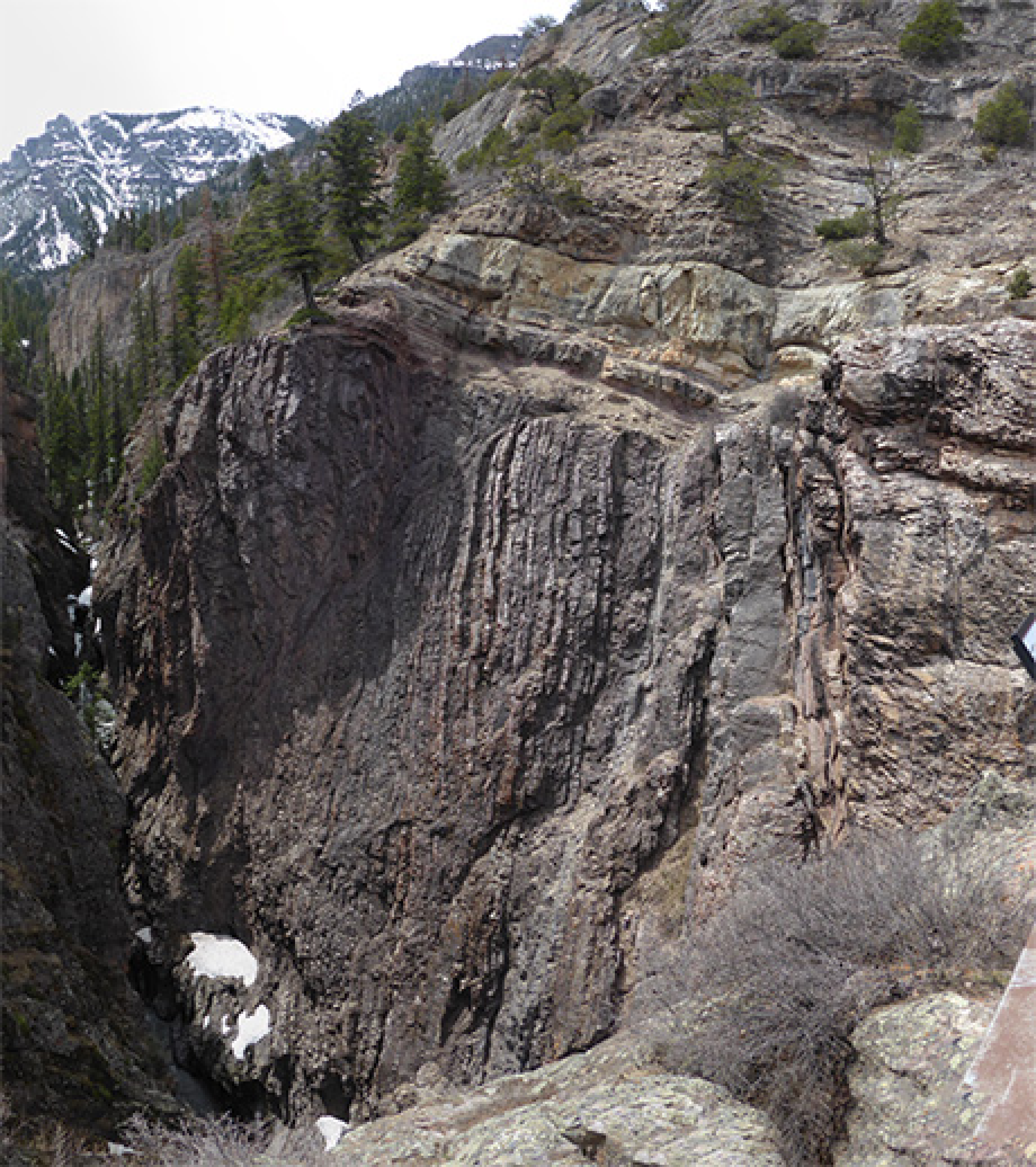}
\caption{Angular unconformities in the Uncompahgre Formation.
Top: South of Silverton, Colorado,
along the route of Durango-Silverton Narrow Gauge Railroad. 
Bottom: Along the side of Uncompahgre Gorge
within the Box Canyon State Park, Colorado.}
\label{Fig4}
\end{figure}

Additional geological features of the Plateau and its environment
which motivate our giant-impact hypothesis include:

\begin{enumerate}
\item The Colorado Plateau is roughly centered on a set of perpendicular wrench (transform) 
faults (Baars and Stevenson 1981) of cross-continent sizes.  Though the initiation of segments
of these faults could be as early as 1.7 Gyr (Baars 2000), the proposed 750 Ma impact 
event, which we will detail later in this paper, likely had enlarged/reactivated these
faults.  The Plateau's boundaries were delineated by high-angle transform faults originating
from the Neoproterozoic era, which are likely to be responsible for the initial severing of 
the entire Plateau crustal block from its surrounding North America craton, but which did 
not lead to immediate uplift of the Plateau in the Precambrian time due (likely)
to the confining pressure from the same surrounding craton.  Many of these 
high-angle faults did not evolve to become thrust faults until the Laramide Orogeny 
of Late Cretaceous/early Tertiary time (Kelly and Clinton 1960). The uplifting of the Colorado 
Plateau itself occurred mainly in the Cenozoic, and the cause of which could be
the combined effects of Laramide Orogeny, the Basin-and-Range extension 
dynamics, and the shallow-angle subduction of the Farallon Plate from the west, 
all of which are capable of re-activating the Proterozoic high-angle faults surrounding 
the Plateau boundary, and, as we will discussed later in the paper, all of which could 
themselves be the consequences of other giant impact events.

\item There exists the so-called Great Unconformity in stratigraphy across the Colorado Plateau, 
reflecting missing sediments of up to 1.2 Gyr, between 1700 Ma and 540 Ma.  The most well-known 
display of this Great Unconformity is in the Inner Gorge of the Grand Canyon of the 
Colorado River (see, for example, Hamblin 2008), but its presence is wide-spread across the entire 
Plateau region, and exposures of the Great Unconformity can be found in especially the fissured and
faulted areas, both along deep canyons as well as along the Plateau's boundary regions. 
Figure \ref{Fig2}, Top Frame, shows the Baker's Bridge Unconformity 
near Durango, Colorado, which is just beyond the Plateau's eastern boundary. 
Figure \ref{Fig2}, Bottom Frame, shows another example of the Great Unconformity
along Arizona's state highway 87.  The Great Unconformity is also present in other parts of 
the US (i.e. Wyoming and New York State), as well as in 
the rest of the world (i.e., Canada, Ireland, and Africa). 
See further discussions in Van Hise (1909/2017); Ward (2001); Share (2012); 
Peters and Gaines (2012); and the references therein. 

\item
Across the Colorado Plateau there exist surface and subsurface detritus 
material (boulders, cobbles, pebbles and re-lithified crushed rocks) which were derived from
either local or nonlocal Precambrian rocks (Barker 1969; Karlstrom 1989; Gonzales et al. 1994;
Condon 1995; and the references therein).  Figure \ref{Fig2}, for example, displays
such re-lithified crushed basement rocks just above the Great Unconformity interface.
In this case the source material for re-lithification was derived locally, and it
clearly shows evidence of having been through partial melting and vitrification.
The basement material below the Great Unconformity, on the other hand, also show evidence
of melting and flow.

\item In the Grand Canyon region, a segment of the so-called Grand Canyon 
Supergroup sedimentation sequence (see, e.g., Timmons,
Karlstrom, and Sears 2003; Hamblin 2008), with 
deposition time between 1200 Ma and 750 Ma, were found to be faulted 
into the basement rock (i.e. the Vishnu Complex) at various locations.  
Where the Supergroup is present, the time gap representing the Great 
Unconformity is reduced correspondingly (i.e., the Proterozoic section
is terminated at 750 Ma, instead of at 1.7 Ga).  It is noteworthy that the top
of the outcropped Grand Canyon Supergroup has the deposition age (750 Ma) coinciding
with that of the rifting of the Rodinia Supercontinent.

\item In western Colorado, around the Plateau's eastern boundary,
the Proterozoic Uncompahgre Formation, consisting of alternating layers of
quartzite and phyllite, forms the cores of the Grenadier 
Range and Snowdon Peak of the Needle Mountains, which is part of the 
western San Juan Mountains.  This Formation, which outcrops also at other locations 
(i.e. along the Animas Canyon near Ouray, Colorado, in various other locales
within Needle Mountains, as well as in the Rico area of Colorado), were often 
found to be severely faulted and folded.  Figure \ref{Fig3} shows upturned Uncompahgre 
Formation along the so-called ``million-dollar Highway'', US-550.  Figure \ref{Fig4}, 
Top Frame, shows the Uncompahgre Formation in a sandwich pattern (two vertically
upturned sections bracketing a central section of normal bedding planes), 
just south of Silverton, Colorado, along the route of the Durango-Silverton 
Narrow Gauge Railroad (Gonzales and Heerschap 2012). 
The formation of this sandwiched sedimentary pattern, which can also be seen in a similar
photograph in Blakey and Ranney (2018), clearly requires  
the action of strong and rapid shear forces, applied in the vertical
direction with respect to the normal bedding planes, rather than from the side
as is the case for the usual tectonic forces.
A related example, in Figure \ref{Fig4}, Bottom Frame, shows
the late-Devonian Elbert formation unconformably overlying the 
upturned beds of the Precambrian Uncompahgre Formation at the Box Canyon
State Park near Ouray.  Note in particular that on the right hand side of this picture
of Box Canyon the Uncompahgre Formation itself has two sections oriented nearly 90 degrees
from each other in close proximity, similar to the phenomenon shown
in Figure \ref{Fig4}, Top Frame.  Similar images of the Box Canyon Unconformity
have previously been shown in Hinds (1936, Plate 4B), in
the Geologic Atlas of the Rocky Mountain Region (1972, p. 35), 
as well as in Baars (1992 p. 31), although these previous images did not highlight the 
perpendicular orientations of the sediments of the Uncompahgre Formation itself.  
These patterns of tight, orthogonally-oriented Uncompahgre sediments in juxtaposition,
occurring over such a large scale (i.e., the sizes of small mountains) could 
not possibly have been produced by traditional types of slow and prolonged orogenic events, 
such as plate subduction or superplume upwelling, and they are most likely the result
of an instantaneous, high-shear-rate, and high-intensity giant impact event.

\item In the northern Needle Mountains area, the Uncompahgre Formation forms
a synclinorium which is wrapped around the circa 1400 Ma Eolus Granite intrusion
(see Zinsser [2006], Figure 2 for the most up-to-date Proterozoic outcrop map of the Needle
Mountains area, based on the original map by Barker 1969, supplemented with data from
Gibson and Simpson [1988]; Gonzales [1997]; Harris et al. [1987]; and
Tewksbury [1981]).  In the past, evidence for both allochthonous (Tewksbury 1985a,b, 1986, 1989) and 
paraautochthonous (Harris et al. 1987; Harris 1990) nature of the Uncompahgre Formation 
in this area had been reported.  The giant impact picture helps to reconcile these two 
points of view, and regards the synclinorium of highly deformed Uncompahgre Formation
as being pushed against and draped around the northern Needle Mountains front
by the impactor traveling from N-NE to S-SW.  The Uncompahgre Formation in this 
area thus acquired allochthonous characteristics despite being paraautochthonous
when first formed.  The giant impact picture also helps to explain the sometimes violent
inter-mixing of the Uncompahgre Formation and Eolus Granite, which we will
show images later.  The pre-existing Eolus Granite intrusion which forms
the core of the Needle Mountains is likely also the reason that the Colorado
Plateau's eastern boundary has an inward dent/kink in the western San Juan Mountain
area, because the more resistant granite (compared to the cover sequence rocks)
likely prevented the chiseling of the otherwise rounded boundary of the impact
crater through the granite block, thus the severed Plateau block went around the
Needle Mountains' western boundary.

\item The maps on p. 38-39 of the Geological Atlas of the Rocky Mountain Region
(1972, or else see Kelley 1979, Figure 1) show that the major and minor fold axes and faults across the Plateau 
have a roughly NW-SE orientation, perpendicular to the proposed impact trajectory.
Such folds and faults were found to also extend to the north beyond the 
Uinta Mountains region, which is usually considered the northern boundary of the
Plateau -- though some geologists, such as Dutton 1882, p. 54, considered the 
Plateau to extend further north into the Teton Range and northern Rockies, and
the paleo-geologic maps of the region confirm such associations
(Blakey and Ranney 2008, 2018).  The evidence that the possible touch-down 
spot of the impactor (or area affected by the initial shock wave of 
the impactor) resides further north in the Rockies includes also
the abundance of pebbles, cobbles, and boulders of younger Precambrian age,
in the Jackson Hole and Teton Mountains area. These detritus do not seem to 
have a corresponding local source (Love et al. 2003, p.59), and they
could be the result of jetting in the rear of the impactor's trajectory.
Similar pebbles, cobbles and boulders were also found abundantly
on the Plateau itself (Condon 1995; Gonzales et al. 1994), composed of 
parent rocks of Precambrian ages.  The cover sequence rocks in the northern
Needle Mountains (part of the western San Juan Mountains) show 50\% shortening
via south-vergent folding and south-directed shearing (Zinsser 2006 and
the references therein).  The rocks in Taos Range also show 
pervasive southward shearing fabrics which postdate the rarer
northward shearing fabrics (Grambling et al. 1989, p. 107).

\item The Proterozoic cover sequences outcropped along the Plateau's boundary 
in all four adjacent states of Arizona, New Mexico, Colorado, and Utah, 
as well as in the Northern Rocky Mountains (i.e., the Belt Series of Idaho, 
Montana, and Wyoming -- see, e.g., McKee 1972 and the references therein)
show similar styles of brecciation and metamorphism.  In Utah (Figure \ref{Fig5}), 
the deformation occurs mostly in the Uinta Mountain Group and its western extension,
the Big Cottonwood Formation (Hansen 2005, p. 73, 76; Bennis-Smith 
et al. 2008, Dehler et al. 2010).  In the western San Juan Mountains of Colorado (Figure \ref{Fig6}), 
the deformation is most pronounced in the Uncompahgre Formation (Baars 2000, 
pp. 115-121).  In Arizona (Figure \ref{Fig7}), deformation is observed in both the 
Paleoproterozoic Mazatzal and in the late-Mesoproterozoic Apache Groups, and, 
in milder form (benefited apparently from fault protection), in the Grand Canyon Supergroup
which has a Neoproterozoic upper layer.  In New Mexico (Figure \ref{Fig8}), the deformation  
occurs in various cover sequence groups in southern Sangre de Cristo 
Mountains (i.e., the Taos Range, Reed 1984; the Pecuris Range, Montgomery 1956; the 
Santa Fe Range, Bauer and Raiser 1995; Erslev et al. 2004, etc.).  
Note that the east side of the Plateau boundary, especially the northeast 
side (i.e., the Colorado Front Range), appear to have suffered a greater degree
of damage as a result of the proposed impact event, which accounts
for the higher degree of uplift, erosion, and metamorphism observed
there, compared to the west and south side.  This could also explain the
older ages of the outcropped Precambrian rocks in the north-east region,
as a result of the higher degree of uplift and erosion of the Precambrian
sediments.  Parts of the top-most layers of the cover sequence could
also have left the Earth during the impact-induced vaporization. 
We also need to remember that the ages of formation of the top-most
exposed Precambrian cover sequence is not the same as the time of
the last major metamorphic event on these rocks.  The latter can be much
more recent due to the rapid removal of the younger layers  of sediments during the
impact event.

\item Exposed Proterozoic rocks of the Plateau and surrounding areas also 
show evidence of dynamical metamorphism likely produced by giant impact 
in the forms of: (1) Shatter cones of varying sizes, some in clusters spanning miles, with
``horsetailing'' striations on rock surfaces (Figures \ref{Fig8}, \ref{Fig9}). 
These macroscopic features have traditionally
been attributed to the shock metamorphism effect from meteorite impact (French 2003);
(2) Severe compression and flattening of pebbles embedded in conglomerates 
(Figure \ref{Fig10}, Top Left Frame); (3) Fusing of impact-shattered rocks from
a variety of sources in a matrix which itself containing melts, to form suevites (Figures \ref{Fig10}, 
Top Right, Bottom Left, and Bottom Right Frames).  See French (2003) for discussion
of the impact origin of suevites; (4) Partial melting of large sections of rock bodies 
to change their nature from quartzite and granite to inhomogeneous igneous rocks 
formed under nonequilibrium conditions (Figures \ref{Fig11}, \ref{Fig12}).  
Cox (2002) discussed psot-deposition alteration of Mazatzal group quartzite in Arizona.
Fackelman et al. (2008) found impact shatter cones and microscopic shock alteration 
to Paleoproterozoic rocks northeast of Santa Fe, New Mexico, within an extended 
section of the Sangre de Cristo Mountains.  Our Figures \ref{Fig8} and \ref{Fig9} 
confirm such discoveries.  The giant impact scenario helps to solve the mystery 
of Santa Fe impact structure which was previously found to be lacking an obvious 
impact crater despite the discovery of an abundance of shatter cones in the area 
(i.e., the whole Colorado Plateau itself is now the proposed impact crater).

\item In the Black Canyon of the Gunnison National Park in Colorado, the Canyon walls which
are made of Precambrian metamorphic rocks are injected with coherent 
pegmatite sills and dikes hundred of meters in length (Figure \ref{Fig13}).  
The age of the injection is likely to be Neoproterozoic, based on
cross-cutting relationships: the pegmatite dikes are observed to intrude the Mesoproterozoic 
Vernal Mesa Quartz Monzonite (dated to be 1.22 Ga in age) in Cedar Point and Chasm view area (Hansen 1965).
The lengths and coherence of these sills and dikes show that they are likely
caused by a single episode of strong and sudden brittle deformation event.
As we know, the 1.1 Ga Grenville Orogeny had not significantly affected the
Black Canyon of the Gunnison area (and even in areas that the Grenville had
affected, the deformation is mostly in the form of ductile deformation.  See
later Figure \ref{Fig23}, Right Frame), and the only other significant tectonic
event between the Mesoproterozoic and the onset of the Cambrian is the
750 Ma breakup of the Rodinia Supercontinent.

\item Love et al. (2003, pp. 38-39) mentioned that on Mount Moran
(altitude 3,842 m), which is part of the northern Teton Range
in Wyoming, the 50 meter-thick black dikes near the summit 
of the peak, which cut across older Precambrian rocks,
are similar to the dikes in Tobacco Root and Beartooth Mountains 
in Montana, which had been dated by S.S. Harlan at the U.S. 
Geological Survey to be about 765 Million years old. Similar 
dikes of estimated Neoproterozoic age can also be
found cutting through the Grand Canyon Supergroup (Blakey and Ranney
2008, p. 8); and in the Canadian Rockies (Figure \ref{Fig20})
and even in the eastern United States where the eastern side
of Rodinia rift occurred following the 750 Ma event.  We will
address these further later in the text.

\item The uplifting of the Colorado Plateau during the Cenozoic
era appears to be in a coherent fashion, in contrast to the haphazard 
tectonic activities in its surrounding area. This indicates that the 
basement block of the Plateau was severed from the surrounding craton of 
the continent, and the Basin-and-Range extensional dynamics of the 
Cenozoic era apparently helped to relieve some of the confining pressure 
from the Plateau's boundary, giving rise to both the Cenozoic igneous 
activities in the Plateau's rim, as well as the rise of the modern version 
of the Rocky Mountains, at the site of the eroded ancestral Rocky
Mountains which had first emerged in the Paleozoic era.  The well-known 
thickened root underneath the Plateau (Keller, Braile, and Morgan 1979) 
could be partly a result of impact-induced granite pluton formation, 
partly due to the need for isostatic equilibrium in response to the uplift 
of the Plateau after being freed along its boundary during the extension
phase in the Cenozoic era (i.e., the raised height of the Colorado
Plateau coupled with its thickened root indicate that the Colorado
Plateau is in independent isostatic equilibrium floating above the mantle, 
supporting the notion that the Plateau is severed from its surrounding
craton).

\item Sedimentary record shows that the entire North America continent,
including the Colorado Plateau region, was in a very flat configuration
at around 600 Ma (Blakey and Ranney 2008, 2018).  It is mostly above this 
flat terrain that the Paleozoic sedimentation was laid.  The flat terrain
and the wide-spread existence of the Great Unconformity between
the Proterozoic and Cambrian indicate that the Great Unconformity
is not likely an erosional feature, because there is scant evidence
of the erosional remnants of such wide-spread scale.  Apart from the
detritus of the impact (for example., the conglomerates composed of
brecciated Precambrian rocks left over in the Plateau region, and the
re-lithified crushed Precambrian rocks), the most
natural explanation of the lack of erosional remnants to account for
the world-wide presence of the Great Unconformity separating the
Proterozoic from the Cambrian, is that the missing layers of sediments
have vaporized and left the confines of the Earth.
\end{enumerate}

These characteristics of the Colorado Plateau and its surrounding area, 
added to the correlation of the timing of the formation of the Great 
Unconformity with the timing of the rifting of the Rodinia supercontinent 
at the western margin of the Plateau, as well as the timing of the 
subsequent rifting at the eastern side of the Rodinia Supercontinent 
to form the Iapetus ocean, point to the likelihood of a giant impact 
event occurring around the 750 Ma centered on the Four Corners area
of the western United States. 

\section{Characteristics of the Proposed Giant Impact Event}

In this section we proceed to constrain the characteristics of 
the proposed giant impact event using the known properties of 
the Colorado Plateau and its environment.

\subsection{Size of the Impactor}

We take the size of the Colorado Plateau itself as the approximate size 
of the impact crater.  The Colorado Plateau has an area of 
approximately 337,000 square kilometers, 
thus a diameter of about $D_{cr} = 640 km$
(if we add to this the area around the western San Juan Mountains, as done in
Kelly 1979, it will increase the total area of the Plateau by about 11\%.
So the gross estimate would remain valid in either case).  
Dence et al. (1977) found that for terrestrial impact craters,
progressively larger craters tend to have progressively 
shallower profiles. They quoted result for the 3.6 km
Steinheim Basin and Flynn Creek crater
(Roddy 1977a,b; Reiff 1977), which has depth-to-diameter
ratio of approximately 1 to 24.  
In our following calculation, we take the impact
crater depth $h_{cr}$ to be 1/40 of its diameter, i.e., $h_{cr}$ = 16 km.
This depth is reasonable considering that the estimated original depth
for the Grand Canyon Supergroup is up to 4 km thick, and the Supergroup
itself, since it consists of faulted blocks, does not contain 
the original bottom contact to the basement
rock, which is 1.7 Gyr old (the oldest rocks in the Grand Canyon Supergroup
is about 1.2 Gyr old).  The exposed Uinta Mountain Group in the 
eastern Uinta Mountains
is up tp 7 km thick at the Colorado-Utah border
(Dehler et al. 2010), which also does not reveal its root connecting
to the basement rock.  Furthermore, we need to consider also
the possibility that the entire Plateau Lithospheric block may have 
sunk slightly
into the Asthenosphere at the time  of the impact, since it is obviously 
severed from the surrounding craton, judging from its structural
integrity and isostasy behavior (Keller et al. 1979).  So our choice of
16 km for the crater size appears to be reasonable, as will be borne
out from the overall energetic considerations presented later in
this paper.

From a simple geometric consideration (Figure \ref{Fig14}),
the impactor's radius $R_{imp}$ is obtained as (taking the radius
of the impact crater $R_{cr} = 1/2 ~ D_{cr} = 320$ km)

\begin{equation}
R_{imp} = {{R_{cr}^2 + h_{cr}^2} \over {2 \cdot h_{cr}}}
=3208 ~ km.
\end{equation}
Therefore the hypothesized impactor is only slightly smaller 
than Mars' size ($R_{Mars}$ = 3386 $km$).  From the energy
and momentum balance considerations we will present next,
the estimated impactor mass also turns out to be comparable
to Mars' mass.  Therefore, a Mars-sized impactor allows a 
self-consistent impact scenario to be obtained, which fits 
also other known facts related to the hypothesized impact event.

\subsection{Energetics of the Impact Event}

According to Goldsmith (2001) and the references therein,
at low to intermediate range of impact velocities, when 
the fusion and vaporization of the bodies involved can be 
neglected, the resisting force to the impactor 
can be expressed by the following empirical formula:

$$
F = - M_{imp} {{d v_{imp, before, \perp}} \over {dt}} 
$$
\begin{equation}
= B_1 v_{imp, before, \perp}^2
+ B_2 v_{imp, before, \perp} + B_3,
\label{eq:2}
\end{equation}
where the three terms on the right hand side are contributions
due to the acceleration of the target material adjacent to the
impactor, the effect of the frictional forces, and the cohesive
strength of the target, respectively, and
$v_{imp, before, \perp}$ represents the vertical component
of the impactor's velocity relative to the target, just before the
encounter occurs, and $M_{imp}$ is the mass of the impactor. 
In this expression the acceleration of the target (i.e.,
the Earth as a whole in our example) is ignored, since
Goldsmith (2001) did not study the kind of giant impact events
planetary collisions represent.

For higher impact intensities (as is more relevant to the Colorado 
Plateau impact event), on the other hand,
it is often postulated that the crater volume is 
proportional to the kinetic energy of the impactor (Melosh 1996).

Dence et al. (1977) found that for large impactors the diameter of 
the excavated crater $D_{cr}$ and the energy of the impact $E_{imp}$ 
follow roughly the relation

\begin{equation}
D_{cr}  (km) = 9.7 \times 10^{-5}  E_{imp}^{1/4} (J)
.
\label{eq:1}
\end{equation}
Take $D_{cr}$ = 640 km, we obtain $E_{imp} \approx 1.7 \times 10^{27}$ joules. 
Our impactor is likely to have collided with the Earth
in an grazing-incidence angle (more on that later), 
judging from the slightly elongated
shape of the Colorado Plateau.  Taking the impactor's vertical velocity 
(relative to the surface of the Colorado Plateau) before the encounter
to be $v_{imp, before, \perp} = 5$ km/sec, and taking the impactor mass 
to be 10\% of the mass of the Earth, or $6 \times 10^{26} g$, which
is just slightly smaller than that of Mars, the potentially-available
impact energy would be (assuming about 91\% of the vertical-motion
kinetic energy of the impactor is dissipated during the impact event,
as our later calculation using equation (\ref{eq:8}) would suggest):

$$
0.91 \times {1 \over 2} \times 6 \times 10^{26} \times  (5 \times 10^5)^2 = 7 \times 10^{37} ergs
$$
\begin{equation}
= 6.8 \times 10^{30} ~ joules,
\end{equation}
which is larger by more than 3 orders of magnitude than
the estimation using the Dence et al. (1977) equation based on the
impact crater size, which we had quoted above. 

However, we need to keep in mind that the excavation of the Colorado
Plateau is not the only usage of the impactor's available kinetic energy,
since the case of a giant impact has many features different from the 
past-studied small to medium-sized cratering events.  
From the global-presence of the Great Unconformity
of the late Precambrian period, which we had briefly commented before, 
it is likely that the entire Earth had lost a significant portion 
of the late-Precambrian sediments through strain-energy-release
induced evaporation (i.e., the Earth will be ringing like a gong
after the impact, just like the case of seismic waves after
an earthquake.  See Stein and Wysession 2003.  Note that the initial
strain waves are likely to be shock waves, rather than elastic waves.
See Melosh 1996, p. 29ff.  These shock waves are responsible not
only for transmitting the part of the impact energy that will eventually
be dissipated within the impactor and the Earth, 
but also for accelerating the Earth and
decelerating the impactor, so that they achieve the momentum
balance condition which we will calculate in the following
subsections).  Furthermore, the impactor itself likely had suffered 
crustal vaporization and other structural damages as well
during to the same impact event, which will also absorb part
of the impact energy.  Some of the strain energy release
would also be dissipated as heat, and in the case of the Earth this
will be along the Plateau's boundaries, in its main body (responsible
for the partial melting and shock metamorphism of the material),
as well as in its bottom interface with the Asthenosphere. 
Still further deposited energy on Earth will be used to accelerate 
the broken pieces of the Rodinia supercontinent, though there
is evidence that there might be a time delay between the time
of the impact and the time of the actual rifting of Rodinia.  

If, on the other hand, we use directly the $7 \times 10^{30} J$ 
potentially-available impact energy used for dissipation
(i.e., we have removed the 9\% impact energy used for accelerating the Earth,
as our later calculations using energy-momentum balance conditions
throughout the impact event will predict),
we obtain an equivalent {\em maximum possible}
cratering {\em diameter} of 4989 km
according to the above Dence et al. (1977) equation.
This is slightly smaller than but of the same order as the
{\em radius} of the Earth at 6371 km.  Note that even though the
Dence et al. (1977) equation (equation \ref{eq:1})
was derived using much smaller terrestrial craters, the
log-linearity of the law does not appear to change for increased
crater size on the large-impactor branch.  This is likely a result
of the fact that the physical foundation of the Dence et al.
equation is that the dissipated energy is used mainly for the
excavation of the crater at the surface of the Earth (which would
have given a $D \sim E^{1/3}$ dependence), and a
smaller portion is used to heat the interior of the Earth,
which resulted in the final $D \sim E^{1/4}$ dependence.
For larger craters, such as would be produced in the current
event on the Earth's and the impactor's surfaces, the fact remains 
that most of the dissipated energy is used to evacuate
the surface sediments (overcoming the binding energy
of the rock material) of the Earth and the impactor, 
only a small portion is used for creating
faults and heating up the interior of the Earth (as well as the impactor).
Melosh (1996) also argued favorably for the extrapolation
of the energy-crater-size condition derived from explosive events
to the planetary scale craters.

The global Great Unconformity due to the hypothesized 750 Ma impact
event is likely to be inhomogeneously distributed, with the energy dissipation
right under the Colorado Plateau the most concentrated, and elsewhere
the shedding of the late-Precambrian sediments varied by degrees depending
on the propagation, reflection, and interference of the shock waves of
impact, similar to the seismic waves produced during normal
earthquakes (Stein and Wysession 2003; Boslough et al. 1994).  
Note that since at the time of the impact event Rodinia is 
mostly a single connected supercontinent,
there would not be the antipodal focusing effect observable in the continental
crust that survive.  Any such antipodal focusing effect would have occurred
to oceanic crust which would have long since been subducted.  
Therefore, we see that our energetic estimation above
for the impact event is at least of the correct order of magnitude
to account for the world-wide presence of Great Unconformity at
the Cambrian-Precambrian interface. 

For comparison, the $\sim$ 65 Ma asteroid impact event proposed by Alvarez
et al. (1980), which possibly had led to the extinction of
the dinosaurs, assumed an impactor diameter of 10 km, and
roughly $4 \times 10^{30}$ ergs of impact energy, which is more
than 7 orders of magnitude smaller in energy than
that potentially dissipated during the formation of the Colorado Plateau.
But then, the 65 Ma event did not lead to the breakup
of a supercontinent, nor a globally-present Great Unconformity,
though its effect may be partly responsible for the subsequent
uplift of the Colorado Plateau as well as the exhumation of the
ancestral Rocky Mountains
during the Laramide Orogeny, due to the initiation of the subduction
of the Kula and Farallon Plates underneath the North American Plate.

\subsection{\label{kine} Kinematics and Dynamics of the Impact Event}

We now estimate the remaining impact-event parameters through employing
energy and momentum conservation relations.

We assume that a fraction $k$ of the initial vertical-motion kinetic energy
of the impactor relative to Earth will be left, after dissipation,
to drive the post-impact combined motion of
the impactor plus the Earth in the vertical direction.
We therefore assume that immediately
post impact, the impactor and the Earth have the same vertical-motion
velocity $v_{out, \perp} = v_{imp, after, perp}$, with respect
to the frame of reference of the original equilibrium orbit
of the Earth, whereas the pre-impact vertical velocity
of the impactor with respect to the Earth 
is $v_{imp, before, \perp}$, as 
had been assumed previously (Figure \ref{Fig14}).

In the following calculations, we will initially ignore the Earth's
orbital velocity of 30.5 km/sec but will in the end make an estimate of 
the effect of the impact on Earth's orbit around the Sun.  This is 
equivalent to assuming that the impactor had obtained certain degree 
of dynamical equilibrium within the Solar System when it reached 
the Earth's location, i.e., it has a circular velocity
similar to that of the Earth, apart from its peculiar
velocity with respect to the Earth.

Due to the large mass of the impactor, the vertical relative
speed $v_{imp, before, \perp}$ could not have been much higher than what
we had assumed below of 5 km/sec
without causing further damage to
both the Earth and to the impactor: We expect
the impactor to have survived the impact event as well,
because the Mogollon Rim in Arizona, where the impactor
likely exited, has its 
corresponding bowl-shaped rim  as well, even though not as
steep as the Plateau rim on the north side, i.e., the Uinta 
Mountains and the Colorado Rocky Mountains where the
impactor made its initial landing.  Furthermore,
as we had commented before, across the Plateau the minor fault lines
(as well as many elongated uplifts and basins)
have directions mostly aligned perpendicular to
the expected trajectory of the impactor across
the Plateau.  This shows that the impactor most likely skidded across the
Plateau from N-NE to S-SW and made an exit around the
Mogollon Rim region, rather than totally
evaporated during the impact process. Effectively,
we assumed that that impactor was a rocky planet,
which had an internal strength similar to
that of the Earth, rather than being a gas planet.

In the following calculations we will also 
ignore the Earth's spin velocity at its surface 
of 0.46 km/sec, because the effect of the Earth's 
spin velocity upon the impact event will depend on the 
exact orientation of the impactor's trajectory
with respect to the Earth's spin, which in turn will depend
on the exact location and orientation of the Rodinia
supercontinent at the time of the impact,
of which we have only imprecise knowledge (i.e., certain
models suggest that in late Precambrian the Rodinia supercontinent
is located near the Equator, and North America is
rotated 90 degrees from its current orientation).  
The magnitude of the Earth's spin velocity will in 
any case not affect the order-of-magnitude nature 
of our following calculations.

Using the equations of momentum and energy conservation, and
assuming the impactor has a fraction $f_{imp}$ of the mass of the Earth,
we have

$$
f_{imp} \cdot M_{earth} \cdot v_{imp, before, \perp} 
$$
\begin{equation}
= (1 + f_{imp}) \cdot M_{earth} \cdot v_{out, \perp}
,
\end{equation}
and

$$
k \cdot {1 \over 2} \cdot f_{imp} \cdot M_{earth} v_{imp, before, \perp}^2 
$$
\begin{equation}
= {1 \over 2} \cdot (1 + f_{imp}) \cdot M_{earth} v_{out, \perp}^2
.
\label{eq:6}
\end{equation}

The solutions of these two equations are, for the common vertical component of the exit velocity with respect to Earth's orbit,
\begin{equation}
v_{out, \perp} = v_{imp, after, \perp} =
{{f_{imp}} \over {1+f_{imp}}} \cdot v_{imp, before, \perp}
,
\label{eq:7}
\end{equation}
and for the fraction $k$ of energy left for driving both planets' motion, 

\begin{equation}
k = {f_{imp} \over {1+f_{imp}}}
.
\label{eq:8}
\end{equation}

Therefore, if we choose $f_{imp}=0.1$ (i.e for the impactor to have 10\% of
Earth's mass, which makes it similar in mass as well in size to Planet Mars), 
then k=0.09, or about 9\% of the impactor's vertical-motion
kinetic energy is used to produce the residual vertical motion
of the Earth plus the impactor.  Therefore, 91\% of the impactor's
vertical motion kinetic energy relative to the Earth is dissipated
during the impact event, as we had previously utilized.

Assume the impactor's transverse velocity is originally
$v_{imp, before, \parallel}$ =
25 km/sec, so that we are dealing with an oblique, or close to
grazing incidence ($11.3^o$ impact angle from the horizon).  
The choice of the larger transverse component of
the impact velocity is first of all so that it takes into account of the
slightly elongated impact crater that the Colorado Plateau is
(Melosh 1996, showed in Figure 5.16 which was taken from
Gault and Wedekind 1978, that until the angle of impact is close to
$10^o$, the crater of an oblique impact will remain fairly round.
The Colorado Plateau itself, despite being slightly elongated, does have
an aspect ratio fairly close to 1.  So our choice of impact
angle of $11.3^o$ is not unreasonable).  Secondly, this choice
also enabled the impactor to have enough post-impact total
velocity to overcome the Earth's gravitational field (the escape
velocity from the Earth is 11.2 km/sec),
as well as possibly to escape from the Solar System
(the escape velocity at the Earth distance from the Sun is
about 42.1 km/sec, which, if we consider adding the
impactor's post-impact velocity to the orbital
velocity of the Earth around the Sun which is about 30 km/sec,
then escape of the impactor from the Solar System becomes a 
possibility for the most favorable impact configuration. 
To this possibility we could also add
the scenarios of the impactor running into one of the outer gas
giants, or else into Venus which apparently had regenerated its
crust after 500 Ma.  A sling-shot acceleration scenario is
also to be considered).

During the 30 seconds or so for it to reach the depth of 150 km, the impactor 
would have skidded within the Colorado Plateau for about
750 km, which is close to the dimension of the Plateau in the 
North-South direction. Of course, the initial touch-down location
of the  impactor could be further north, because when the contact
between the impactor and Earth was initially established mostly 
crushing and vaporization of crust is expected, rather than the
immediate dislodging of the section of the crust from the surrounding
craton.

If we assume a 10\% loss of the impactor's velocity component
parallel to the Earth's surface due to friction (i.e., it should
be reduced from 25 km/sec to 22.5 km/sec during the impact process), 
the total velocity of 
the impactor before and after the impact (with respect to
Earth's original circular velocity) are $v_{imp, before} =
25.5$ km/sec and $v_{imp, after} = 22.5$ km/sec, and for the Earth, 
the excess velocity it gained is on the order of $v_{out} = 0.45$ km/sec,
using equation \ref{eq:7}. 
From these results we derive that the coefficient of restitution 
for the impact event is about 88\%.

\subsection{Effect on Earth's Orbit Around the Sun}

Prior to the impact, assume the Earth is on an orbit around
the Sun similar to its current orbit.  The mass of the Sun
is $1.989 \times 10^{33}$ g, the Sun-Earth distance is
$R_{sun-earth}=1.5 \times 10^{13}$ cm, and using
a gravitational constant $G=6.674 \times 10^{-8} cm^3g^{-1}s^{-2}$,
we obtain the gravitational attraction force between the
Sun and the Earth:

$$
F_{sun-earth}
= 6.67 \times 10^{-8} {{5.98 \times 10^{27} \times 1.989 \times 10^{33}}
\over {(1.5 \times 10^{13})^2}}
$$
\begin{equation}
= 3.6 \times 10^{27} ~ dyne
.
\end{equation}

On the other hand, the average impact force $F_{imp}$ can be estimated from:

\begin{equation}
F_{imp} \cdot \Delta t = M_{imp} \Delta v_{imp, \perp}
,
\end{equation}
where $\Delta v_{imp, \perp}$ is the differential velocity of
the impactor before and after the impact, with respect to the
Earth's original orbit.  Here we assume the majority of the momentum
transfer is used for altering the bulk motion of the impactor
and the Earth (as our calculation in \S\ref{shear} next will show).

Taking once again the impact duration to be about 30 seconds,
we obtain $\Delta v_{imp, \perp} = v_{imp, before, \perp} - v_{imp, after,
\perp} = 4.55$ km/sec, therefore

$$
F_{imp}= {{ 6 \times 10^{26} \cdot 4.55 \times 10^5} \over 30} 
$$
\begin{equation}
= 9.1 \times
10^{30} ~ dyne
,
\end{equation}
which is significantly larger than the gravitational attraction between
the Sun and the Earth.  Thus, the Earth will indeed
be accelerated to the terminal velocity $v_{out, \perp} = 0.45km/sec$
which we had determined in \S\ref{kine}.

The circular velocity of the Earth's orbit around the Sun,
on the other hand, is around 30.5 km/sec.  The ratio of the
after-impact velocity to pre-impact velocity is thus on the
order of 1.01.  The eccentricity of the Earth's orbit is
about 0.0167, and it varies historically between $0.0034 - 0.058$.
Therefore, this hypothesized impact event would contribute to the
Earth's eccentricity an amount well within its normal range of variation.

\subsection{\label{shear} Shear and Dislodging of the Colorado Plateau and of the Rodinia Supercontinent}

As we have obtained previously, the average impact force, assuming
a 30 second impact duration, is about $10^{31}$ dyne.  From the integrity of the Colorado Plateau,
it is likely that the entire Plateau block is at least partially
severed from its surroundings at the time of the impact.  
Taking the thickness of the partially severed Plateau to be 
$H_{CP}$ = 150 km (i.e., roughly the average thickness of 
the Lithosphere: here we ignore the subsequent growth in
thickness of the Plateau's root post impact, especially during
the Cenozoic uplift of the Plateau when the severed Plateau mass
had to achieve isostatic equilibrium by offsetting the raised height with
the growth of a thickened root), we obtain that the {\em maximum
available} shear stress across the boundary of the Colorado Plateau due to 
the impact event is on the order of 

$$
\sigma_{CP, max} = {{F_{imp}} \over { 2 \pi R_{CP} H_{CP}}}
$$
\begin{equation}
=3 \times 10^{15} dyne
\cdot cm^{-2}.
\end{equation}

For the Rodinia supercontinent, assuming its severed length to be about 
2000 km, and once again taking the severing depth to be $H_{Rodinia}$ = 
150 km, we can similarly estimate the {\em maximum available}
shear stress for rifting the Rodinia supercontinent through the impact force:

$$
\sigma_{Rodinia, max} = {{F_{imp}} \over { L_{Rodinia} H_{Rodinia}}}
$$
\begin{equation}
= 3 \times 10^{15} dyne \cdot cm^{-2}.
\end{equation}

Therefore, we see that the maximum-available shear stresses for these 
two processes are comparable.

Ohnaka (2013, p.75) presented a linear relation between 
the shear failure strength $\tau_{p0}$ for dry Westerly granite at room 
temperature, versus the normal stress $\sigma_n$,
which is the same as the confining pressure, as (equation unit in MPa):

\begin{equation}
\tau_{p0} = 135.7 + 0.75 \sigma_n. 
\end{equation}

For higher ambient temperatures, the slope and intercept of the above
equation both decrease progressively.  At the deep end of the Lithosphere 
of 150 km, with temperature around 1000 K, the slope to use is around 0.37 
and the intercept is about 50 for the above equation, 
according to Figure 3.11 of Ohnaka (2013).  The pressure at the depth of 
150 km is about 4500 MPa (i.e., assuming about 30 MPa increase in
confining pressure per km increase in depth). 
Therefore, the shear failure strength
at the boundary of the Lithosphere and Asthenosphere is about 
1700 MPa (1.7 GPa), or $1.7 \times 10^{10} dyne/cm^2$, which is more than 5
orders of magnitude smaller than the maximum shear stress we had calculated
above for the Colorado Plateau and for Rodinia during the 750 Ma
event, {\em if we assume} all the impact force is used to produce 
the shear.  

However, the assumption that the total impact force is
equal to the shear force is {\em not at all} reasonable: The Earth,
as we know, was (and is) totally unsupported in 
space apart from the gravitational
attraction of the Sun which keeps it in orbit.  Since the gravitational
force from the Sun is much smaller than the impact force, this means
that the majority of the impact force will be used to accelerate the Earth 
to the terminal velocity which we had calculated above according to 
momentum conservation.  Of the total impact force, only a very small fraction 
acts {\em differentially} at the boundaries of the Colorado Plateau,
due likely to the differential propagation time of the strain/shock waves
arriving at the different locations across the Plateau's boundary. 
It is this differential stress propagation that provided the shear force which
led to the severing of the Plateau from the parent craton, as well as 
the rifting of the Rodinia supercontinent.

Furthermore, since the Colorado Plateau (as well as the Rodinia 
supercontinent) is not entirely severed at the time of the impact,
the local shear force actually present during the impact
event should be on the order of the shear strength of the material times 
the area of the shearing surface (which we had calculated above to be
$2 \pi R_{CP} H_{CP}$ for the Plateau, and $L_{Rodinia} H_{Rodinia}$
for the boundary of the Rodinia supercontinent).  Therefore, the actual
shear force should be on the order $10^5$ times smaller than the total 
impact force, i.e., the total shear force (on Rodinia or else
on the Colorado Plateau) would be on the order of $10^{26} dyne$
(taking the high end of the estimate to account for possible long-range
correlations which give rocks in a single craton addition strength).
Through the action of the initial shear force,
an initial crack will be produced at the Plateau's boundary,
as well as along the eventual rifting lines of Rodinia (both east
and west of the resulting central craton).  Once the critical 
Griffith crack length is exceeded along these initial weaknesses, the cracks
can also self-propagate both forward (as in the case of Rodinia)
and downward (as in the case both of Rodinia and the Plateau).

With the shear strength of the Plateau region thus calculated to be 
about 1.7 GPa, which, incidentally, also coincides with the 
maximum oceanic crustal strength quoted in Price (2001, p.50), we compare
it to the peak shock pressure estimated for the Santa Fe impact structure 
(Fackelman et al. 2008), which is about 5-10 GPa.  Also for comparison, 
the peak shock pressure for the simulation presented by Boslough 
et al. 1994 is 6 Mbar, or 600 GPa, 0.1 second after impact for an 
assumed 10 km diameter, 20 km/sec vertical incident-velocity 
impactor, but it soon diffuses and dissipates into 10 bar, or 1 MPa mean 
stress level during the shock-wave's subsequent propagation into the Earth's 
interior.  We thus see that our estimated shear strength is at least not 
in conflict with the inferred peak shock pressure for Santa Fe impact structure
(which we will later show to be likely just a component of the 750 Ma
impact crater): Since shock metamorphism is caused by 
the direct impact force, rather than by the differential
propagation of strain waves, we expect the former to be bigger in magnitude.

Looking back at equation (\ref{eq:2}), we see that the impact process can
in fact be viewed through this equation in a broad-stroked fashion:  
The first term is mainly responsible for the acceleration of the Earth
(and the deceleration of the impactor), if we re-interpret
the qualifier "target material adjacent to the impactor" to mean the target
(and the impactor) as a whole. The third term overcomes
the structural strength of the Earth (as well as the impactor)
to rift Rodinia and to dislodge the Colorado Plateau, as well as to produce the 
series of stress/strain/shock waves that created the Great Unconformity 
and possibly destroyed the surface layer of the impactor as well.
The second term, the friction force, is partly (apart
from the strength of the material) what stopped the Colorado
Plateau from sinking into the mantle completely.  It also encapsulates
the effect of shock metamorphism, heating, etc..  Of course, the
actual physical processes are distributed and convoluted, and will
certainly benefit from future more sophisticated modeling and analyses.
Our order of magnitude calculations in the current paper serve 
to establish an initial hierarchy of organization, which helps to 
disentangle the complicated processes of the 
proposed impact event, and to establish its basic plausibility.

\section{Origin of the Impactor \label{section4}}

For a Mars-sized impactor to come close to Earth's orbit, the inevitable 
question is: Where did the impactor originate?  Although large planetary 
impactors have been hypothesized in scenarios of the formation of our
own moon, or else to account for the tilt of the spin axis of Uranus, these
events were currently postulated to have occurred during the first few hundred 
million years since the formation of the Solar System.  At 750 Ma, the Solar 
System would have already gone through more than 3.8 Gyr of evolution,
thus it should have long since come into a dynamically stable configuration, 
at least for the giant planets (Laskar 2001).  Thus, we have to expand our
vision to outside of our own Solar System.

We argue that a more plausible mechanism for the generation of the impactor(s) 
is the encounter of our Solar System with the Galactic spiral density wave crests 
as the Sun orbits around the Milky Way Galaxy.  Recently-advanced theories of 
the dynamics and evolution of spiral galaxies (Zhang 1996, 1998, 1999, 2016, 2017;
Zhang and Buta 2007, 2015; and the references therein) indicate that 
galactic spiral arms are sites of gravitational collisionless 
shocks, due to an intrinsic azimuthal phase-offset between the 
density and potential perturbation patterns of a galactic spiral density
wave mode (Figure \ref{Fig15} and Figure \ref{Fig16}).  This inherent phase offset 
leads to a temporary local gravitational instability at the spiral arms
(Figure \ref{Fig16} Frame b), and a significant reduction of the effective 
mean-free-path for particle scattering at the spiral-arm instability, owing to the 
collective/cooperative motions of the streaming stars participating
in the support of the spiral density wave mode (Zhang 1996).  
This cooperative behavior changes the disk galaxy from
a collisionless system (Binney and Tremaine 2008) to an effective collisional (or scattering)
system, with the collisional/scattering mean-free-path approximately
equal to the width of the modal instability structure,
which is on the order of 1 kilo-parsec (or 3 kilo-lightyears)
for the Solar Neighborhood (Zhang 1996).

The presence of local gravitational instability and potential-density phase shift 
indicates that the spiral arms of density wave modes are themselves propagating 
fronts of collisionless shocks (Figure \ref{Fig16}, especially Frame d).  
Since the streaming matter (i.e. stars and gas) in a galaxy disk generally rotates 
differentially (in the inner galaxy the matter rotates with higher angular speed 
than in the outer galaxy), and a single density wave mode usually rotates with a 
constant pattern speed which is intermediate between the angular speeds of the inner 
and outer galaxy disk matter (the complications of several nested density wave modes 
in a single galaxy, which can have multiple pattern speeds, were discussed in Zhang 
and Buta 2007, 2015, and shown here in Figure \ref{Fig15}c, \ref{Fig15}d -- yet within 
a broad range of the galaxy's radius there is generally still only a single mode and 
a single pattern speed of the wave mode), within the so-called Corotation Radius 
of the mode (where the wave and the disk matter rotate at the same speed) 
the streaming disk matter experiences excess compression
while crossing (or over-taking) the spiral arms (Zhang 1996;
Figure \ref{Fig15}b).  This gravitational-instability effect due to the
nonlocality of the field (i.e., potential and density not being 
aligned as a result of the presence of the phase shift)
can both directly perturb the stellar orbit, and also trigger
the formation of massive stars.  These massive stars evolve quickly and die
in spectacular explosions as supernovae.  The blastwave of
a supernova (a rapidly expanding shock wave of material consisting of
the majority of the mass of the original exploding star)
can further perturb the orbits of nearby stellar and planetary objects,
or else can lead to the formation of new objects out of the debris
material, which may acquire significant peculiar velocities.
All of the above-mentioned perturbative effects can lead
to the formation of rogue exoplanets, which have the potential
to invade into the inner confines of our Solar System.  One possible candidate
for such an invader in our present Solar System is dwarf planet Pluto, which has a moderately
eccentric and inclined orbit, and which has its spin axis in gross misalignment with respect to the 
orbital plane of the majority of the giant planets (except Uranus) in the Solar System.  
Other possible invisible intruders that lie outside the observed edge of our 
Solar System may manifest as gravitational perturbations to the outer giant planets' orbits, as
recent observations seem to indicate.

In Figure \ref{Fig17}, we present a compiled graph of previously published data on 
both the largest extinction events on Earth during the Phanerozoic Eon (red histogram, 
with relative intensity scales), as well as known major tectonic events throughout Cambrian 
and Precambrian periods (blue lines, scale not calibrated).  We see from this
figure that the approximate period of the largest extinction events during the 
Phanerozoic Eon (especially the clustering of these events) is about 250 million
years.  This period is also close to the spacing of the major tectonic events in 
the last segment of the Precambrian.  Our 750 Ma event continues this quasi-periodic
trend, even though at this Precambrian epoch there was not yet hard-bodied fossil 
record of complex organisms.  The spacings of the blue lines are seen to increase 
to about 300-400 Ma towards earlier Precambrian, which is expected if the Solar system 
lied further out in the Milky Way Galaxy at that time, and migrated slowly inward
with time with reduced period (Zhang 2017).

In what follows, we show that the approximate period of 250 million years 
for Earth's major tectonic and extinction events in the Phanerozoic and
late Precambrian, is similar to the present period that the Solar System encounters 
a Galactic spiral density wave arm or spur.
We assume that the Milky Way has a two-armed spiral structure (Fux 1997, 1999 
and the references there in), which is common along all of the observed disk galaxies 
(the occasionally reported 4-armed spiral structure for our Galaxy may simply have 
counted the inter-arm spurs as major spiral arms, or else have counted the response 
of the gas to a two-arm spiral potential forcing, and these spurs could lead to the minor
peaks in the extinction/tectonic event plot.  Other peaks of the extinction plot 
may also be related to the Suns periodic crossing of the Galactic plane 
in the vertical direction).  Using a pattern speed for the Galactic spiral 
structure of $\Omega_p = 13.5$ km/sec/kpc, and the circular speed of stars
at the Solar neighborhood of $\Omega = 220 km/sec/8.5kpc$ (assuming the
Sun's location in the Galaxy is 8.5 kpc from the Galactic Center, and its
circular speed around the center of the Galaxy is 220 km/sec, Binney
and Tremaine 2008), we obtain (for two-armed spiral) that $2 (\Omega - \Omega_p) = 
2 \pi / 250 Ma$, or that the spiral-crossing period is similar to
the Galactic Year at the Solar radius.  These facts lend support to
the idea that the Galactic environment might have provided periodic
sources of impactors to power both the major extinction events, as well as 
major plate-tectonic events, on Earth.  Other Solar-System giant planets might have 
experienced similar giant impact events as well.  For example, Venus is known to have 
a young crust of around 500 million years, and its spin is in the opposite direction 
to its orbital motion.  Another well-known example is Uranus, which has its spin axis 
tilted 90 degrees from the plane of the Solar System (which is itself roughly aligned with
the plane of Galactic rotation).  Mars and Saturn have moderate levels of spin-axis tilt,
similar in amount to the tilt of Earth's spin axis.

Because the spiral arms or spurs in galaxies have finite width,
multiple impactors produced during the same spiral-arm crossing episode
of the Solar System can invade Earth's orbit during a short (from a
geological standpoint) period of time.  This may explain the
near coincidence in time of the occurrences of the Decaan Traps in India
and the Chicxulub Crater in the Yucatan, which are both dated to the K/T 
(or K/Pg) boundary but are spatially separated; or else the near coincidence 
in time of the formation of Emeishan Traps in China and the Siberian Traps 
in Russia during the late-Permian/early-Triassic period, with period also 
coinciding with the end-of-Permian mass extinction event and the completion 
of the assembly of the Pangaea supercontinent (Stow 2010; Wignall 2015).
On the extinction-intensity plot (Figure \ref{Fig17}), this trend shows 
up as closely grouped major extinction events, around especially 250 Ma 
and 500 Ma.  

One potential issue with this proposal is the phase of the Sun relative to the
spiral arms of our Galaxy, which, depending on the model used,
may not always put the Sun near a major spiral arm at the current epoch.
This phase offset could be accounted for in several ways. (1).
There is the time needed to form new massive stars after an arm crossing,
and for stellar evolution to carry the massive stars into the stage
to produce new supernova. (2). Depending on the peculiar velocity
acquired by the rogue planet, it takes time for it to reach the outer
confines of our Solar System.  (3). It takes time to gradually dissipate 
the noncircular component of the exoplanet's velocity when it starts to 
participate in the Solar System's motion. This dissipation is possible because
of the known result that a system naturally evolves towards the
configuration of lowest energy, which, for a disk configuration,
is that of a circular orbit for given amount of angular momentum
(Lynden-Bell and Kalnajs 1972). (4).  Over an even longer time period, the 
inner and out disk will exchange angular momentum as well, so inside 
corotation the mean orbital radius of a planet will decay secularly,
and outside corotation the orbital radius of a planet generally
increases.  Depending on the particular entry parameters of the rogue planet,
these dissipative and secular evolutionary effects may bring
it gradually to near-Earth orbit with a grazing impact condition.
This scenario, of course, may not be realized for every intruder
into the Solar System.  Even if a possible invader, such as Pluto,
had reached the outer Solar System, it may take tens or event hundreds of
millions of years for it to cause a giant impact event on Earth
(or on another giant planet of our Solar System), if at all.

For now, We can at least take face value of the statistical correlation of the period 
of the major extinction events, as well as the period or half-period of 
the supercontinental cycle, with the period of the Galactic spiral-arm-crossing
at the Sun's orbital radius, and say that there is now 
a plausible source for large impactors from our Galactic environment, and
the supply of these impactors can occur with a period similar to the
observed period of major extinction and tectonic events on Earth.

The recent discovery by European Space Agency's Gaia satellite that
Gliese 710, a star which is about 60\% as massive as the Sun,
is likely to have a close encounter (to the distance of the Oort Cloud)
with our Solar System in the next 1-2 million years (Bailer-Jones et al. 2018); 
as well as another inferred possible close encounter of our Sun with the 
binary system Scholz's star 70,000 years ago (Mamajek et al. 2015), shows that
we indeed cannot regard galaxies, especially our own Milky Way, as collisionless
systems (in galactic dynamics, a collisionless galactic system includes no
binary close encounter or scattering, as well as no head-on collision),
and the spiral density wave modes in galaxies are the most likely provider of
gravitational perturbations to invalidate the collisionless assumption. 

\section{Further Supporting Evidence}

\subsection{Age Determination, Contact Relationship, and Deformation History
for Precambrian Rocks in the Colorado Plateau Area}

Much of the Colorado Plateau interior is covered with sedimentary rocks
of the Phanerozoic Eon, which makes discerning of the effect of
earlier processes difficult.  However, there exist significant
outcrops of Precambrain rocks, mostly along major faults in the Plateau's
boundaries, as well as along isolated interior faults (such as the Grand Canyon 
Inner Gorge).  See, for example, the Geologic Atlas of the Rocky Mountain Region (1972).

For the purpose of substantiating the 750 Ma impact event, we would like
to identify the top layers of the surviving lithology right after
the impact, in order to gather evidence for impact-induced alteration
and movement in these Precambrian rocks.

The outcrops we are most interested in, first of all, consist of
the Proterozoic quartzite, slate, and other meta-sedimentary rocks of
the so-called ``cover sequence'', which invariably lack
hard-body fossils.  Accurate dating of such
meta-sedimentary rocks on the Plateau had been carried out most
successfully for the Chuar Group of the Grand Canyon Supergroup 
(Karlstrom et al. 2000), as well as for the Uinta Mountain Group and 
its western extension the Big Cottonwood Formation (Dehler et al. 2005, 2010).  
Neoproterozoic ages of 740-770 Ma have been obtained for the
youngest sediments of both of these groups. 

Erslev et al. (2004) have dated K-feldspar in pegmatite dikes in the 
southern Sangre de Cristo Mountains, and have found components with ages 
between 600-800 Ma, which they attributed to the breakup of the
Rodinia. 

For the Uncompahgre Formation of the western Colorado, 
on the other hand, less certain age assignments have been            
obtained.  Although Paleoproterozoic depositional ages have been obtained
for certain horizons of the Uncompahgre Formation
(see, e.g. Karlstrom et al. 2017 and the references therein),          
signals of younger Precambrian ages in fact have been found when analyzing the samples
of Uncompahgre Formation of the Needle Mountains (Dean 2004;
Wu 2007).  However, in these instances the data points showing younger ages were
treated as spurious and incompatible with the age constraints set by previous
workers (i.e., Barker 1969), who first proposed that the Uncompahgre
Formation must be older than the intruding Eolus Granite (circa 1.4 Ga in age)
with which it is in contact with in the Needle Mountains area of Colorado.
Barker, to be sure, did highlight a so-called ``Uncompahgre disturbance'' event, 
responsible for the intense folding and attendant low-to-high rank metamorphism of 
the Needle Mountains rocks, which occurred at an unspecified time 
after the deposition of the Uncompahgre Formation.

From our previous discussions, we see that this ``Uncompahgre 
disturbance'' event discussed by Barker is likely the giant impact 
event at the 750 Ma, which is seen to have scrambled together fragments 
the basement and the cover sequence rocks, as well as the Eolus Granite. It also
produced conglomerates (i.e. in the Vallecito Reservoir area) 
consisting of material from all the Precambrian rocks.  Therefore, the
Uncompahgre disturbance necessarily happened AFTER the intrusion of
the Eolus Granite at 1.4 Ga.

Since the proposed event had thrown entire mountains of Uncompahgre Formation 
as well as parts of the Irving Formation of the basement rocks upon and around
the Needle Mountains Eolus plutons (as evidenced by the upturned morphology
of rocks making up the entire Grenadier Range in Needle Mountains) , we no longer have the need to interpret
the contact between these country rocks and the Eolus Granite as intrusive, though
this does not exclude the possibility that certain segments of the contact
could be the original intrusive contact.  We want (and need only) to demonstrate 
the fault contact nature of portions of Uncompahgre Formation with Eolus, 
so as to assert that higher horizons of Uncompahgre Formation COULD BE younger than
the Eolus intrusion age of 1.4 Ga, even if these horizons in Uncompahgre Formation
had since been lost through either impact vaporization or later erosion, or else lay
in remote eegions of the Colorado Plateau awaiting further dating analyses.
This expectation is only reasonable given that the Plateau region was flat during
the time of the proposed impact (Blakey and Ranney 2008, 2018), so the Neoproterozoic 
sediments discovered in Arizona and Utah SHOULD have existed also in Colorado, 
and in New Mexico.  

We stress that the reviously observed gradient in metamorphic grade of 
Uncompahgre Formation and Irving Formation in proximity to
the Eolus contact (Noel 2002; Deen 2004; Wu 2007), which was used partly
to establish the intrusive-contact nature between Eolus and Uncompahgre,
could just as well be a result of dynamics/pressure-induced metamorphism 
(or else a result of dynamics-induced thermal effect, rather than 
granite-intrusion-induced thermal effect), due to the impactor's forcing (i.e.
ramming) of these transported country rocks into contact with the preexisting Eolus.
The syclinorium shape of the Uncompahgre Formation north of the Eolus Granite
Plutons supports such a scenario, as is the determination (Dean 2004; Wu 2007) that the metamorphism
of Uncompahgre in this region occurred at the same time as its structural change
(i.e., the intense folding to conform to the shape of the outer contours of the Eolus pluton). 
This newly proposed scenario also alleviated the difficulty in interpreting
the Noel (2002) observation that the $S_3$ foliations of Irving Formation,
which are in the eastern contact of Eolus Batholith with the Vallecito 
River Valley, appear to have developed during contact metamorphism 
(since they are defined by sillimanite and hornblende), 
yet are east-west striking and subvertical, and transect the contact with Eolus 
pluton at a high angle, thus cannot have developed due to ballooning of the pluton.
If such foliation developed through impact-induced dynamical metamorphism
during the impactor's traverse from north to south, these observed features
would be naturally explained.

Furthermore, since Dean (2004) and Wu (2007) had
both demonstrated the correlation of macroscopic deformation and
microscopic metamorphism of the Uncompahgre rocks in this region,
we expect the same gradient in the degree of metamorphism to hold also for a segment of
the Uncompahgre Formation synclinorium next to the Twilight Gneiss 
in the northwestern region of the Needle Moubtains (Zinsser 2006, Figure 2).
The laboratory demonstration of this prediction will confirm the role of dynamical
metamorphism without the need for Eolus granite to
create contact aureole and thermal metamorphism.  
The ``slapping on'' interpretation of the origin of some of the country rocks 
around Eolus Granite revives the original proposal of Cross et al. (1905) 
that some of the contacts between Eolus and country rocks are faults.
The possible ductile shear contact between Uncompahgre Formation and
Irving formation in western Needle Mountains was also the
original motivation of Tewksbury (1985a) to interpret Uncompahgre 
Formation as allochthonous. Grambling et al. (1989., p. 91) report
cases in southern Sangre de Cristo mountains where the contact
between Mesoproterozoic granites and country rocks show no
contact aureole, and one apparently shearing contact separates rocks
which differ by 6 kbar in peak metamorphic pressure.

A significant portion of the distortions of the Uncompahgre
and Irving Formations in Needle Mountains is north-south shortening, east-west
extension (Harris 1990; Gonzales et al. 1996; Noel 2002; Zinsser 2006).
This is irrespective of whether the country rocks are located to the
north, south, east, or west of the Eolus Granite formations.
This shows that the disturbance that created these distortions
cannot be due to the ballooning of the Eolus granite, as
many previous studies have assumed, but rather is on a scale
much larger than the Needle Mountains.  Furthermore,
in the Uncompahgre Gorge region, the distortion to the lithologies and the
type of dynamical metamorphism in cover sequence rock appear to continue that of the 
Needle Mountains (Harris 1990; Wu 2007).  Since the Gorge is many tens
of kilometers away from the influence of the Eolus intrusion,
it is impossible to attribute these signatures of dynamical
metamorphism (i.e., two distinct sizes of quartz grains, with
the smaller ones attributed to the newer dynamical disturbance)
to the intrusion of Eolus Granite.  On the other hand, the
giant impact scenario can easily account for the similarity
of distortions in cover sequence rocks in these two areas.

The giant impact scenario can also account for the polyphase deformation scenario 
porposed in Harris et al. (1987) and Harris (1990):  During the impactor's initial touch-down,
it vaporized the top-most layers of sediments, and also created a kind of 
domino-effect and pushed cover-sequence rocks northward in spreading layers,
in what Harris et al. (1987) called ``north-directed, thin-skinned thrusting''.
Subsequently, the impactor traversed southwestward, which
generated vertical folds, and pushed the rocks southward against 
the Eolus Granite, which further distorted these folds.
Harris (1990) emphasized the continuity of the three
phases of deformation found in the cover sequence.  Harris et al. (1987)
also considered the possibility that some of the deformation could have
occurred post Eolus intrusion. 

We quote here also a passage from Baars (2000, p.117) in his discussion
of the Uncompahgre Formation in the western San Juan Mountains:
``Sometime after the sediments were deposited, but before Paleozoic
events began the sedimentary layers were altered somewhat to form
quartzites, which are highly cemented sandstones, and slates,
which are slightly metamorphosed shales.  The degree of alteration was
only slight when compared to the older basement complex, but the thick
deposits of sedimentary rocks are in rather poor condition when compared
to the younger Paleozoic strata.
The younger Precambrian sequence is called the Uncompahgre Formation
for exposures of the highly folded strata in Uncompahgre Canyon
immediately south of Ouray.  The formation may be approximately
the same age as the Grand Canyon series of red beds, which are
sandwiched between the Vishnu Shist and Tapeats Sandstone in the depth
of the Grand Canyon.  However, many geologists who have studied
the two areas believe that the Uncompahgre is somewhat the older,
as it is more highly metamorphosed than the Grand Canyon
sequence.  This may have resulted from a more severe history of
tectonic activity (mountain building) in the San Juans, however, and
may reflect nothing about the relative ages of the two units".

So, it is clear that some previous-generation geologists had in fact speculated
on the possibility that the metasedimentary rocks across the Colorado Plateau
are contemporaneous, which makes sense because these sediments
originated from the same shallow-sea environment, and reached
similar deposition depth of several kilometers (Baars 2000, p. 117),
before apparently suffering the 750 Ma impact event which led
to their varying degrees of metamorphism and folding (the exact
degree of metamorphism would depend in part on their relation to an
existing or newly created fault, as well as on the location with
respect to impactor's trajectory and surface topography), as well as
material loss through vaporization.  Subsequently, 
the upturned and folded quartzites and slates also suffered differing
degrees of erosion, during repeated faulting actions throughout the
phanerozoic eon (Baars 2000, p. 119-121).  For example, the uplift and
erosion experienced in the Colorado Front Range, along the Plateau's
northeastern boundary, are significantly more than in the Plateau's
western boundary.  

The post-Eolus giant impact event also helps to resolve the apparent conflict between 
the need to create synmagmatic strain field involving N-S contraction 
and E-W extension to account for the deformation characteristics of the Proterozoic 
country rocks around Needle Mountains pluton (Gonzales et al. 1996), and the
internal characteristics of many of the 1.4 Ga plutons in southern
Laurentia (of which Eolus is a typical member), i.e., being of A-type chemistry or being 
anorogenic, which implies that these plutons should have been emplaced in an extensional, or 
rifting, environment (Anderson 1983; Hoffman 1989).  The impact scenario
can also easily account for the fact that many of the Mesoproterozoic
anorogenic rocks in the southern Rocky Mountains area are moderately to
intensely foliated: ``The foliations are solid-state deformational fabrics,
and in places the foliated middle Proterozoic plutons are separated
by shear zones from supracrustal rocks lacking any contact metamorphic
overprint.  One is left with the suspicion that a major component
of ductile deformation occurred more recently than 1500 Ma in areas as
widely separated as central Colorado and Central New Mexico'' (Grambling
and Tewksbury 1989).  We would certainly agree!

\subsection{Breakup of Rodinia Supercontinent}

We have previously commented that on the western part of the Rodinia 
Supercontinent, the 750 Ma impact event likely caused the rifting of 
Australia and East Antarctica from the proto-North-America continent 
Laurentia (Moores 1991). The hingeline of this separation on the
North American Continent is located in Idaho, Oregon, eastern Washington, 
as well as in western Canada along the western front of the Canadian Rockies.  
It is also responsible for the dislodging of the Colorado Plateau from 
its surrounding craton, as well as for creating a multitude of weaknesses 
in the Plateau's basement, causing the eventual formation of 
the Colorado River system, among other things.  In corresponding regions
of Tasmania, Australia, Proterozoic sediments similar to that in the Grand
Canyon region of the U.S. had been identified in recent years
(Mulder et al. 2918).

The damage to the Rodinia Supercontinent, furthermore, is not limited to
its western regions.  On the eastern side of the Rodinia supercontinent, 
close to the present Appalachian Mountain chains, evidence of a similar
continental breakup is also abundant.  In Figures \ref{Fig18}, \ref{Fig19},
the Neoproterozoic suevite/conglomerate formations in the Grandfather Mountain 
region of North Carolina are shown, with evidence for impact melting and 
fusing prominently displayed. Plutonic rocks with ages dated to be close to
750 Ma have also been extensively studied (King and Ferguson 1960;
Su et al. 1994; Goldberg et al. 1986; Ownby et al. 2004).
Figure \ref{Fig20} shows possible impact shock metamorphism signature
in Precambrian rocks off the Skyline Drive in Shenandoah Mountains area of Virginia.

Additional evidence of the breakup of Rodina is abundant from the
west side of the original supercontinent.  Figure \ref{Fig21}, Top Frame
shows a giant block of Neoproterozoic igneous intrusion 
(dated to about 750 Ma, see Gadd 2008) into metamorphosed Archean
rocks (Bottom Frame, and part of the metamorphic effect may have occurred
during the 750 Ma event), near Toad River Bridge along the Alaska Highway, 
in British Columbia, Canada.

The breakup of the Rodinia Supercontinent eventually led to the
formation of two subsets of supercontinents Gondwana and Laurentia, as well
as the formation of the proto-Atlantic Ocean, the Iapetus Ocean
(Blakey and Ranney 2018).  These breakups, apart from owing to
weaknesses in the Lithosphere created by the impact, may also had been affected
by the global isostasy effect which leads to the major redistribution of landmass,
aided by the low viscosity zone between the lithosphere and asthenosphere
(more on that in \S\ref{broad}), which could account for the apparent delay between the timing of
shock metamorphism and igneous activity signatures (0.7-0.8 Ga) and the timing
of the actual rifting (0.6-0.7 Ga) of the different regions of the Rodinia 
supercontinent.

The coordinated rifting of the entire Rodinia Supercontinent at around 750 Ma,
especially the synchronized igneous activities across the continent, shows that
the impact force involved must be great enough to affect the whole Earth simultaneously,
another piece of strong evidence in support of a giant impact scenario.

\subsection{Formation of the Great Unconformity}

The dislodged Colorado Plateau did not start its uplift right away, 
due to the confining pressure from its surroundings. 
Therefore, immediately after the impact a flattened landscape should have existed
at the Plateau region, due mostly to impact-induced evaporation of material.
This flattened landscape in late Precambrian and early Cambrian was confirmed
in the paleo-geological maps of Blakey and Ranney (2018) for that time period.
Apart from residual crushed material post impact there was no tall mountains to erode, 
and no significant re-sedimentation after 750 Ma, until the beginning of the Cambrian 
(540 Ma) for most regions, though in some area detritus classified as Cryogenian 
(circa 635 Ma) was also found.  This explains the widespread occurrence of the 
Great Unconformity, the exact age gap of which depends on the locale such unconformity is
observed (i.e., the actual age gap for each region depends on how much of the 
Precambrian cover sequence or even basement rocks were removed during the impact event, 
and how soon the surrounding topography and other environmental conditions
work together to allow regional sedimentation to re-start).  

As we had mentioned before in this paper, in many regions of the Colorado Plateau,
the Cambrian sediments immediately above the Precambrian top layer appear
to be re-lithified remains of charred and melted Precambrian fragments
derived locally (see, e.g. Figure \ref{Fig2}).  Thus there is some leeway in the 
accurate assignment of ages to these recycled local rocks, since the component material 
had presumably been sitting there from 750 Ma until the epoch when re-lithification 
had bounded the fragments, according to the scenario presented in this paper.
Other forms of presumed impact detritus in the Plateau area and its surroundings
are pebble, cobbles, boulders and various forms of conglomerates
made from Precambrian material sourced either locally or from a distance region
(Condon 1995).

The varying degrees of age and metamorphism for the upper layers of the exposed
Precambrian metasediments across the Plateau and in the northern
Rocky Mountains likely reflect their varying locations 
during the impact event, with the fault-protected units
(i.e., the Grand Canyon Supergroup), or area away from the direct impact
crater (i.e. the Belt series in the norther Rockies) avoiding high-grade 
metamorphism as well as impact-induced evaporation, whereas most of
the Uncompahgre Formation of the San Juan Mountains apparently suffered both a 
higher degree of impact shock metamorphism (Baars 2000, pp. 117-119), as well as 
uplift-induced subsequent erosion.  On the other hand, near the Flaming Gorge area
in Utah, the degree of metamorphism of the Uinta Mountain Group rocks within the
same horizon can change drastically within a short distance, due possibly to the 
differing positions of the rocks with respect to the local faults and
shock wave propagation (the Uinta Mountain front lies in the area where
the impactor is expected to have made the initial substantial contact with
the Earth's crust).

Further north from the Colorado Plateau, in Wyoming, younger Precambrian 
quartzite cobbles and pebbles litter the basin of Teton Village and 
Jackson Hole community.  Love et al. (2003, p. 59) stated that these quartzite
pebbles do not have a local source, and they appeared similar to the metamorphosed 
Proterozoic sedimentary rocks in southwest Montana and nearby parts of Idaho.  
It is possible that in these northern neighborhoods of the Colorado Plateau,
some of the quartzite pebbles may have been thrown out of the impact
crater (i.e., the Colorado Plateau), which would explain the
nonlocal nature of the quartzite pebbles in the Jackson
Hole area.  Similar types of pebbles are also found on the Plateau 
itself (Condon 1995), for example in the Vallecito Creek area of Needle Mountains. 
One might question whether the current distance of the
Jackson Hole area to the boundary of the (traditionally defined)
Colorado Plateau might be too far for this scenario to be possible.  
Here we must take into account that the Basin-and-Range extension
dynamics since the Cenozoic Era has led to enlargement of the distance 
between landmarks by about a factor of two, and the Yellowstone-Teton-Snake 
River Basin area has been shown to be fully affected by such extension 
dynamics, just as in the neighboring Nevada.  Once this is taken into 
account, the distances involved become entirely within the dimensions 
of impact-excavated crater outer rims (Dence et al. 1977; 
Melosh 1996).  Furthermore, Dutton (1882)'s original consideration to 
include part of the Rocky Mountains north of the Uinta Uplift into the 
definition of the Colorado Plateau may reflect the fact that 
this northern addition to the Plateau (put into its right place before the
Basin and Range extension dynamics had stretched it northwestward)
could be the initial touch-down area of the impactor. 
The fact that in an oblique impact event the uprange jet in the rear
of the impactor is much more powerful than the down range jet in
front of the impactor (Melosh 1996, p.49) could account for the fact that
the most numerous cobbles and pebbles of Neoproterozoic age
are found in the northern direction of
the Plateau (i.e. in Wyoming).

\subsection{Spin-Axis Tilt and Polar Wander}

It is well known that the Earth's spin axis is tilted from the
orbit plane of the Earth in the Solar System by about 23.5$^o$.  
From angular momentum considerations, the Earth's spin axis is expected 
to be aligned with its orbital angular momentum axis at birth, so the
tilt of the spin axis of the Earth is likely produced by later
catastrophic events, such as giant impacts.

If this has been the case, the currently measured spin-axis tilt would be
the cumulative effect of multiple past impacts.  Since the directions
of these past impact events are uncorrelated, the additional contribution to
the total tilt amount by each impact event should sometimes reinforce
one another and sometimes cancel.  Still, we want to look into
whether a giant impact of the magnitude proposed in the current
paper would produce an incremental tilt smaller than or on the order of the 
currently-observed spin axis tilt of the Earth, since if the proposed event
had instead produced an exorbitant amount of additional tilt, then 
the likelihood of such a scenario is more in doubt.

The spin of the Earth corresponds to a surface rotational
speed at the equator of 0.46 km/sec.  As a result of the proposed 750 Ma
impact event, the impactor's velocity in the tangential direction (i.e.,
parallel to the surface of the Earth) was reduced by about 10\% of its initial
value (i.e. to 22.5 km/sec from the initial 25 km/sec, see \S\ref{kine}).
Since the impactor's mass is about 10\% of the Earth's mass,  
we expect the Earth's surface velocity change to be on the order of
0.2 km/sec, from momentum transfer consideration.  Thus the proposed event is of
the correct magnitude to be able to produce a 23 degree tilt of
the Earth's spin axis, with the quantitative result depending
on the actual orientation of the impactor's trajectory with
respect to the Earth's spin direction at the epoch of impact
(i.e., we need to consider vector addition
of the angular momentum of the original spin and the 
excess spin due to impact.  An accurate account will also need to consider
the mass distribution of the whole Earth).  The reasonableness of the
proposed impact scenario is nonetheless reinforced by the past
numerical simulations of the formation of our Moon (Benz et al. 1986,1987,1989;
Cameron and Benz 1991; Reufer et al. 2012; Canup 2012; Wyatt et al. 2016;
and the references therein), which normally
assumed a Mars-sized impactor on the proto-Earth, and which also succeeded
in reproducing the current tilt of the Earth's spin axis.  Of course, many
such impact events were likely to have occurred in the Earth's history,
and the current tilt axis of the Earth reflects the net effect
of all such impact events in the Earth's history.  Another point
of reference is that Neptune has its spin axis almost at 90 degrees
to its orbital angular momentum plane around the Sun.  This is usually
reproduced in numerical simulations by an assuming an Earth-sized impactor
which hit Neptune early in the history of the Solar System. 

Besides changing the Earth's spin-axis tilt and producing true polar drift in space, 
the past impact events were likely to be responsible for producing the (apparent) 
polar wandering phenomena of the Earth.  Wegener himself had already commented on 
the possible correlation between polar wandering and the redistribution of the 
continental mass in his monograph (Wegener 1929).

\subsection{``Snowball Earth'' and Origin of Certain ``Dropstones''}

Giant impact events are expected produce dusty atmosphere which prevents 
sunlight from reaching the Earth's surface for prolonged periods of time .  
The well-known correlation of ``Snowball Earth'' (continental-scale glaciation)
episodes with mass extinction episodes lends support to the idea that giant impacts
could be the simultaneous trigger of both Snowball Earth and major 
mass extinction events.

A well-known episode of Snowball Earth event occurred after the
breakup of Rodinia, between the time period of 720 Ma and 635 Ma. 
On Antelope Island within the Great Salt Lake of Utah, evidence of
this great continental glaciation period have been well preserved
(Hayes et al. 2013 and the references therein).
Previously, many of the conglomerate formations have been interpreted
as ``dropstones'' deposited by advancing and retreating glaciers
of the Cryogenian era (i.e. between 720 - 635 Ma), after the Rodinia
breakup.  These conglomerate formation are also found further north
up to regions of Canada.  However, closer examination shows that a 
significant portion of these conglomerates are likely to be giant-impact 
induced, rather than formed as a result of glacial deposit and compactification.  
An example in favor of this interpretation is shown in Figure \ref{Fig22}, 
where clear signature of impact melt is displayed within the matrix of
a giant slab of Neoproterozoic conglomerate located in the northern
end of the Antelope Island State Park, Utah.

\section{\label{broad}Broader Implications}

\subsection{Multiple Giant Impacts in the Earth's History}

The proposed giant impact event at 750 Ma centered on the Colorado
Plateau is only one example of such giant impacts which might have
occurred multiple times during Earth's history, starting from
the well-established Formation-of-the-Moon episode which was 
commonly believed to have occurred during the Earth's infancy. 
As noted early on by T.C Chamberlin, and quoted in Van Hise and Leith (1909): 
``The groups Paleozoic, Mesozoic, and Cenozoic are not, in the
conception of some of us, defined by a distinct kind of life,
as in the case pf some of the minor horizons, for the life
fits better the idea of a gradation than of a distinct separation.
These great divisions are rather, as I see it, at least,
great historic movements fundamentally dependent on dynamic
events, than paleontological divisions.  Originally they were
supposed to be separated by universal catastrophes to life, and
their distinctness was due rather to the intervention of the
catastrophe than to the different quality of the life, which is
merely seized upon as a characteristic suited to nomenclature.''
The episodic nature of the major tectonic and extinction events 
graphed previously in Figure \ref{Fig17} lends support to such
multiple-giant-impact viewpoint.  The well-known distribution
pattern of sediments across stage boundaries, i.e., starting from
coarse conglomerates, refining gradually upward, is also an indirect
support for the giant impact interpretation.

A well-known and well-studied example of a major impact event
is the Sudbury Impact Crater in Ontario, Canada (see the many contributions in 
Dressler and Sharpton 2000), which has an estimated age of 1.85 Ga.
The original (before modification) size of the crater was estimated to be 130 km.
The Sudbury impact crater is the third-largest on Earth, 
after the 300 km Vredefort crater in South Africa, and the 150 km 
Chicxulub crater under Yucatán, Mexico (these rankings were of course made
before the proposed Colorado Plateau impact crater with diameter of 640 km).  
Shatter cones and shock-altered
quartz crystals were found in the Sudbury Basin area. 
The original crater had gone through tectonic alterations since
its initial formation, the most notable of which is the circa
1.1 Ga Grenville Orogeny.  Even though the impact-altered rocks
and the tectonically-altered rocks partially overlap in their
geographical location, the signatures of these two kinds of
processes are distinctly different.  The Left and Right Frames of
Figure \ref{Fig23} show examples of rocks in the Sudbury area, that
were affected predominantly by each one of the above two processes. 
We can see clearly that the rock alterations in the Colorado Plateau
area, which we had presented previously in especially Figure \ref{Fig3} and 
Figures \ref{Fig5} - \ref{Fig9}, more closely resemble 
the impact-altered rocks in the Sudbury area, rather than the tectonically 
altered rocks (produced during the Grenville Orogeny) in the
eastern outskirts of the Sudbury Basin.
In essence, the difference is partly that of brittle deformation
versus ductile deformation, and partly impact-induced partial
melting and shock metamorphism, which do not show up in tectonically
induced deformation.

Another giant impact event, also with a Mars-sized impactor, had previously been proposed
for the Pacific Ocean region during the mid-Jurassic epoch (H. Zhang
1998, 2016; H. Zhang et al. 2013; Li et al. 2014), with evidence
across the Pacific Rim both in the Asian countries, especially
in the east coast of China (i.e., the Yan-Shan Movement), as well 
as in the west coast of the US and Canada (i.e., the formation of 
the Franciscan Melange and the initiation of Nevadan Orogeny). 

The Jurassic giant-impact event in the Pacific, in addition to
causing widespread deformation on land across the Pacific Rim,
was likely responsible also for initiating the shrinking of the Pacific Ocean 
(through the breakup of its original oceanic plate and the subsequent regrowth 
of the ocean floors), which in turn might have been responsible in
part for the opening up of the Atlantic Ocean, which continues to this day,
and for the initiation of the Basin and Range movement in western US
as well as in eastern China.

We propose therefore that the kinematic effect of a giant impact, as well as 
the gravitational equilibrating effect of the displaced continental material, 
may serve as major driving forces in the post-impact evolution of the Earth's 
tectonic plates, in addition to the traditionally attributed slab pull, 
ridge push, as well as other thermal and dynamical driving forces (see, for
example, Price 2001 and the references therein).  This 
newly realized importance of {\em global isostasy} (i.e. the tendency for 
continental plates to achieve global dynamical equilibrium by moving in the 
horizontal directions on a spherical Earth's surface),
means that we may have come full circle in understanding the origin
of Earth's geomorphology:  Instead of plates being pushed around by
convection currents from underneath, the gravitational and kinematic
interactions of plates from above, as a result of giant-impact-induced
mass and energy re-distribution, may be the major driving engine for
continental drift.

\subsection{Mantle Convection and Plumes}

What then about the currently-popular mantle convection and mantle plume picture?
The history of the emergence of this picture is well documented in Sullivan (1991),
chapter 5 ``The Mantle Controversy''.  The entire picture was adopted not because
of overwhelming evidence in support of it, but rather, giving the overwhelming 
evidence in support of plate motion, and the lack of a suitable candidate mechanism 
for driving this motion, the mantle convection was forced onto the stage in an act 
of desperation.

In fact, even at the time of inception of the mantle convection proposal, much evidence 
went against it, as quoted in Sullivan (1991).  Gordon MacDonald, in particular,
presented ample evidence for a rigid interior of the Earth, and that as a result
much larger temperature gradient needed than observed is needed to drive any 
global-scale mantle convection.  Taking ``realistic values'' for the plasticity of 
the interior, MacDonald said, ``we see that the thermal requirements of convection 
in the mantle are so severe that the hypothesis can be dismissed''. 

Sullivan quoted also a paper by Bullard et al. (1956), early proponents of the mantle
convection mechanism for generating the heat flow which was observed from ocean floor:
``In brief'', they wrote, ``it seems not impossible that the objections to the
existence of convection currents in the mantle can be overcome by reasonable
assumptions about the properties of the material of which it is composed.
That the mantle does really possess these properties is not self-evident.
In fact, the main reason for supposing that it does is the desire to have
convection currents to account for the oceanic heat flow.''

Decades later, we are nowhere closer to confirming the properties of the
mantle that would support a continent-wide convection current.  The mantle plume
variation to account for the heat emanating from certain openings in the Earth's
crust, on the other hand, can be countered by a ``chicken and egg'' type of argument,
i.e., some of the heat flows above triple junctions could be results of
continental rifting, rather than the cause of it.

Given the likelihood of giant impacts in the Earth's history, and their
effectiveness in driving both the initial rift and the subsequent redistribution
of landmass through gravitational force, the lack-of-viable-driving-mechanism
is not longer a viable argument in favor of mantle convection and its variations
to serve as major driving forces.  Even if some redistribution of mantle
material does occur, this is now seen to be more likely a result of the
redistribution of continental mass, not the cause of it.  Even the heat flow
observed through the continental and oceanic crust, could have been generated
by the frictional force of the continental motion, initially set off by the
giant impacts.

\section{Conclusions}

We have proposed and demonstrated that a diverse range of geomorphological
features of the Colorado Plateau can be naturally accounted for if
they are the results and aftermath of a giant impact event hypothesized
to have occurred around 750 Ma.
These features include: (1) The surprising structural integrity of 
the Plateau during the past 600 Ma, despite the vigorous igneous and 
orogenic activities at its boundaries; (2) The occurrence of the 
so-called ``Great Unconformity'' throughout the Plateau region, as well 
as elsewhere in the world, surrounding the period of the
proposed impact event; (3) The wedged insertion of large chunks of 
the late-Precambrian Grand Canyon Supergroup sedimentation sequence into the
basement rocks of the Vishnu Complex, as well as the similarity in age
of the upper-most deposition layer of the Supergroup (the Chuar Group) to the
timing of rifting of the Rodinia supercontinent around 750 Ma from the Plateau's 
western edge, as well as the subsequent rifting of Rodinia along the present
east coast of North America continent; (4) The Plateau-wide presence of 
metamorphosed late-Precambrian sediments which often display evidence of melt and 
shock metamorphism; (5) The thick basalt dikes cutting through basement rocks on 
the Plateau, as well as further north in the central and northern Rocky Mountains, 
which can be dated to about 750 Ma; (6) The Plateau-wide presence of detritus with 
Neoproterozoic upper age, which formed the compositional material of the Paleozoic 
conglomerates and re-lithified sediments after the Great Unconformity; (7) The 
presence of large quantities of pebble and cobbles of Neoproterozoic age in the 
Jackson Hole, Wyoming area, which have no local source, and which are in the down
range direction of the excavation jet of the hypothesized impactor's trajectory.

We further demonstrated quantitatively that a Galactic spiral-density-wave induced, 
Mars-sized rogue exoplanet is likely to be the impactor colliding with the Earth 
at about 750 Ma, which caused the severing of the Plateau boundary from its surrounding
craton, and ultimately led to the formation of the Colorado Plateau we see today after 
subsequent plate tectonic evolution.  The correlation of the density-wave spiral-arm-crossing 
period of our Solar System to the period of the major extinction events on Earth, and 
to the period or half period of the supercontinent cycle, also supports a Galactic 
origin to the Earth's major tectonic cycles.

\section*{Afterword}

The work described in this paper was inspired by the pioneering studies
of Professor Hongren Zhang (1934-2016, the author's late father), former 
editor-in-chief of the Episodes Journal of the International Union of Geological 
Sciences (IUGS) from 1997-2004, and former President of the IUGS from 2004-2008.
Since the early 1970s, Prof. H.  Zhang gradually formed and evolved ideas on 
a possible giant impact origin for the eastern China's geomorphology, and his first 
article outlining his ideas on this topic was published in 1998, as a summary of
an invited address to the Chinese Geological Society general assembly in 1997. 
Afterwards, there was a hiatus of ten years when Prof. H. Zhang was fully occupied 
with the responsibilities of the IUGS. It was only after his retirement in 2008 
from the IUGS post that he was able to focus on this work, together with several
collaborators, during the final eight years of his life.

The idea of a possible giant impact origin for the Colorado Plateau came to 
the present author independently, during an Oct. 2016 field trip to the 
southwestern United States, following her decision to carry on her late 
father's unfinished quest on the general validity of the giant-impact mechanism
for the initiation of many major episodes of Earth's plate-tectonic motion,
a subject the present author had had many illuminating discussions
with Prof. H. Zhang during the past few decades, even though her 
previous field of research was focused on astrophysics.  Since she had taken
an early retirement from formal employment to pursue independent research,
she had the freedom to devote a significant amount of time to learning
the geology necessary for this work, and to go on field trips.

When the first draft of the current paper was near completion at the end of 2017, 
the author came upon quite by accident a popular-science article on the 
Smithsonian Air and Space Magazine website, by Paul D. Spudis (Spudis 2015), 
a senior staff scientist at the Lunar and Planetary Science Institute in 
Huston TX, which speculated on the possibility that the Colorado Plateau 
might be an ancient impact scar.  Spudis (2015) did not offer quantitative 
calculations, nor field studies, since the article was mainly spurred by 
a family vacation trip of his through the area, which did not afford field 
studies.  The epoch of the impact event within the 4 billion years of 
Precambrian was also not specified in the Spudis article, nor the size 
of the impactor.  However, Spudis (2015) did mention that the idea of 
a possible large impact origin for the Colorado Plateau had been in the air for 
some time, and he first learned it years earlier from his own professor 
Carleton Moore, then the Director of Center for Meteorite Studies at 
the Arizona State University.  

So far the author has been unable to find a mention of the idea of a 
(planet-sized) giant-impact event for precipitating the formation of the 
Colorado Plateau in any scholarly publications.  Prof. Hongren Zhang's 
previous studies in China, despite being the inspiration for this work, did 
not bear directly on the regional geologies of the Colorado Plateau 
-- Prof. H. Zhang had in fact not visited the Colorado Plateau during 
his lifetime.  The present author is also solely responsible for
the quantitative dynamical analyses of the proposed impact event, and
for the proposal that that the episodic nature of the major
extinction events and the supercontinent cycle is likely a result of
the periodic crossing of the Solar System into Milky-Way's major
spiral arms.  These galactic-dynamics connections are enabled
in part by her own original studies of density-wave induced
secular evolution of spiral galaxies in the past three decades,
summarized in a recent research monograph published by De Gruyter
(Zhang 2017). 

The author acknowledges helpful interactions with David Gonzales, Brian Jersak, 
Hailong Li, Hongjie Qu, Meng Wang, Michael Williams, and Bryant Wyatt
at the various stages of this work.

\section*{References}
Alvarez, L.W.,  Alvarez, W., Asaro, F., Michel, H.V., 1980.
Extraterrestrial cause for the Cretaceous-Tertiary extinction.
Science, vol. 208(4448), pp. 1095-1108

Anderson, J.L., 1983. Proterozoic anorogenic granite plutonism of
North America, in Medaris, L.G., and others, Eds., Proterozoic Geology.
Geological Society of America Memoir 161, pp. 133-154

Baars, D.L., 1992. The American Alps: The San Juan Mountains of the
Southwest Colorado.  University of New Mexico Press, Albuquerque

Baars, D.L., 2000. The Colorado Plateau: A Geological
History. University of New Mexico Press, Albuquerque

Baars, D.L., Stevenson, G.M., 1981. Tectonic evolution of the
Paradox Basin, in: D.L. Weigand, ed., Rocky Mountain Association
of Geologists Guidebook, pp. 23-31

Bailer-Jones, C.A.L., Rybizki, J., Andrae, R., Fouesneau, M. 2018.
New stellar encounters discovered in the second Gaia data release,
Astronomy and Astrophysics, vol. 616, pp. A37-A49

Barker, F., 1969. Precambrian geology of the Needle Mountains, southwestern
Colorado. United States Geological Survey Professional Paper 644-A

Bauer, P.W., Raiser, S., 1995.  The Picuris-Pecos fault --Repeatedly
reactivated, from Proterozoic (?) to Neogene.
Annual NMGS Fall Field Conference Guidebook No. 46,
pp. 111-115

Bennis-Smith, M.B., Sprinkel, D.A., Stewart, R., West, L., Elder, T.,
Bostick-Ebbert, N., 2008. A Field Guide to the Flaming Gorge-Uinta
National Scenic Byway. Donning, Virginia Beach

Benz. W., Slattery, W.L., Cameron, A.G.W., 1986. The origin of the Moon and the
single impact hypothesis I.  Icarus, vol. 66, pp. 515-535

Binney, J., Tremaine, S., 2008.  Galactic Dynamics, second ed., Princeton
University Press, Princeton

Blakey, R.C., Ranney, W.D., 2008. Ancient Landscapes
of the Colorado Plateau.  Grand Canyon Association, Grand Canyon

Blakey, R.C., Ranney, W.D., 2018. Ancient Landscapes
of Western North America.  Springer, Cham

Boslough, M.B., Chael, E.P., Trucana, T.G., Crawford, D.A.,
Campbell, D.L., 1994.  Axial focusing of impact
energy in the earth's interior: a possible link to flood
basalts and hotspots, in: Proceedings of the Conference on New
Development Regarding the KT Event and Other Catastrophes
in Earth History, Eds. G. Ryder, S. Gartner, Fastovsky

Bullard, E.C., Maxwell, A.E., Revelle, R., 1956. Heat flow
through the deep sea floor.  Adv. Phys., vol.3, pp. 153-181

Cameron, A.G.W.,  Benz. W., 1987. The origin of the Moon and the
single impact hypothesis II.  Icarus, vol. 71, pp. 30-45

Cameron, A.G.W., Benz. W., Melosh, H.J., 1989. The origin of the Moon and the
single impact hypothesis III.  Icarus, vol. 81, pp. 113-131

Cameron, A.G.W., Benz. W., 1991. The origin of the Moon and the
single impact hypothesis IV.  Icarus, vol. 92, pp. 204-216

Canup, R.M., 2012. Forming a Moon with an Earth-like composition
via a giant impact.  Science, vol. 338(6110), pp. 1052-1055

Condon, S.M., 1995, Geology of pre-Pennsylvanian rocks in the Paradox
Basin and adjacent areas, Southeastern Utah and Southwestern Colorado.
US Geological Survey Bulletin 2000-G, Washington DC

Cox, R., Martin, M.W., Comstock, J.C., Dickerson, L.S.,
Ekstrom, I.L., Sammons, J.H., 2002. Sedimentology, stratigraphy, and
geochronology of the Proterozoic Mazatzal Group, central Arizona.
GSA Bulletin vol. 114, no. 12, pp. 1535-1549

Cross, C.W., Howe, E., Irving, J.D., Emmons, W.H., 1905. 
Description of the Needle Mountains quadrangle, 
Colorado: USGS Geologic Atlas, Folio 131

Dean, R.L. III, 2004. Aureole structure of 1.4 Ga plutons in
southern Colorado and their tectonic implications.
Master Thesis, University of Texas at El Paso

Dehler, C.M., Fanning, C.M., Link, P.K., Kingsbury, E.M.,
Rybczynski, D., 2010.
Maximum depositional age and provenance of the Uinta
Mountain Group and Big Cottonwood Formation, northern
Utah: Paleogeography of rifting western Laurentia.
GSA Bulletin, September/October 2010, vol. 122, no. 9/10;
pp. 1686-1699

Dehler, C.M., Sprinkel, D.A., Porter, S.M., 2005. 
Neoproterozoic Uinta Mountain Group of northeastern Utah: 
Pre-Sturtian geographic, tectonic, and
biologic evolution, in Pederson, J., and Dehler, C.M., Eds., 
Interior Western United States: Geological Society of 
America Field Guide 6, p. 1–25

Dence, M.R., Grieve, R.A.F., Robertson, P.B., 1977. 
Terrestrial impact structures: Principle characteristics
and energy considerations, in: Impact and Explosion Cratering, 
Eds. Roddy, D.J., Pepin, R.O., Merill, R.B.
Pergamon, New York, pp. 247-275

Dickinson, W.R., 1989. Tectonic setting of Arizona through geologic
time, in: Jenny, J.P., Reynolds, S.J., Eds., Geologic
Evolution of Arizona, Arizona Geological Society Digest, Tucson,
vol. 17, pp. 1-16

Dressler, B.O., Sharpton, V.L., Eds., 2000. 
Large Meteorite Impacts and Planetary Evolution II.
Geological Society of America Special Paper 339

Dutton, D.E., 1882. Report on the geology of the high plateaus
of Utah.  U.S. Geological Survey Second Annual Report, pp. 49-166

Erslev, E.A., Fankhauser, S.D., Heizler, M.T.,
Sanders, R.E., Cather, S.M., 2004.
Geological Society of America Field Guide 5,
pp. 15-40

Fackelman, S.P., Morrow, J.R., Koebert, C., McElvain, T.H., 2008.
Shatter cone and microscopic shock-alteration evidence for
a post-Paleoproterozoic terrestrial impact structure near
Santa Fe, New Mexico, USA.  Earth and Planetary Science Letters, 
vol. 270, pp. 290-299

French, B.M., 2003. Traces of Catastrophe: A Handbook of Shock-Metamorphic
Effects in Terrestrial Meteorite Impact Structures.  Lunar and Planetary
Science Institute Contribution No. 954

Fux, R., 1997. 3D self-consistent N-body barred models of the Milky Way. 
I. Stellar dynamics, Astronomy and Astrophysics, vol. 327, 983-1003

Fux, R., 1999. 3D self-consistent N-body barred models of 
the Milky Way. II. Gas dynamics. 
Astronomy and Astrophysics, vol. 345, pp. 787-812

Gadd, B., 2008. Canadian Rockies Geology Road Tours, Corax Press, Jasper AB

Gibson, R.G., Simpson, C., 1988. Proterozoic polyphase deformation
in basement rocks of the Needle Mountains, Colorado: Geological Society
of America Bulletin, vol. 100, pp. 1957-1970

Goldberg, S.A., Butler, J.R., Fullagar, P.D., 1986.  The Bakersville dike swarm:
Geochronology and petrogenesis of late Proterozoic basaltic magmatism in the
southern Appalachian Blue Ridge.  American Journal of Science, vol. 286, pp. 403-430

Goldsmith, W., 2001. Impact: The Theory and Physical Behavior
of Colliding Solids. Dover, New York

Gonzales, D.A., 1997.  Crustal Evolution of the Needle Mountains
Proterozoic Complex, Southwestern Colorado.  PhD Dissertation, University
of Kansas, Lawrence

Gonzales, D.A., Conway, C.M., Ellingson, J.A., Campbell,
J.A., 1994.  Proterozoic geology of the western and southeastern
Needle Mountains, Colorado.  USGS Open-File Report 94-437

Gonzales, D., Heerschap, L., 2012. Geology of the Durango-Silverton Train Route.
 D\&SNGRR and Fort Lewis College

Gonzales, D.A., Karlstrom, K.E., Siek, G.S., 1996. Syncontractional crustal anatexis
and deformation during emplacement of $\sim$1435 Ma plutons, western Needle
Mountains, Colorado.  Geology, vol. 104, pp. 215-223
 
Grambling, J.A., and Tewksbury, B.J. 1989.
Preface. In Proterozoic Geology of the
Southern Rocky Mountains.  Geological Society of America Special Paper 235,
Eds. J.A. Grambling and B.J. Tewksbury,
pp. v-vi

Grambling, J.A., Williams, M.L., Smith, R.F., Mawer, C.K., 1989.
The role of crustal extension in the metamorphism of Proterozoic rocks
in northern New Mexico. In Proterozoic Geology of the
Southern Rocky Mountains.  Geological Society of America Special Paper 235,
Eds. J.A. Grambling, B.J. Tewksbury, pp. 87-110 

Hamblin, W.K., 2008.  Anatomy of the Grand Canyon.
Grand Canyon Association, Grand Canyon

Hanson, W.R., 1965. Black Canyon of the Gunnison: Today and Yesterday.
USGS Geological Survey Bulletin 1191, Washington DC

Hansen, W.R., 2005. The Geological Story of the Uinta Mountains,
second edition. Falcon, Guilforo

Harris, C.W., Gibson, R.G., Simpson, C., Eriksson, K.A., 1987.
Proterozoic cuspate basement-cover structure, Needle Mountains, Colorado.
Geology, vol. 15, pp. 960-953

Harris, C.W., 1990. Polyphase suprastructure deformation in metasedimentary
rocks of the Uncompahgre Group: Remnants if an early Proterozoic
fold belt in southwest Colorado. Geological Society of America
Bulletin, vol. 102, pp. 664-678

Hayes, D.S., 2013. Neoproterozoic Snowball Earth, Before and
After.  PhD Dissertation.  Utah State University, Logan, Utah

Hinds, N.E.A., 1936. Contributions to Precambrian Geology of Western
North America.  II. Uncompahgre and Beltian Deposits in Western
North America.  Carnegie Institute of Washington

Hoffman, P.F., 1989.  Speculations on Laurentia's first gigayear (2.0 to 1.0 Ga).  i
Geology, vol. 17, pp. 135-138

Hunt, C.B., 1956. Cenozoic geology of the Colorado Plateau.
U.S. Geological Survey Professional Paper 279

Karlstrom, K., 1989, Early recumbent folding during
Proterozoic orogeny in central Arizona,
in: Proterozoic Geology of the Southern Rocky Mountains.
Geological Society Special Paper 235. Eds. J.A. Gambling, B.J. Tewksbury,
pp. 155-171

Karlstrom, K.E., Bowring, S.A., Dehler, C.M., Knoll, A.H., Porter, S.M.,
Des Marais, D.J., Weil, A.B., Sharp, Z.D., Geissman, J.W., Elrick, M.B., 
Timmons, J.M., Crossey, L.J., Davidek, K.L., 2000.
Chuar Group of the Grand Canyon: Record of breakup of Rodinia,
associated change in the global carbon cycle, and ecosystem
expansion by 740 Ma. Geology, vol. 28(7), pp. 619-622
 
Karlstrom, K.E., Gonzales, D.A., Heizler, M., Zinsser, A., 2017. 
$^{40}$Ar/$^{39}$Ar age constraints on the deposition and metamorphism of the
Uncompahgre Group, southwestern Colorado, in New Mexico Geological Society 
68th Annual Fall Field Conference Guidebook, pp. 83-90

Keller, G.R., Braile, L.W., Morgan, P., 1979.
Crustal structure, geophysical models and contemporary
tectonism of the Colorado Plateau. Tectonophysics, vol. 61, pp. 131-147

Kelley, V.C., 1979.  Tectonics of the Colorado Plateau
and new interpretation of its eastern boundary.
Tectonophysics, vol. 61, pp. 97-102

Kelley, V.C., and Clinton, N.J. 1960.  Fracture systems and tectonic
elements of the Colorado Plateau.  New Mexico University Pubs. Geology,
no. 6

King, P.B., Ferguson, H.W., 1960.  Geology of northeasternmost Tennessee.
U.S. Geological Survey Professional Paper 311

Laskar, J., 2001. Solar System: Stability. Encyclopedia of Astronomy
and Astrophysics, ed. P. Murdin, article id 2198. Bristol:
Institute of Physics Publishing

Li, H., Zhang, H., Qu, H., Wang, M., 2014. Initiation, the
first stage of the Yanshan (Yenshan) movement in Western Hills,
Constraints from Zircon U-Pb dating,
Geological Review, vol. 60(5), pp. 1026-1042

Love, J.D., Reed, J.C.Jr., Pierce, K.L., 2003.
Creation of the Teton Landscape. Second edition, Grand Teton
Natural History Association, Moose

Lynden-Bell, D., Kalnajs, A., 1972. On the generating mechanism
of spiral structure. Monthly Notices of the
Royal Astronomical Society, vol. 157, pp. 1-30

Mamajek, E.E., Barenfield, S.A., Ivanov, V.D., Kniazev, A.Y., Valsanen, P.,
Beletsky, Y., Boffin, H.M.J., 2015.  The Astrophysical Journal Letters, vol. 800(1), pp. L17-L20

McKee, B., 1972. Cascadia: The Geologic Evolution of the Pacific
Northwest.  McGraw-Hill, New York

McPhee, J., 1998. Annals of the Former World. Farrar, Straus and Giroux, 
New York

Melosh, H.J., 1996. Impact Cratering: A Geologic Process,
Oxford University Press, Oxford

Montgomery, A., 1956, Precambrian geology of the Picuris Range, north central
New Mexico. New Mexico Geological Society 7th Annual Fall
Field Conference Guidebook pp. 143-146

Moores, E.M., 1991. Southwest
U.S.-East Antarctica (SWEAT) connection:
A hypothesis, Geology, vol. 19(5), pp. 425-428

Mulder, J., Karlstrom, K., Halpin, J.A., Merdith, A.S., Spencer, C.J., Berry, R.F., McDonald, B., 2018.
Rodinian devil in disguise: Correlation of 1.25-1.10 Ga strata between Tasmania and Grand Canyon.
Geology, vol. 46, pp. 991-994

Noel, M.E., 2002.  Structure and Metamorphism of the Eolus Granite, Needle Mountains,
Colorado: Implications for Regional Metamorphism and Proterozoic Tectonics in Southern
Colorado.  Master thesis.  University of Texas at El Paso.

Ohnaka, M., 2013. The Physics of Rock Failure and Earthquakes.
Cambridge University Press, Cambridge

Oreskes, N., 2003. Plate Tectonics: An Insider's History
of the Modern Theory of the Earth.  Westview Press, Boulder, Colorado

Ownby, S.E., Miller, C.F., Berquist, P.J., Carrigan, C.W.,
Wooden, J.L., Fullagar, P.D. 2004, U-Pb geochronology and geochemistry of
a portion of the Mars Hill terrane, North Carolina-Tennessee: Constraints
on origin, history, and tectonic assembly.  In Tollo, R.P., Corriveau, L.,
McLelland, J., Bartholomew, M.J., Eds., Proterozoic Evolution of the
Grenville Orogen in North America.  Boulder, Colorado, Geological
Society of America Memoir, vol. 197, pp. 609-632

Peters, S.E., Gaines, R.R., 2012. Formation of the
``Great Unconformity'' as a trigger for the Cambrian
explosion. Nature, vol. 484, pp. 363-366

Price, N.J., 2001. Major Impacts and Plate
Tectonics: A Model for the Phanerozoic Evolution
of the Earth's Lithosphere. Routledge, New York

Raup, D., Sepkoski, J., 1982. Mass extinctions in the marine 
fossil record, Science, vol. 215, pp. 1501–1503. DOI:10.1126/science.215.4539.1501.

Reed, J.C.Jr., 1984.  Proterozoic rocks of the Taos Range, Sangre de Cristo
Mountains, New Mexico, in: New Mexico Geological Society Guidebook, 35th Field
Conference: Rio Grande Rift (Northern New Mexico),
Baldridge, W.S., Dickerson, P.W., Riecker, R.E.; Zidek, J., Eds.,
pp. 179-185

Reiff, W., 1977. The Steinheim basin - An impact structure, in:
Roddy, D.J., Pepin, R.O., Merill, R.B., Eds.,
Impact and Explosion Cratering, Pergamon, New York, pp. 309-320

Reufer, A., Meier, M.M.M., Benz, W., Wieler, R. 2012, Icarus, vol. 221, pp. 296-299

Roddy, D.J., 1977a. Pre-impact conditions and cratering
processes at the Flynn Creek crater, Tennessee, in:
Roddy, D.J., Pepin, R.O., Merill, R.B., Eds.,
Impact and Explosion Cratering, Pergamon, New York, pp. 277-308

Roddy, D.J., 1977b. Tabular comparisons of the
Flynn Creek crater, United States, Steinheim impact crater, Germany,
and Snowball Explosion crater, Canada, in:
Roddy, D.J., Pepin, R.O., Merill, R.B., Eds.,
Impact and Explosion Cratering, Pergamon, New York, pp. 125-161

Rohde, R.A.,  Muller, R.A., 2005. Cycles in fossil diversity, 
Nature, vol. 434, pp. 209-210

Sepkoski, J., 2002, A compendium of fossil marine animal genera 
in: Jablonski, D., Foote, M., Eds., Bull. Am. Paleontol. no. 363.
Paleontological Research Institution, Ithaca, NY

Share, J., 2012. The Great Unconformity and the late Proterozoic-Cambrian time interval: 
Part II - The rifting of Rodinia and the ``Snowball Earth'' glaciations that followed, 
http://
\newline written-in-stone-seen-through-my-lens.blogspot.com/2012\_02\_01\_archive.html

Signor, P., Lipps, J., 1982. Sampling bias, gradual extinction 
patterns and catastrophes in the fossil record, in: Geologic 
Implications of Impacts of Large Asteroids and Comets on the Earth, 
I. Silver, P. Silver Eds., Geol. Soc. Amer. Special Paper 190, 
Boulder, Colorado, pp. 291-296

Spudis, P.D., 2015. Could the Colorado Plateau be an ancient impact scar?
www.airspacemag.com/daily-planet/could-colorado-plateau-be-ancient-impact-scar-180956994

Stein, S., Wysession, M., 2003.  An Introduction to Seismology, Earthquakes,
and Earth Structure. Blackwell Publishing, Malden

Stow, D., 2010. Vanished Ocean: How Tethys Reshaped the World. Oxford
University Press, Oxford

Su, Q., Goldberg, S.A., Fullagar, P.D., 1994. Precise U-Pb zircon ages of Neoproterozoic
plutons in the southern Appalachian Blue Ridge and their implications for the
initial rifting of Laurentia.  Precambrian Research, vol. 68, pp. 81-95

Sullivan, W., 1991.  Continents in Motion.  2nd edition. 
American Institute of Physics, New York

Tewksbury, B.J., 1981. Polyphase Deformation
and Contact Relationships of the Precambrian
Uncompahgre Formation, Needle Mountains, Southwestern Colorado.
PhD Dissertation, University of Colorado at Boulder

Tewksbury, B.J., 1985a. Revised interpretation of the age of
allochthonous rocks of the Uncompahgre Formation, Needle Mountains,
Colorado: Geological Society of America Bulletin, vol. 96,
pp. 224-232

Tewksbury, B.J., 1985b. Similarity in deformation style as
evidence for correlation between the type section and
Needle Mountains portion of the Proterozoic
Uncompahgre Formation, southwestern Colorado:
Geological Society of America Abstracts with Programs, vol. 17, p. 267

Tewksbury, B.J., 1986. Conjugate crenulation cleavages in the
Uncompahgre Formation, Needle Mountains, Colorado: Journal
of Geology, vol. 8, no. 2, pp. 145-155

Tewksbury, B.J., 1989. Proterozoic geology of the Needle Mountains:
A summary, in: Proterozoic Geology of the Southern Rocky Mountains,
Eds. J.A. Gambling, B.J. Tewksbury, pp. 65-73

Timmons, J.M., Karlstrom, K.E., Sears, J.W., 2003.
Geologic structure of the Grand Canyon Supergroup, in:
Grand Canyon Geology, 2nd edition, Eds. S.S. Beus, M. Morales,
pp. 76-89

Twento, O., 1980.  Precambrian geology of Colorado, in: Colorado Geology,
Eds. H.C. Kent, K. W. Porter.  Rocky Mountain Association
of Geologists, pp. 37-46

Van Hise, C.R., Leith, C.K., 1909/2017. Pre-Cambrian 
Geology of North America.  Government Printing Office, Washington DC.  
New edition in Forgotten Books Classic Reprint Series

Ward, J., 2001. http://www.vindicator.ca/about/
\newline ballyDonegal/unconformity.html

Wegener, A., 1929. Die Entstehung der Kontinente und
Ozeane.  The first Dover edition of the English translation (by J. Biram)
of the 4th revised edition of this book, entitled The Origin 
of Continents and Oceans, was published in 1966

Wignall, P.B., 2015.  The Worst of Times: How Life on Earth Survived
Eighty Million Years of Extinctions. Princeton University Press, Princeton

Wilson, J.T., 1963. A possible origin of the Hawaiian islands.
Canadian Journal of Physics, vol. 41(6), pp. 863-870

Wu, K., 2007. Proterozoic Tectonic Evolution of the Needle Mountains,
Colorado: Integrated Structural and Petrologic Analysis of Crustal
Formation and Evolution.  PhD Thesis. University of Texas at El Paso

Wyatt, B.M., Petz, J.M., Sumpter, W.J., Turner, T.R., Smith, E.L., Fain, B.G.,
Hutyra, T.J., Cook, S.A., Hibbs, M.F., Goderya, S.N., 2016. 
Creating an Isotopically Similar Earth-Moon System with Correct Angular Momentum from a Giant Impact.
arXiv:1611.02769

Zhang, H., 1998, Yanshan Event, Acta Geologica Sin1ca, vol. 72(2),
pp. 103-111

Zhang, H., 2016, The Phasing of Yanshan Movement
and its several important questions, Acta Geologica Sinica, vol. 90(9),
pp. 2176-2180

Zhang, H., Zhang, Y., Cai, X., Qu, H., Li, H., Wang, M., 2013,
The triggering of Yanshan Movement: Yan Shan Event,
Acta Geologica Sinica, vol. 87(12), pp. 1779-1790

Zhang, X., 1996. Secular evolution of spiral galaxies. I. A collective
dissipation process.  The Astrophysical Journal, vol. 457, pp. 125-144

Zhang, X., 1998. Secular evolution of spiral galaxies. II.
Formation of quasi-stationary spiral modes.
The Astrophysical Journal, vol. 499, pp. 93-111

Zhang, X., 1999. Secular evolution of spiral galaxies. III.
The Hubble Sequence as a temporal evolution sequence.
The Astrophysical Journal, vol. 518, pp. 613-626

Zhang, X., 2016. N-body simulations of collective
effects in spiral and barred galaxies,
Astronomy and Computing, vol. 17, pp. 86-128

Zhang, X., 2017. Dynamical Evolution of Galaxies.
De Gruyter, Berlin/
\newline Boston

Zhang, X., Buta, R.J., 2007.
The potential-density phase shift method for determining
the corotation radii in spiral and barred galaxies,
The Astronomical Journal, vol. 133, pp. 2584-2606

Zhang, X., Buta, R.J., 2015.
Galaxy secular mass flow rate determination
using the potential-density phase shift approach:
Application to six nearby spiral galaxies,
New Astronomy, vol. 34, pp. 65-83

Zinsser, A., 2006.
New Stratigraphy, Polyphase Deformational History 
and Basement-Involved Thrust Belt Model for the 
Proterozoic Uncompahgre Group and Vallecito Conglomerate, 
Needle Mountains, Colorado.  Master Thesis, University
of New Mexico

\clearpage

\pagebreak

\pagebreak

\pagebreak

\pagebreak

\clearpage
\begin{figure}
\vspace{10cm}
\includegraphics{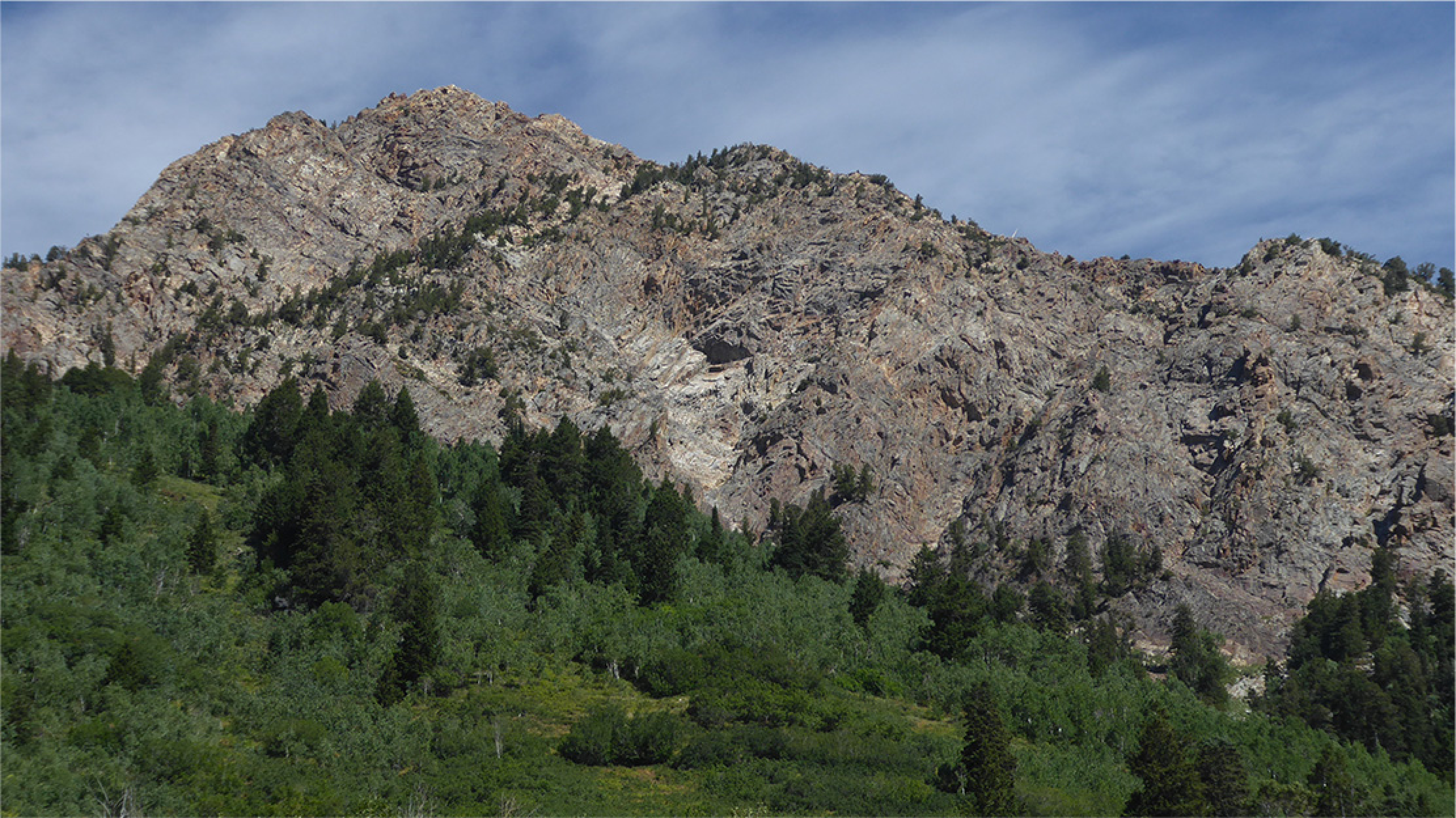}
\caption{Melted and brecciated Precambrian rocks in
Little Cottonwood Canyon area, Utah.}
\label{Fig5}
\end{figure}

\clearpage
\begin{figure}
\vspace{10cm}
\includegraphics{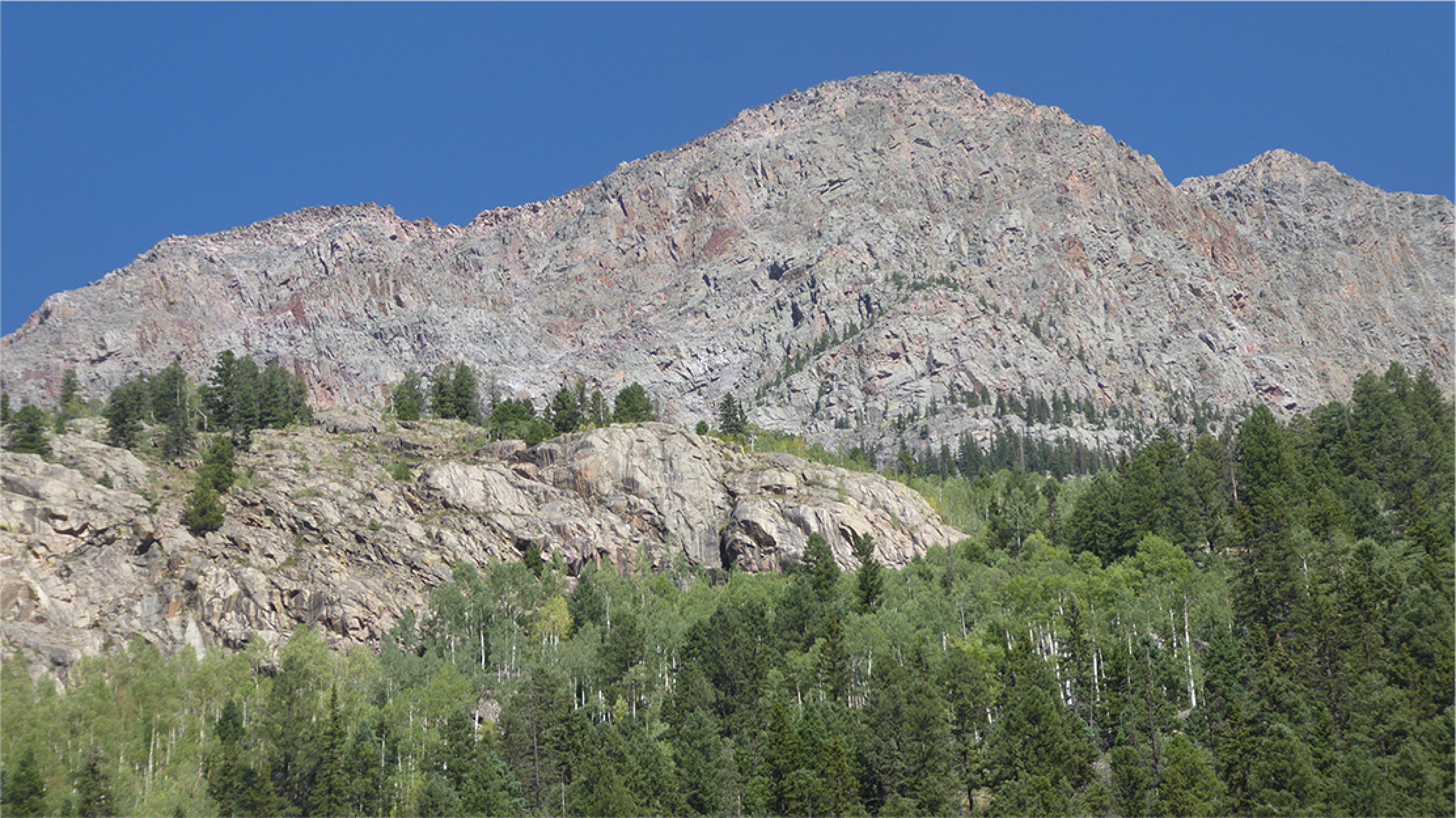}
\caption{Melted and brecciated Precambrian rocks along the route of Durango-Silverton
Narrow Gauge Railway, south of Silverton, Colorado.}
\label{Fig6}
\end{figure}

\clearpage
\begin{figure}
\vspace{10cm}
\includegraphics{Fig7Copy.eps}
\caption{Melted and brecciated Precambrian rocks along AZ-87, south of Pine, Arizona.}
\label{Fig7}
\end{figure}

\clearpage
\begin{figure}
\vspace{10cm}
\includegraphics{Fig8Copy.eps}
\caption{Melted and brecciated Precambrian rocks along NM-475 near Santa Fe, New Mexico.
The rocks in this area form shatter-cone clusters.}
\label{Fig8}
\end{figure}

\begin{figure}
\vspace{18cm}
\includegraphics{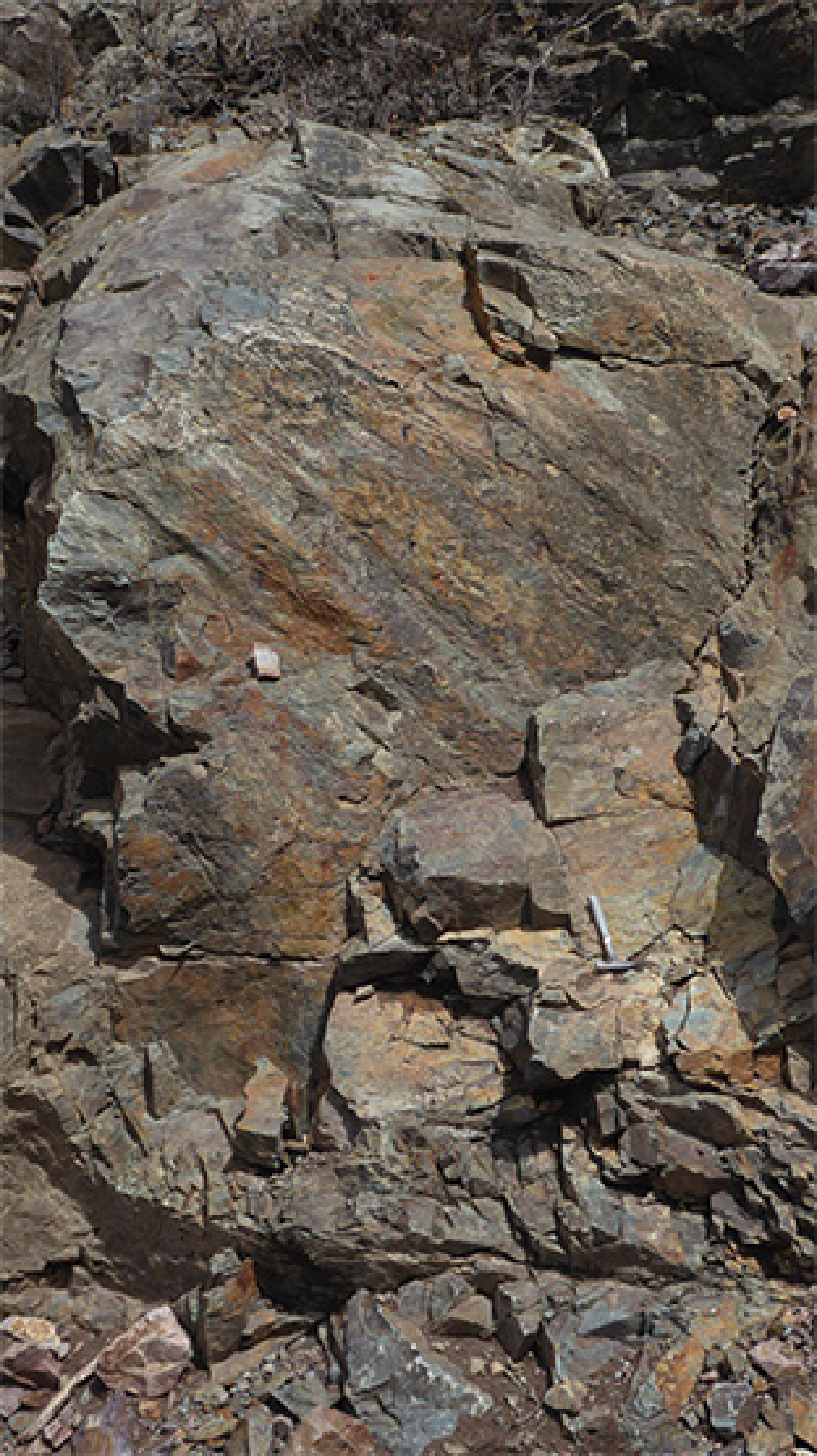}
\includegraphics{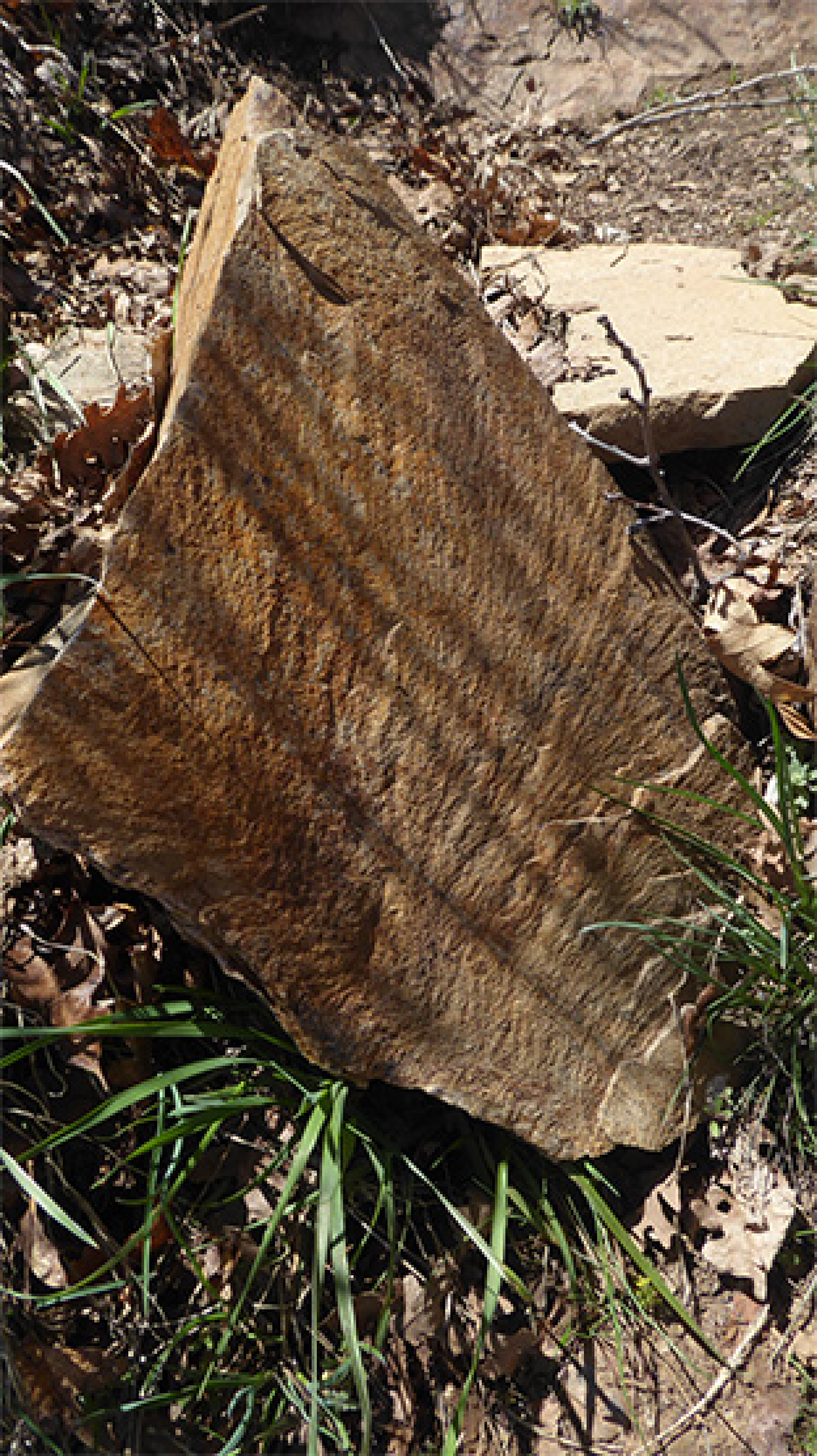}
\includegraphics{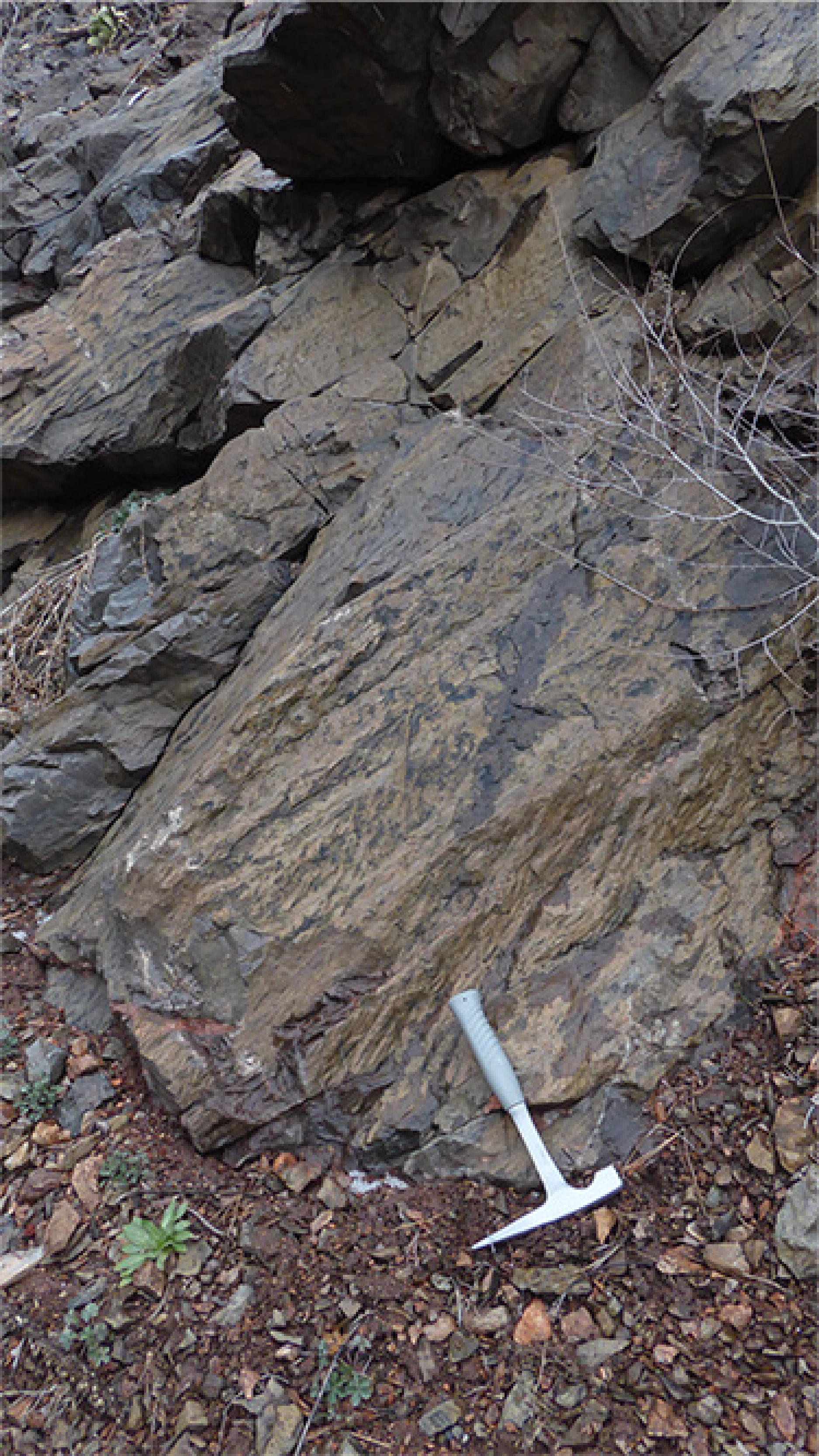}
\includegraphics{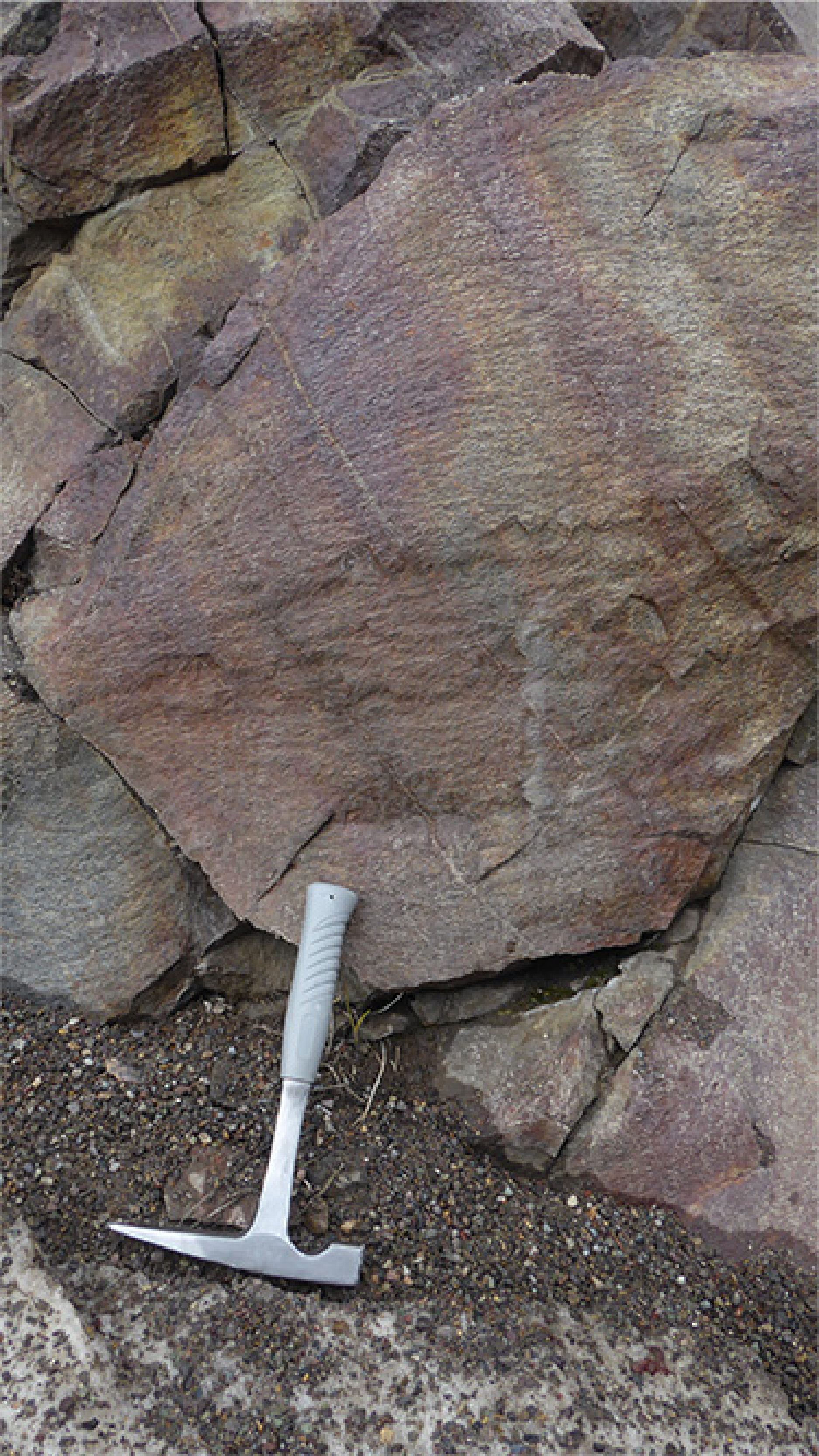}
\caption{Signatures of shock metamorphism in Precambrian cover sequence rocks of the
Colorado Plateau.  Top Left: Uncompahgre Formation along US-550 in San Juan
Mountains.  Top Right: Rocks in Black Canyon of the Gunnison National Park.
Bottom Left: Santa Fe Impact Structure rocks.  Bottom Right: Uncompahgre Formation in Rico Mountains.}
\label{Fig9}
\end{figure}

\clearpage

\begin{figure}
\vspace{10cm}
\includegraphics{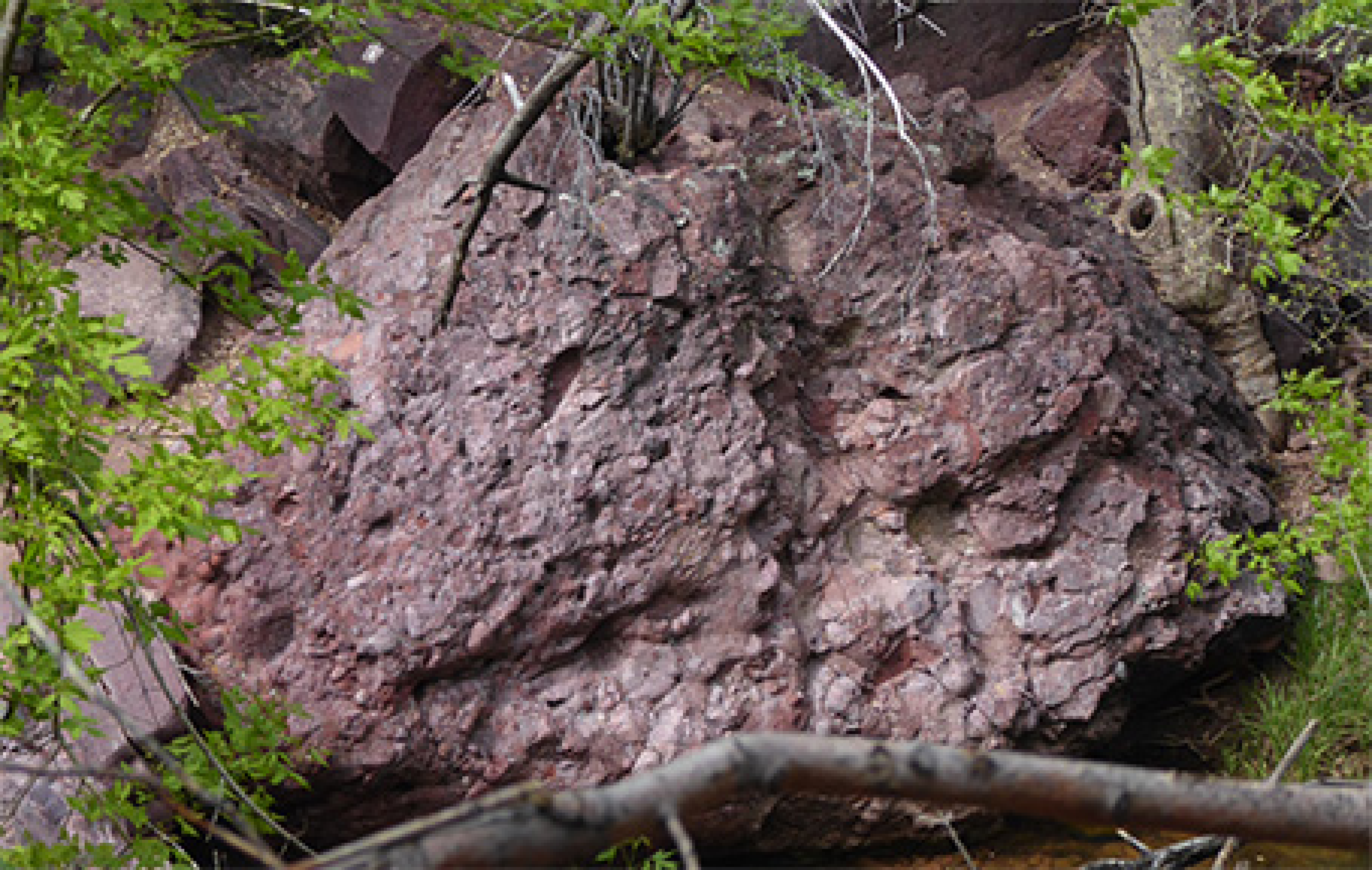}
\includegraphics{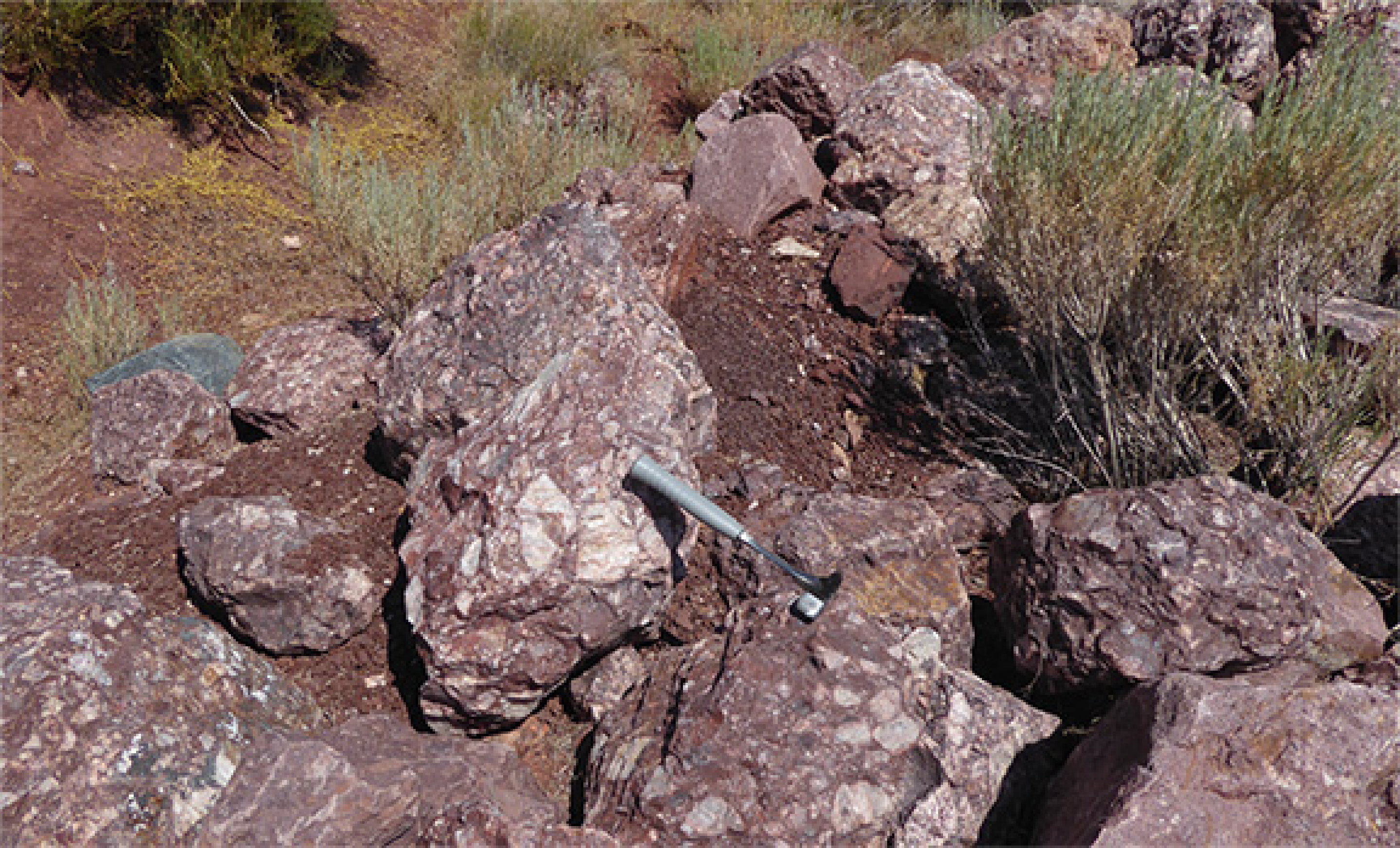}
\includegraphics{Fig10cCopy.eps}
\includegraphics{Fig10dCopy.eps}
\caption{Compressed and partially melted conglomerate and suevite formations. 
Top Left: Marquenas Formation in Vadito Group in southern Pecuris Mountains area, New Mexico.
Top Right: Conglomerate of Precambrian rocks in Black Canyon of the Gunnison National Park, Colorado.
Bottom Left: Mazatzal Formation in Tonto Bridge State Park, Arizona.
Bottom Right:  Uinta Mountain Group along Browns Park Road (UT-1364), northeastern Utah.}
\label{Fig10}
\end{figure}

\begin{figure}
\vspace{16cm}
\includegraphics{Fig11aCopy.eps}
\includegraphics{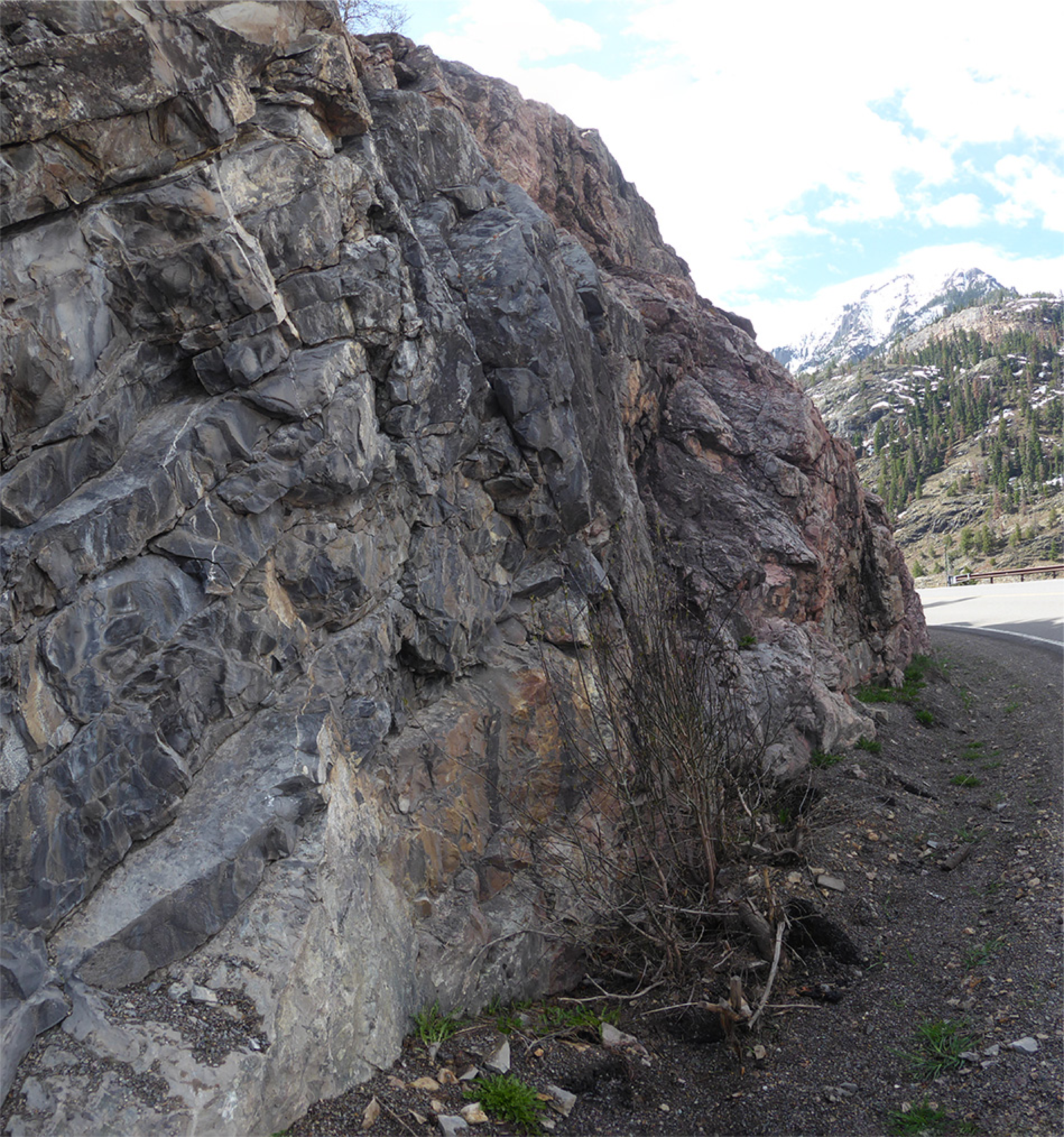}
\includegraphics{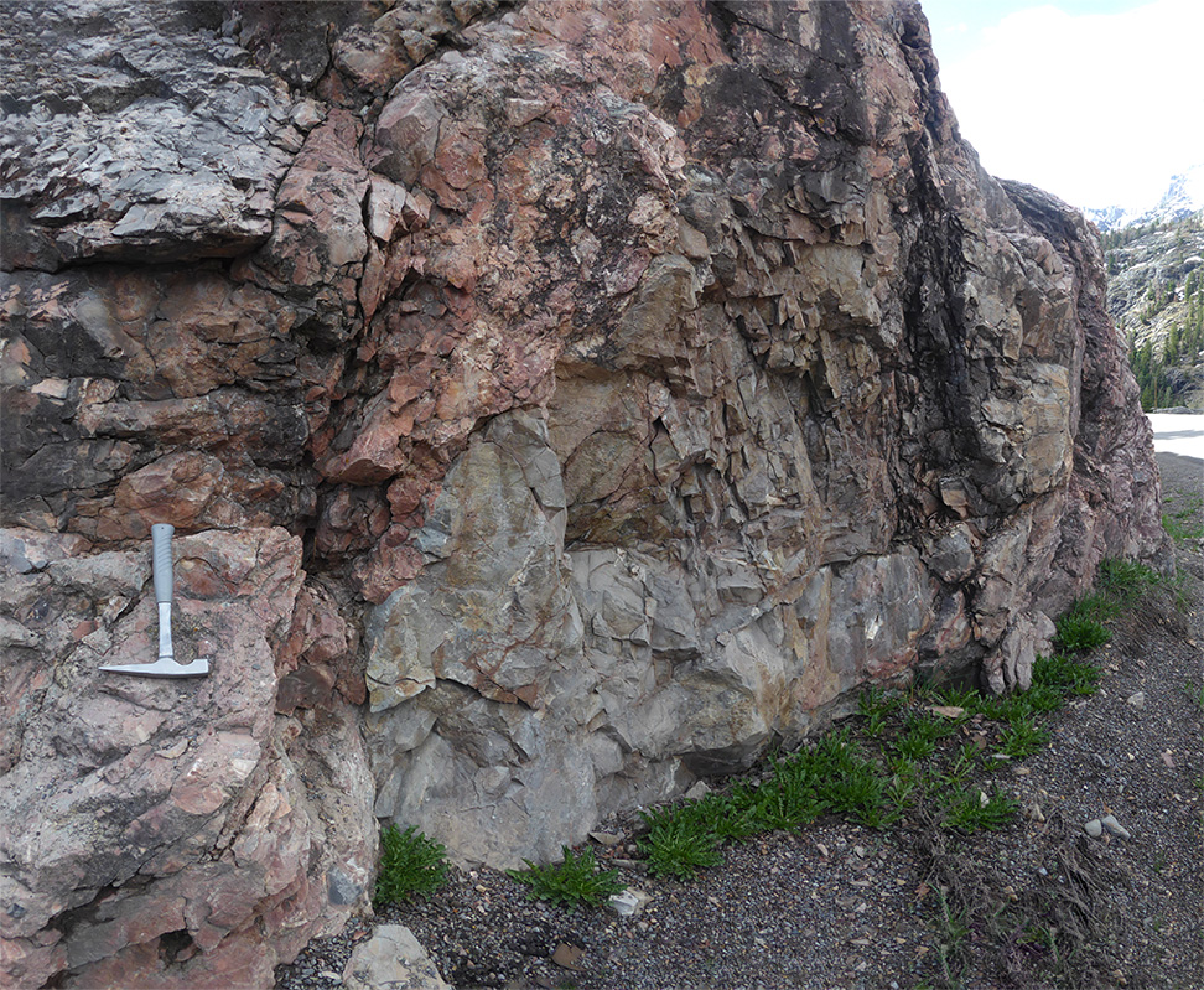}
\caption{Migmatite of Uncompahgre Formation south of Ouray, Colorado.
Top frame: Fish-eye view.  Bottom two frames: close-up view of
select areas, with the left frame showing more severe melting. 
}
\label{Fig11}
\end{figure}

\pagebreak

\begin{figure}
\vspace{18cm}
\includegraphics{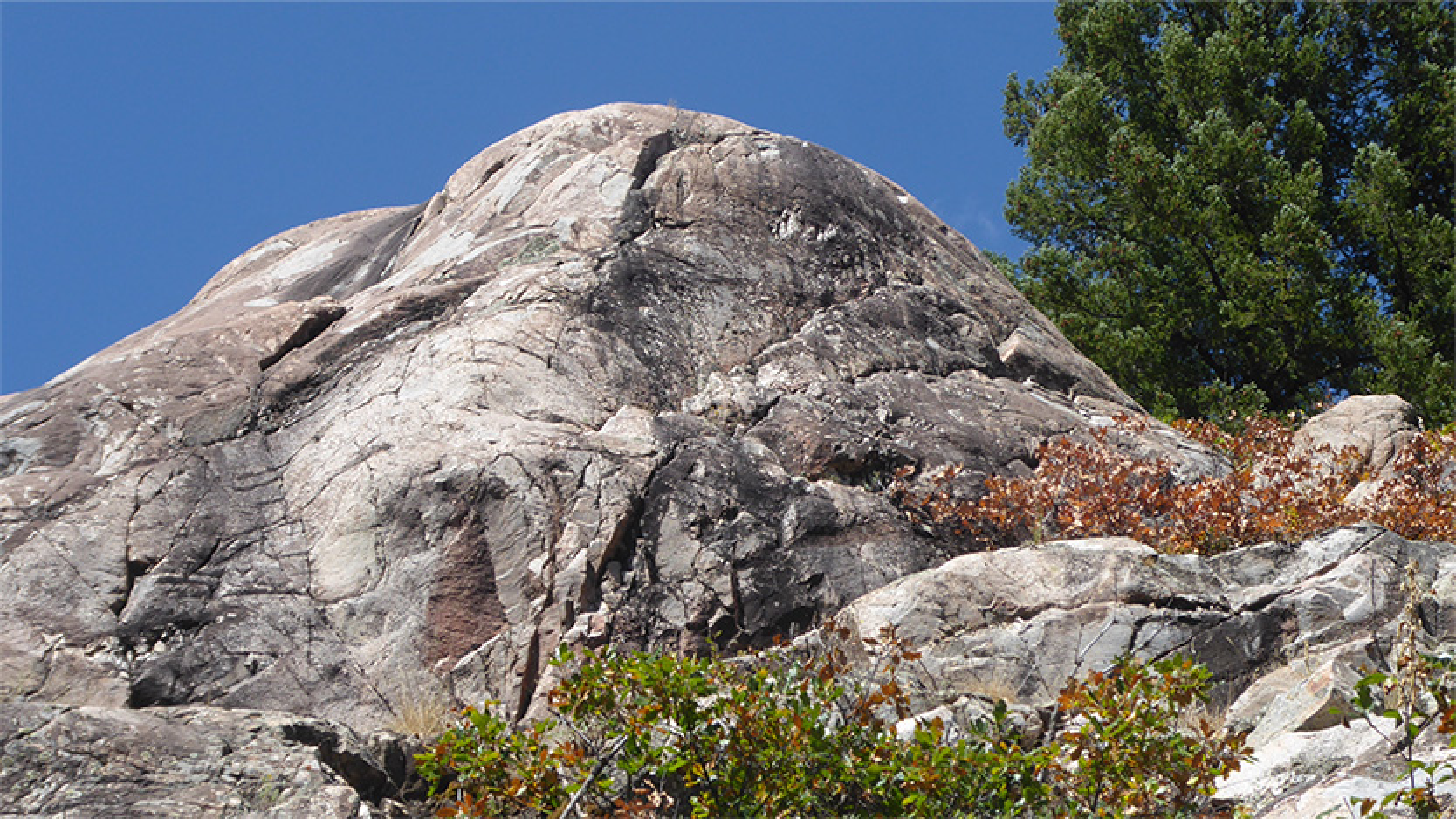}
\includegraphics{Fig12bCopy.eps}
\caption{Top: Partially melted Eolus Granite formation just off the
Purgatory Flats Trail in southwestern Needle Mountains, Colorado.
Bottom: Migmatite of Eolus Granite and Uncompahgre Formation
along US-550 north of Durango, Colorado.}
\label{Fig12}
\end{figure}

\pagebreak

\begin{figure}
\vspace{18cm}
\includegraphics{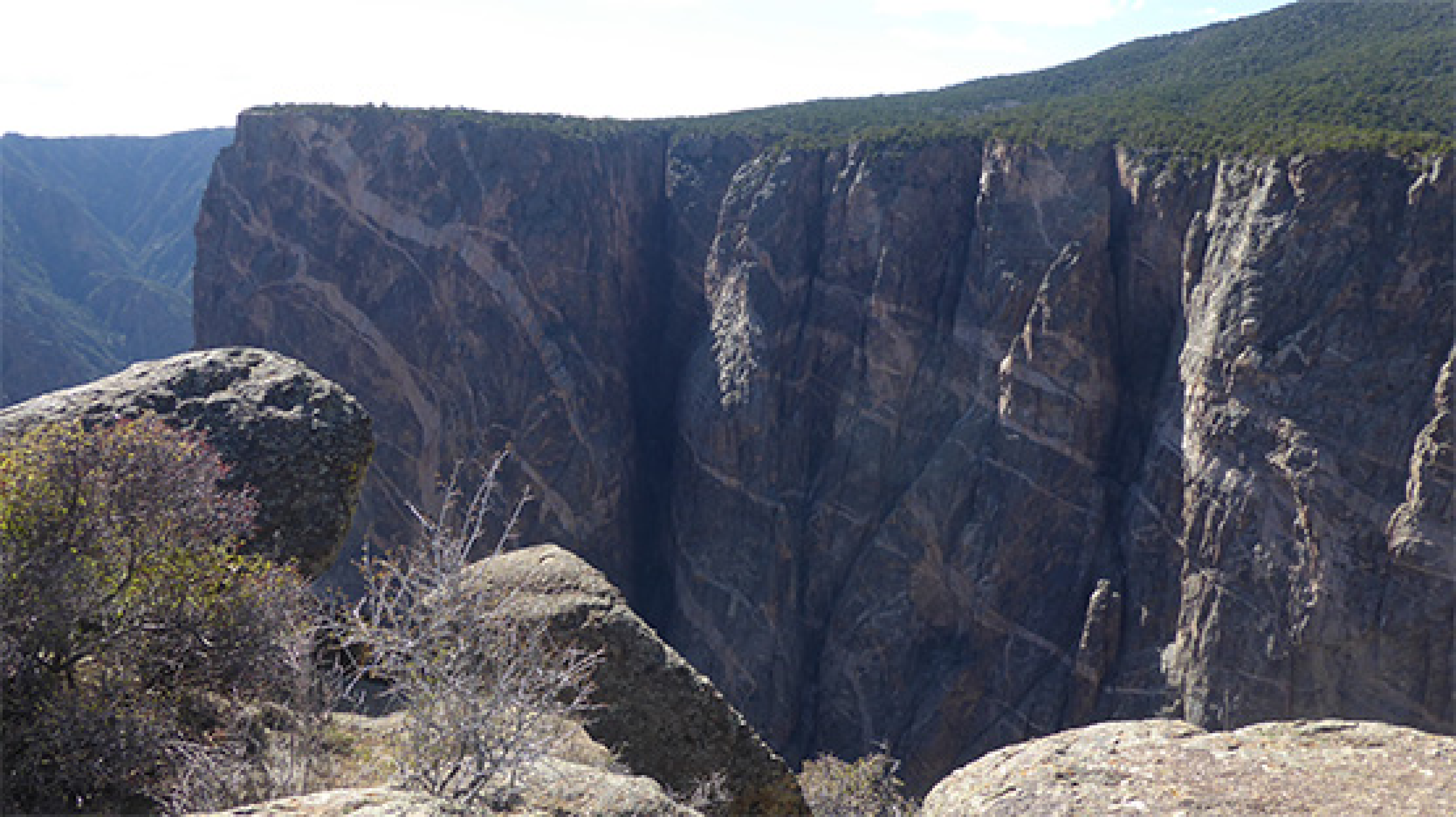}
\includegraphics{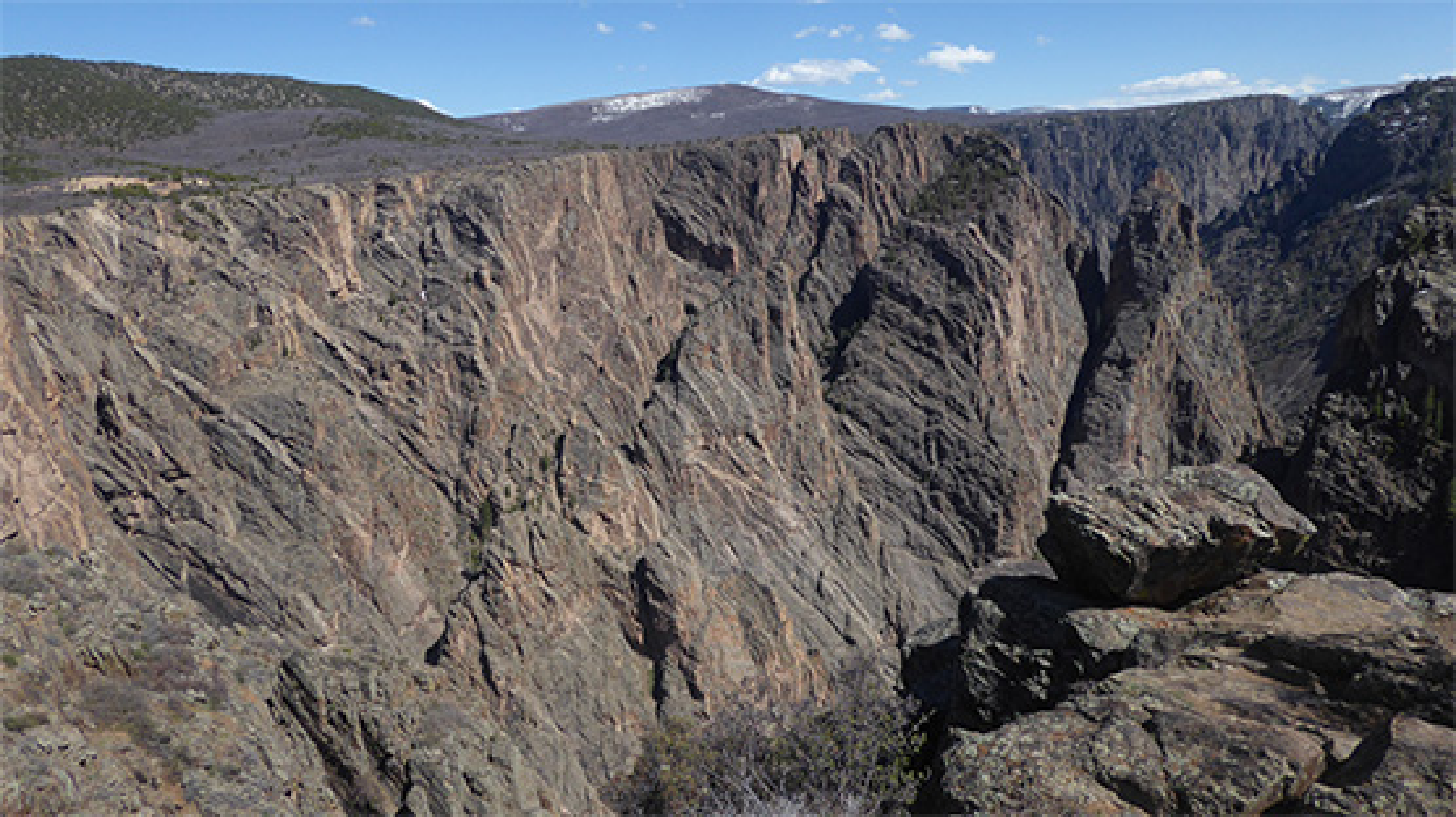}
\caption{Pegmatite sills and dikes on the walls of the Black Canyon of the Gunnison River.} 
\label{Fig13}
\end{figure}

\begin{figure}
\vspace{14cm}
\includegraphics{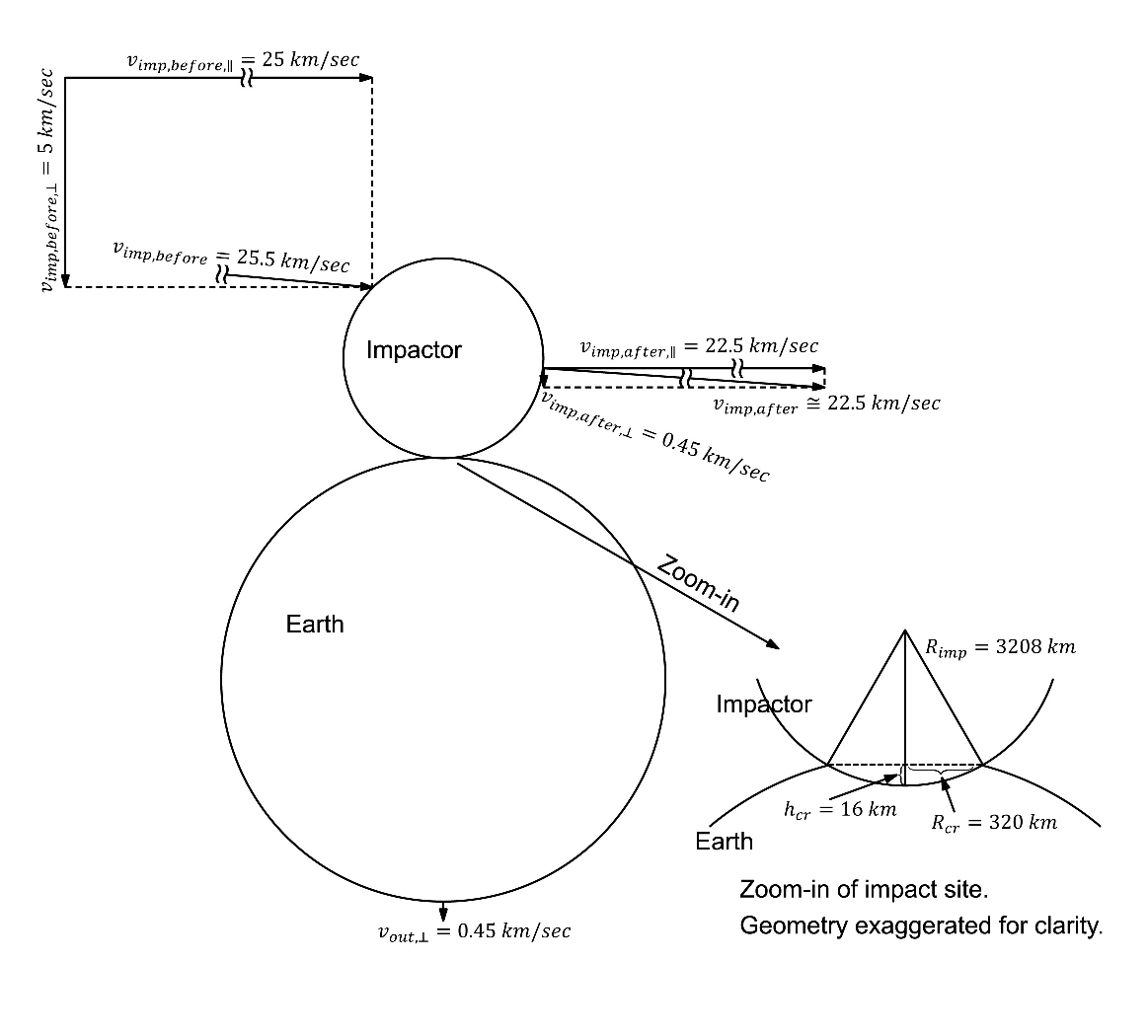}
\caption{Schematics of the impact event. Earth's orbital
and spin motion are not shown in the figure, and are also not included
in the impact calculation.}
\label{Fig14}
\end{figure}

\pagebreak

\begin{figure}
\vspace{12cm}
\includegraphics{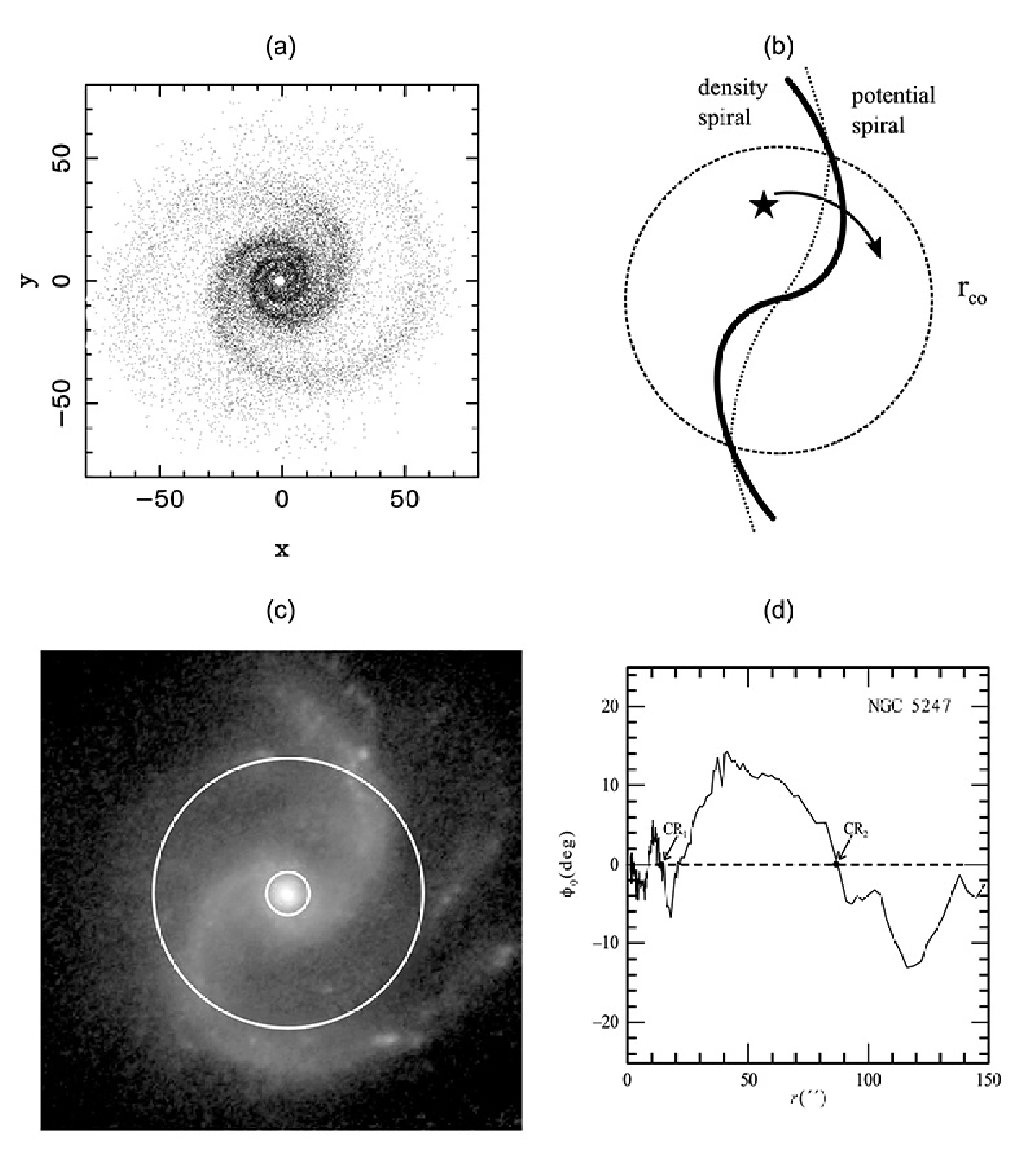}
\caption{(a) Close-up view of an N-body simulated spiral galaxy
possessing a spontaneously-formed density wave mode.  Note the
density enhancement at the inner edge of the spiral arms inside
the corotation radius of 30, indicating
the presence of collisionless shocks in this pure-particle
simulation.  (b) Schematic of potential (dashed) and density (solid) 
spirals in a disk galaxy
possessing intrinsic density wave mode of spiral type.
The position of a typical star (such as our Sun)
inside the Corotation Radius ($r_{co}$, where the density wave and
the orbiting stars rotate with the same angular speed) is also marked.
Our Sun, which resides inside the corotation radius of the
Milky Way, overtakes the spiral density wave periodically
in its orbital motion around the center of the Milky Way.
(c) An observed spiral galaxy, NGC 5247, with two corotation circles
superimposed on grey-scale galaxy image.  The existence
of two corotations indicates that this galaxy has two resolved
nested density wave patterns, with the inner pattern  
 rotating faster than the outer spiral pattern.
(d) The potential-density
phase shift values calculated using the galaxy image, with the two
corotation radii positions indicated.
(adapted from Zhang 1996,2017; Zhang and Buta 2007).}
\label{Fig15}
\end{figure}

\pagebreak

\begin{figure} 
\vspace{12cm}
\includegraphics{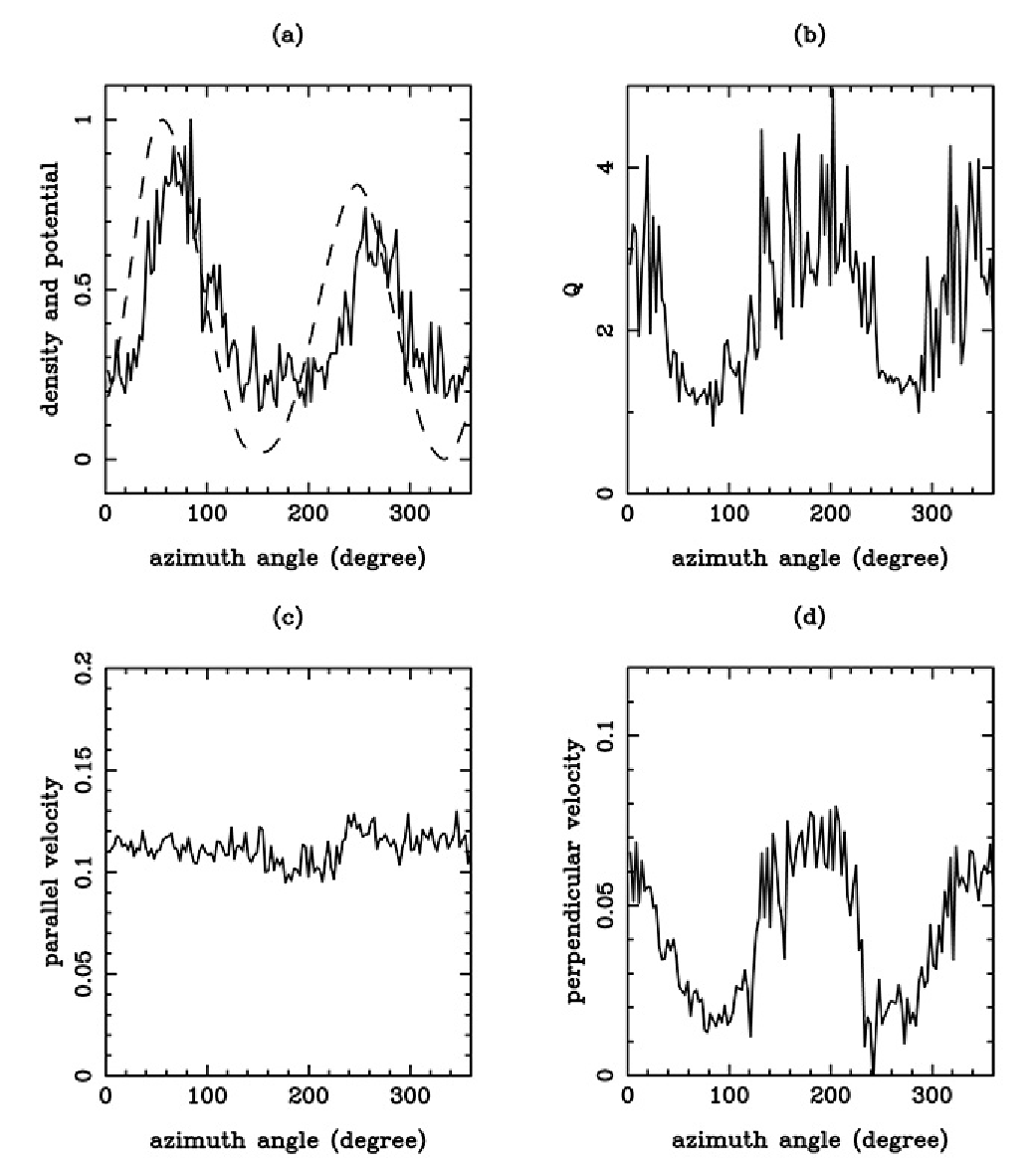}
\caption{Signature of spiral-arm collisionless shock in N-body 
simulation of a spiral galaxy. 
At a galactic radius inside the so-called corotation radius
(where the density wave spiral pattern rotate at the same speed
as the underlying disk matter), we plot the
the azimuthal distributions of the following parameters, respectively:
(a) Surface density (solid line) and negative potential (dashed line).
Note that the potential is phase-shifted in azimuth from the density.
(b) The gravitational instability indicator Q parameter.
Note the sharp decrease in Q parameter at the location of spiral
arms, indicating the formation of temporary local gravitational
instability, which generates an effective mean-free-path
that is about 1 kpc (or 3 thousand light years) in size
for the Solar neighborhood, about the width of the Milky Way
spiral arms.
(c) Velocity component parallel to the spiral arm.
(d) Velocity component perpendicular to the spiral arm. Note the sharp jump
from the supersonic to the subsonic velocity at the location of the
spiral arm, which is a clear indication of the gravitational
collisionless shock.  The sonic
velocity is about 0.04 in the normalized unit
for this particular simulation.
The two quasi-periodic cycles are due to the two-armed spiral pattern
which emerged in the simulation.  Adapted from Zhang (1996, 2017).
}
\label{Fig16}
\end{figure}

\clearpage

\begin{figure}
\vspace{8cm}
\includegraphics{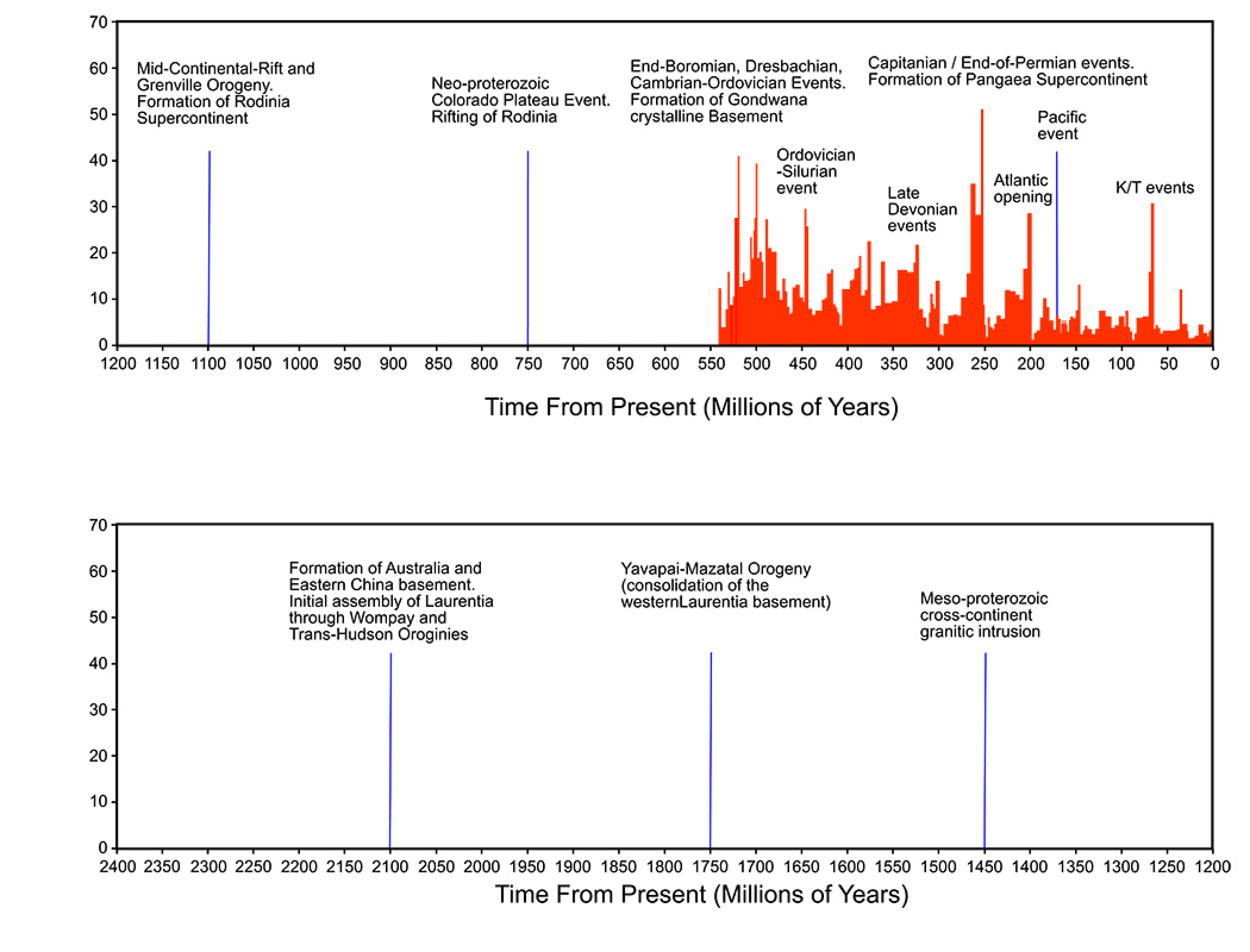}
\caption{Major extinction events in the Phanerozoic Eon of
the Earth's history (red histogram, based on
Extinction\_Intensity.svg on wikipedia, with source
papers: Raup and Sepkoski
(1982); Rohde and Muller (2005);
Sepkoski (2002); Signor and Lipps (1982)); as well as the average epochs of
other major plate-tectonic events
(blue lines, vertical scale and line thickness uncalibrated).}
\label{Fig17}
\end{figure}

\pagebreak

\begin{figure}
\vspace{18cm}
\includegraphics{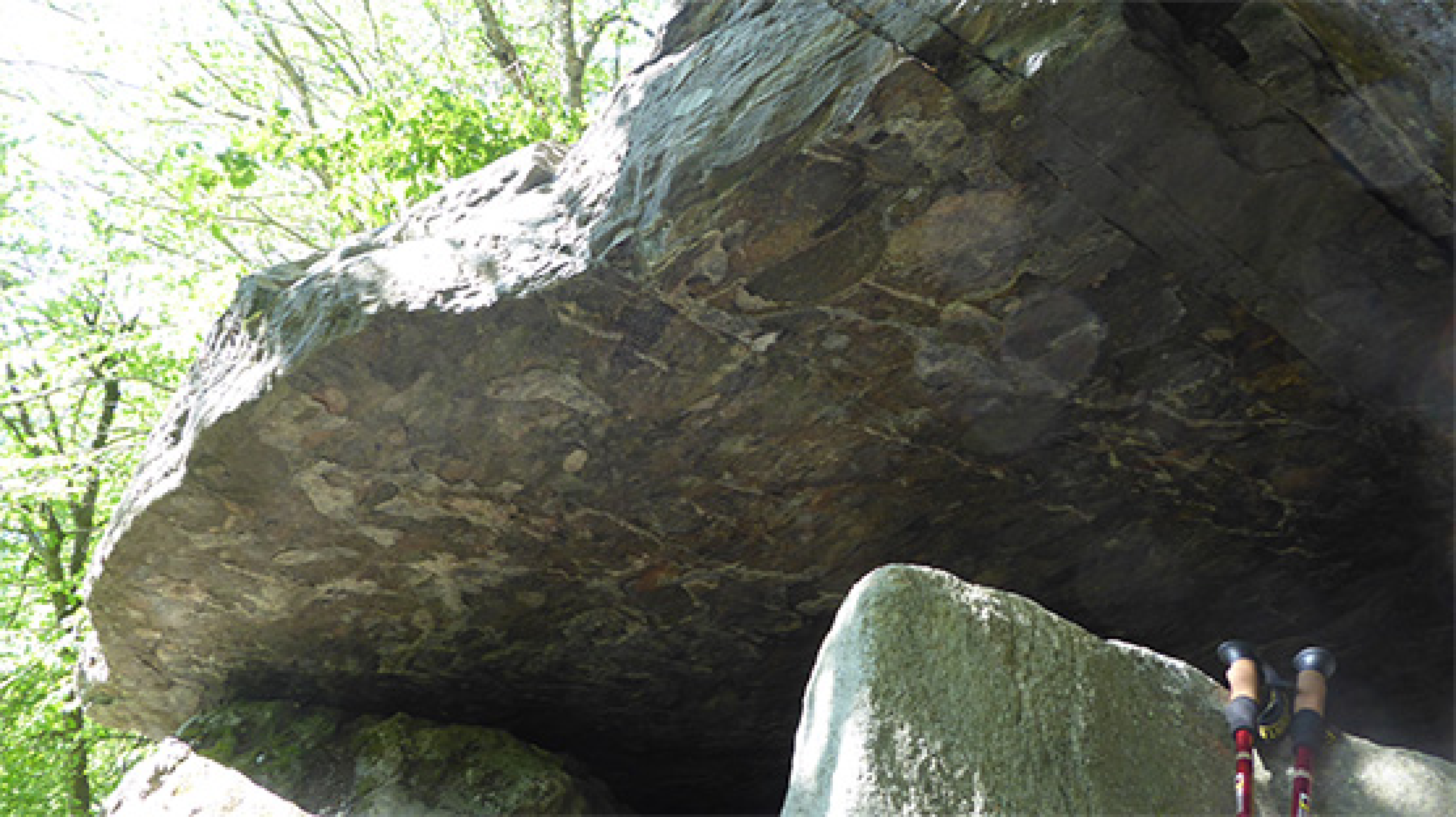}
\includegraphics{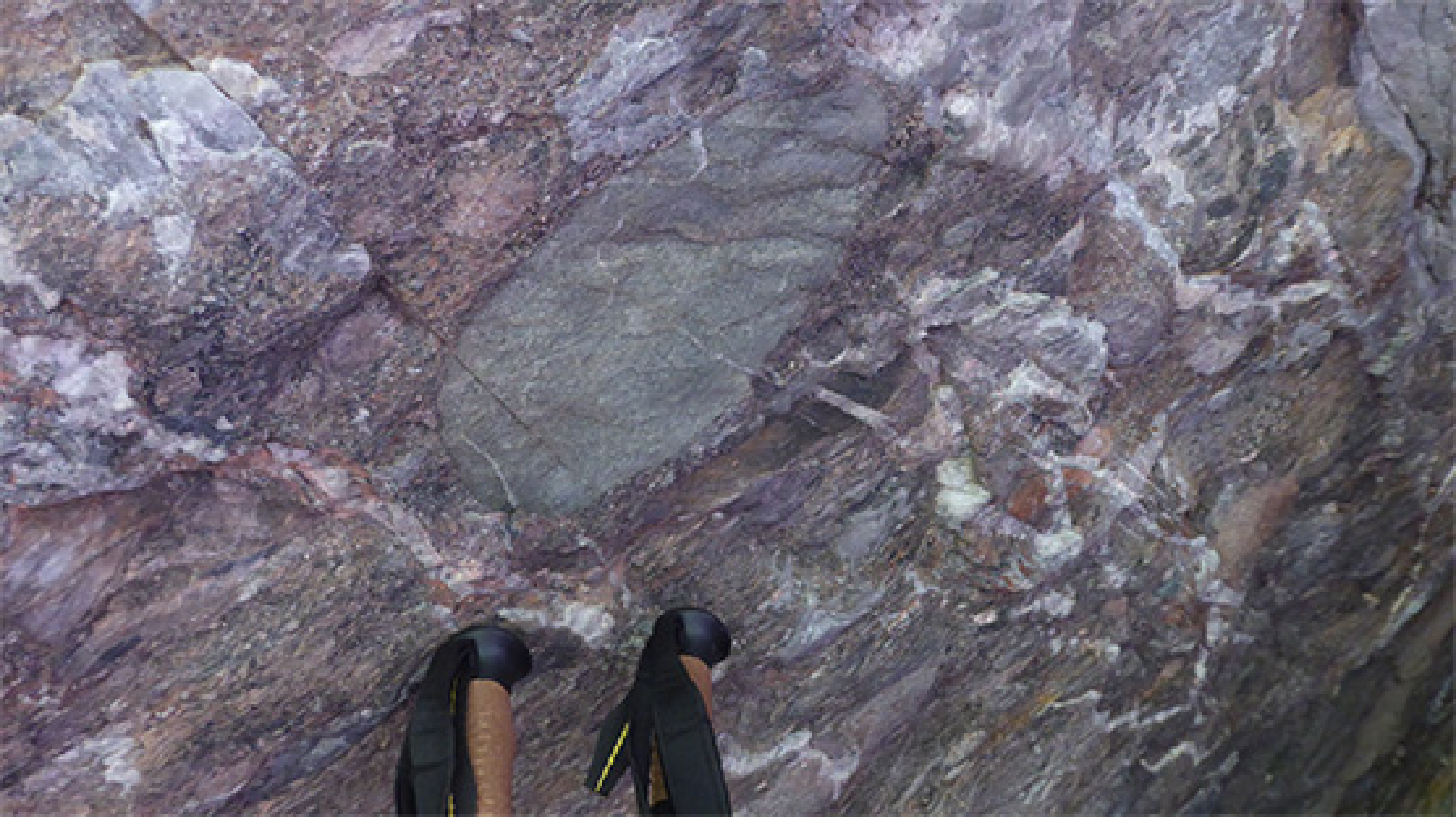}
\caption{Neoproterozoic suevite/conglomerate slab in Grandfather Mountain State Park.
Bottom frame is a close-up view.  Hiking poles for scale.}
\label{Fig18}
\end{figure}

\pagebreak

\begin{figure}
\vspace{18cm}
\includegraphics{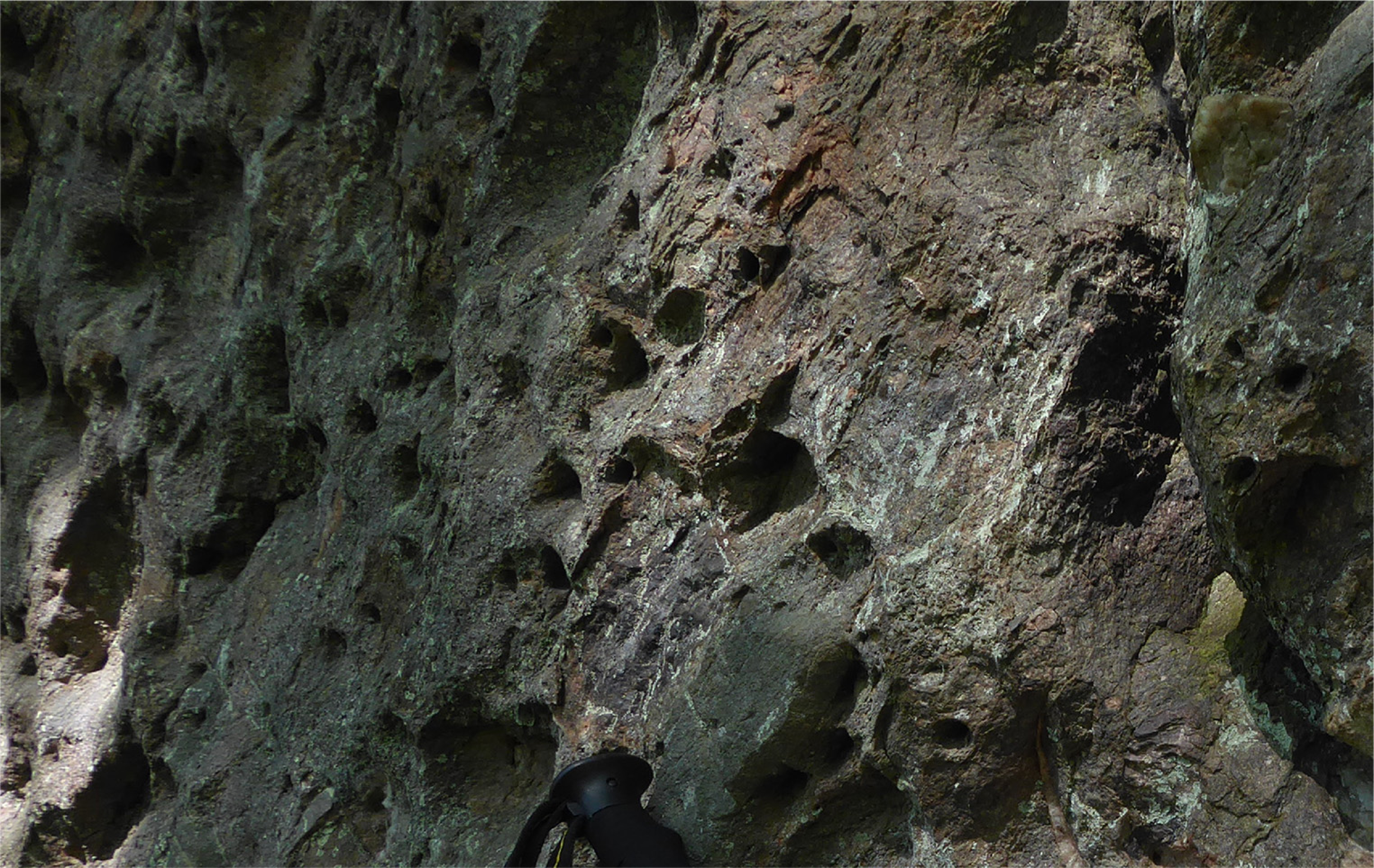}
\includegraphics{Fig19bCopy.eps}
\caption{Pockmarked and partially melted conglomerate rock wall in the 
Grandfather Mountain State Park, North Carolina. Hiking pole tip in foreground
for scale.} 
\label{Fig19}
\end{figure}

\clearpage

\begin{figure}
\vspace{10cm}
\includegraphics{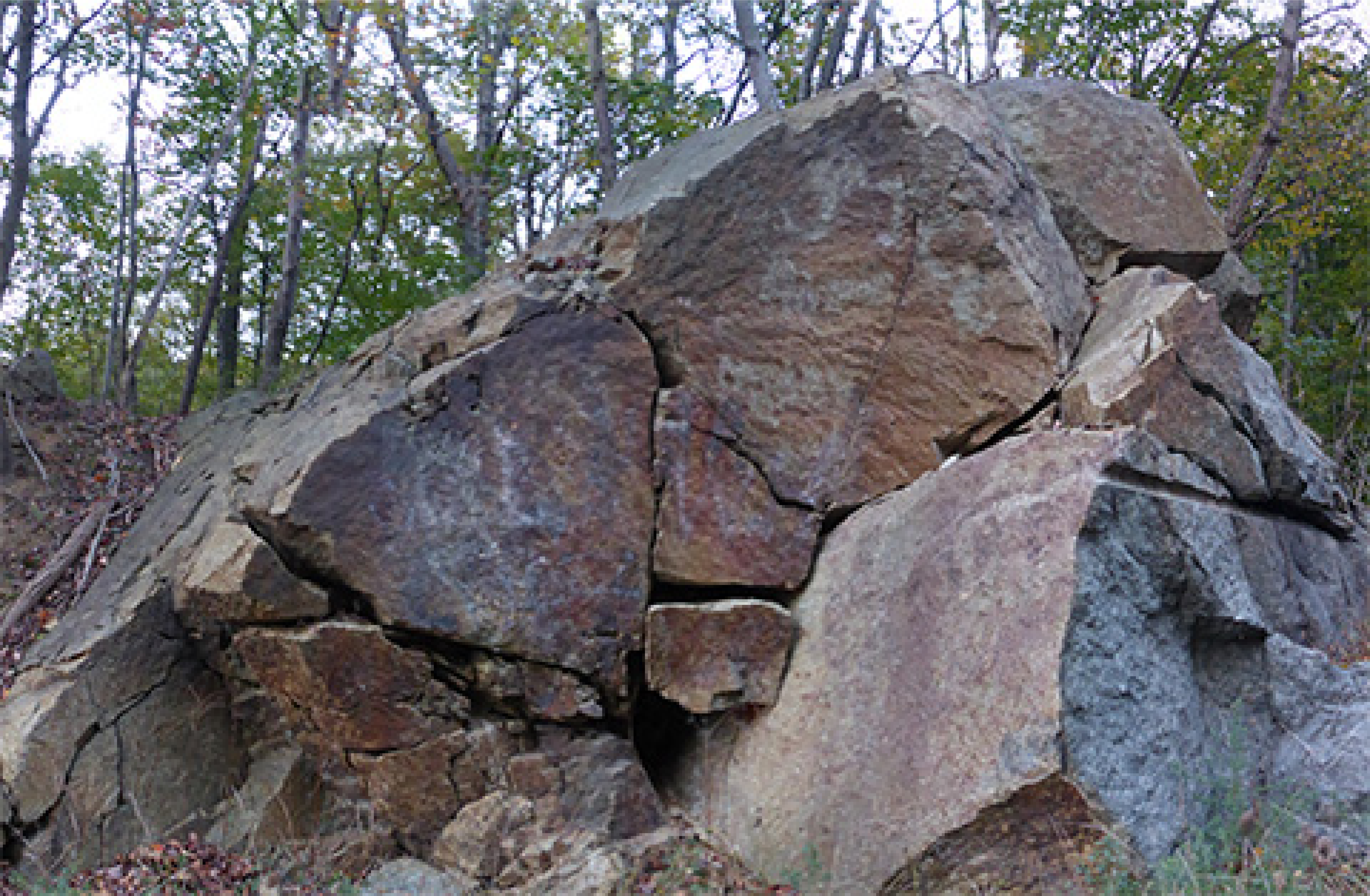}
\caption{Evidence of shock metamorphism near the entrance to the
Shenandoah National Park, off highway US-33 in Virginia.}
\label{Fig20}
\end{figure}

\clearpage

\begin{figure}
\vspace{18cm}
\includegraphics{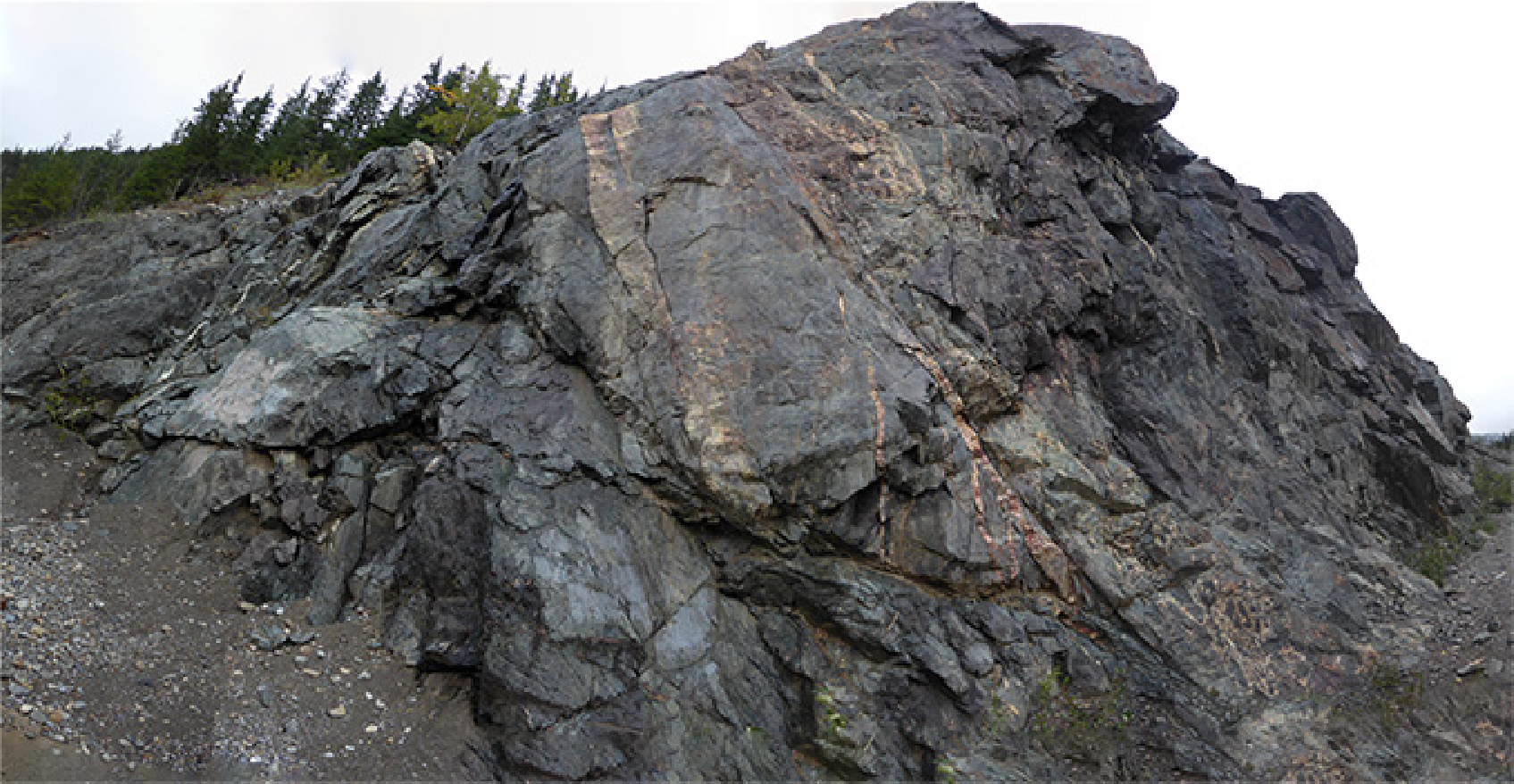}
\includegraphics{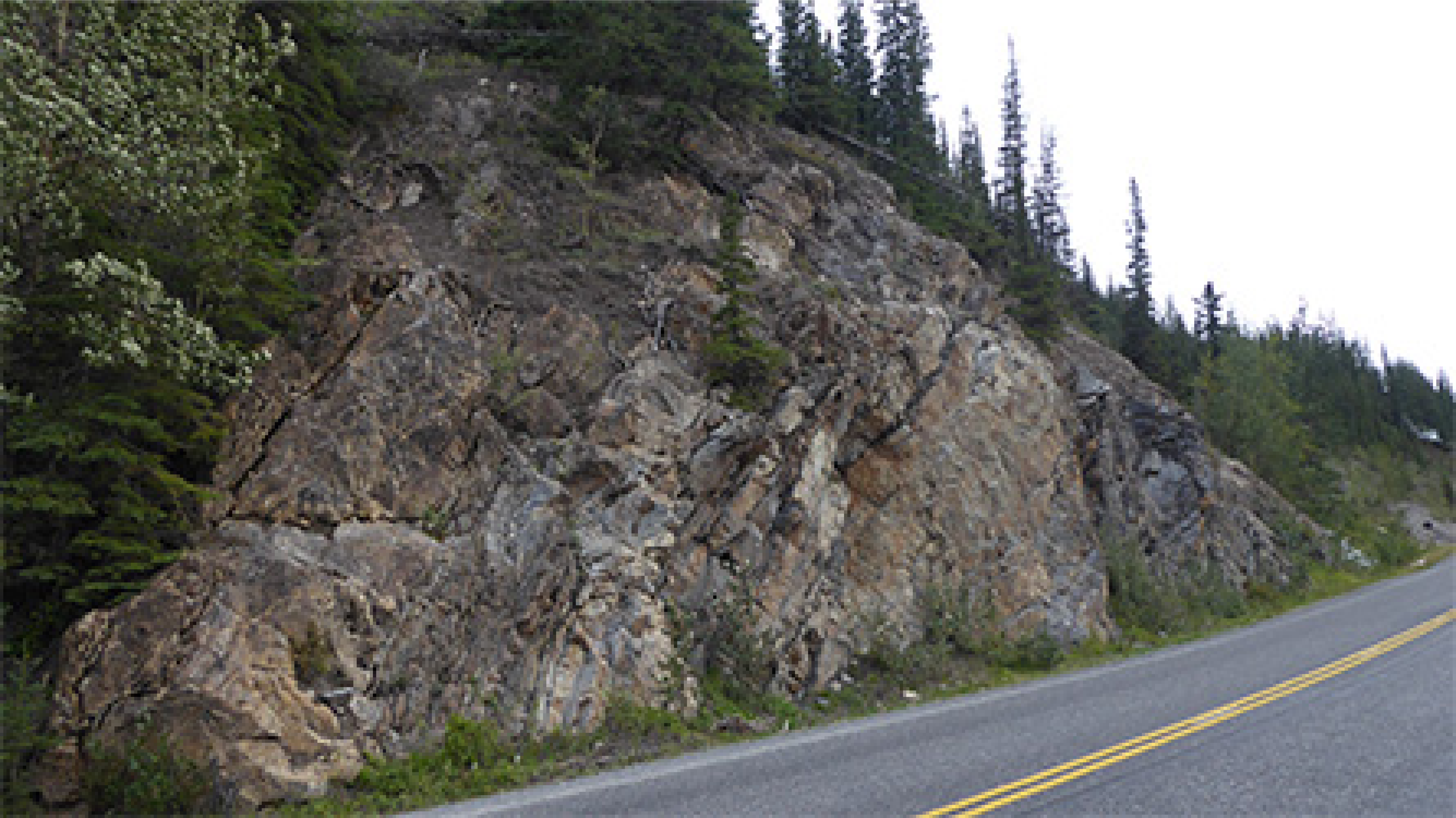}
\caption{Precambrian metamorphic rocks (bottom) and Neoproterozoic
igneous intrusion (top) near Toad River Bridge, Alaska Highway, Canada.} 
\label{Fig21}
\end{figure}

\clearpage

\begin{figure}
\vspace{10cm}
\includegraphics{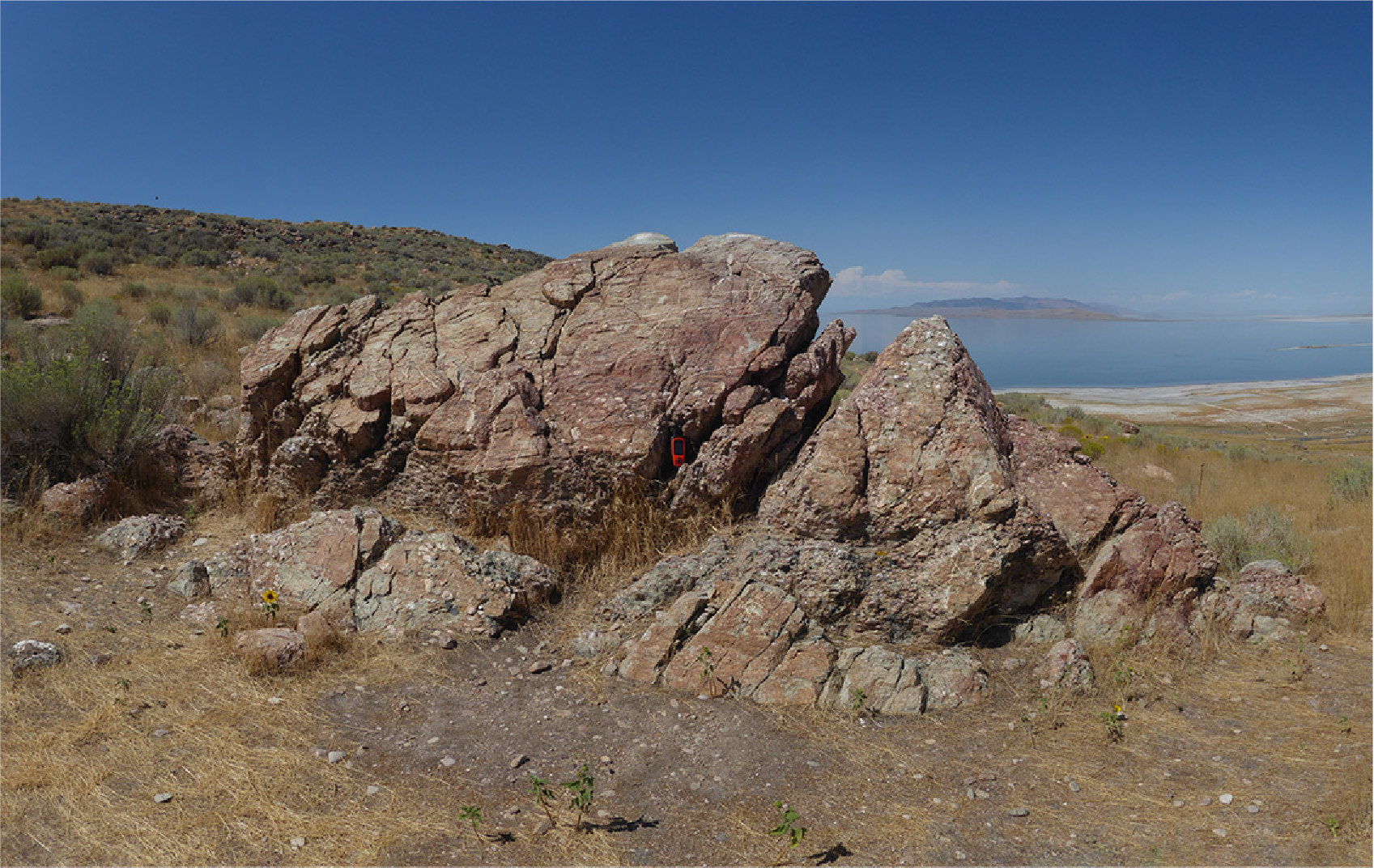}
\caption{Neoproterozoic conglomerate on Antelope Island within
the Great Salt Lake, Utah.  2-inch wide InReach communication
device (with red trim) at the center of image for scale.}
\label{Fig22}
\end{figure}

\clearpage

\clearpage
\begin{figure}
\vspace{10cm}
\includegraphics{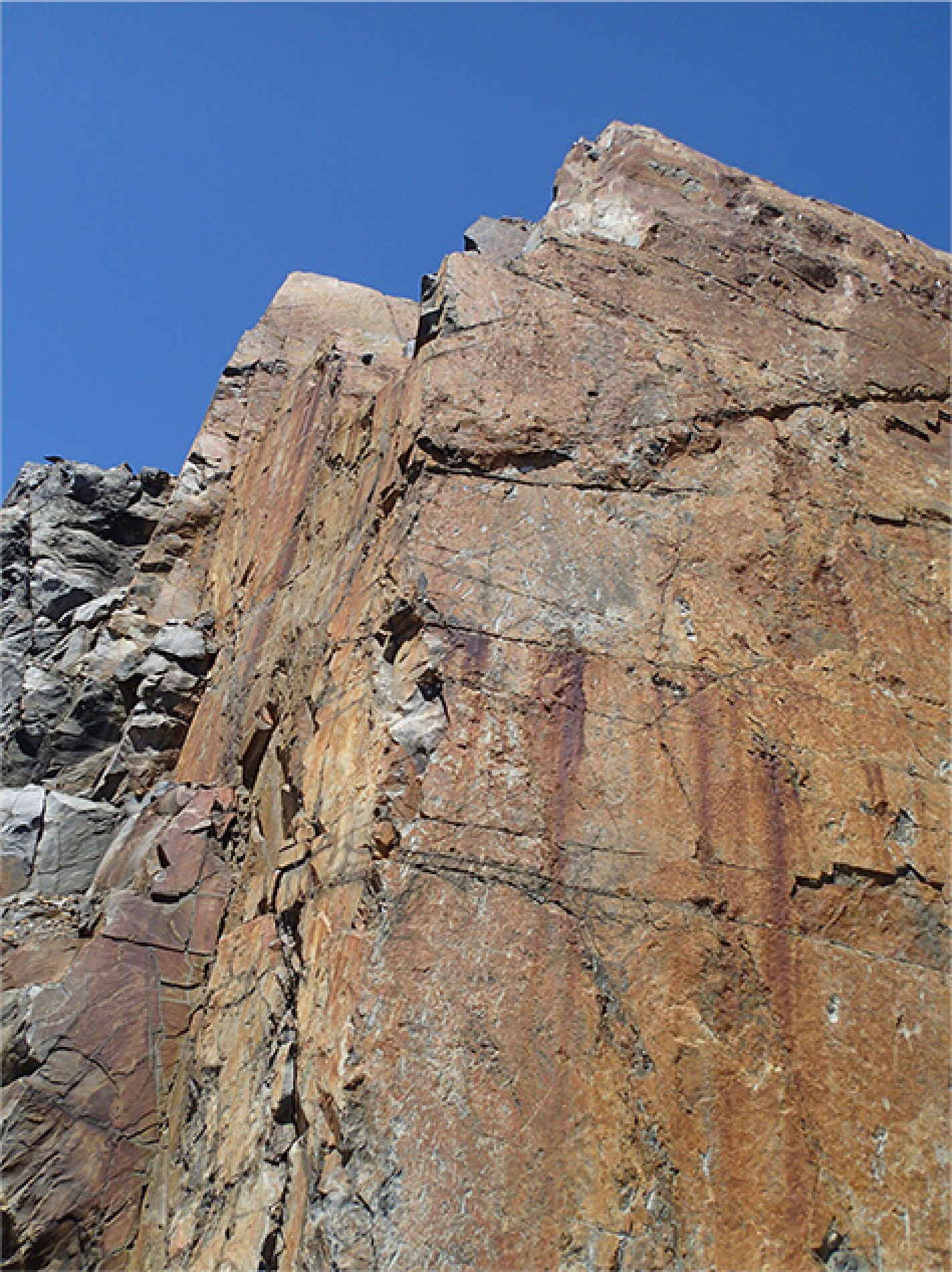}
\includegraphics{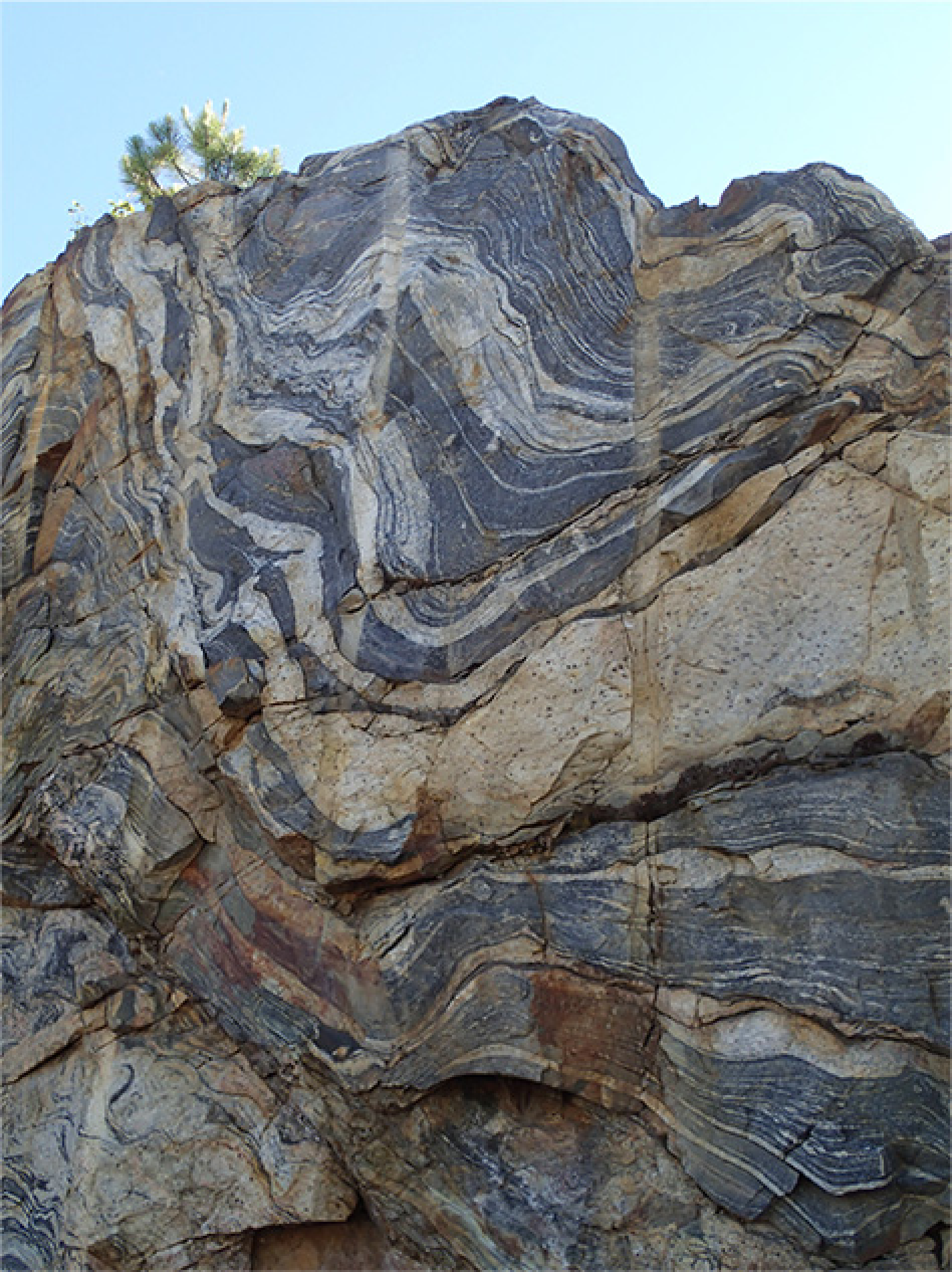}
\caption{Comparison between the 1.85 Ga impactite rocks (Left), and the 1.1 Ga Grenville-deformed rocks (Right).
Sudbury Impact Structure and Environs, Ontario, Canada.}
\label{Fig23}
\end{figure}
\end{document}